\pdfoutput=1
\documentclass[useAMS,usenatbib]{mn2e}
\usepackage{natbib}
\usepackage{graphicx}
\usepackage{times}
\usepackage{subfigure}
\usepackage{url}
\usepackage{amsmath}
\usepackage{amssymb}
\usepackage{longtable,lscape}
\usepackage{cite}
\usepackage{color}
\usepackage{placeins}
\usepackage{float}
\usepackage{gensymb}
\usepackage{hyperref}
\def\hii{\mbox{H\,{\sc ii}}}
\def\feii{\mbox{[Fe\,{\sc ii}]}}
\def\h2{H$_2$}
\def\tab{Table~}
\def\fig{Fig.~}
\def\g12{G12.42+0.50}

\def\sun{$_\odot$}
\def\um{$\rm \mu m$}
\def\arc{$\arcsec$}

\title[Extended green object G12.42+0.50]{Initial phases of high-mass star formation: A multiwavelength study towards the extended green object G12.42+0.50}

\author[Issac et al.]{Namitha Issac$^1$\thanks{E-mail: namithaissac.16@res.iist.ac.in}, Anandmayee Tej$^1$, Tie Liu$^{2,3}$, Watson Varricatt$^4$, Sarita Vig$^1$
\newauthor 
Ishwara Chandra C.H.$^5$, Mathias Schultheis$^6$ \\
$^1$Indian Institute of Space Science and Technology, Thiruvananthapuram 695 547, Kerala, India\\
$^2$Korea Astronomy and Space Science Institute, 776 Daedeokdaero, Yuseong-gu, Daejeon 34055, Republic of Korea \\
$^3$East Asian Observatory, 660 N. A\' ohoku Place, Hilo, HI 96720, USA \\
$^4$Institute for Astronomy, UKIRT Observatory, 660 N. A\' ohoku place, Hilo, HI 96720, USA \\
$^5$National Centre for Radio Astrophysics (NCRA-TIFR), Pune, India \\
$^6$Laboratoire Lagrange, Universit\' e C\^ote d'Azur, Observatoire de la C\^ote d'Azur, CNRS, Blvd de l'Observatoire, F-06304 Nice, France \\}

\begin{document}

\date{}

\pagerange{\pageref{firstpage}--\pageref{lastpage}} \pubyear{}

\maketitle

\label{firstpage}

\begin{abstract}
We present a multiwavelength study of the extended green object, {\g12} in this paper. The associated ionized, dust, and molecular components of this source are studied in detail employing various observations at near-, mid- and far-infrared, submillimeter and radio wavelengths. Radio continuum emission mapped at 610 and 1390~MHz, using the Giant Meterwave Radio Telescope, India, advocates for a scenario of coexistence of an UC {\hii} region and an ionized thermal jet possibly powered by the massive young stellar object, IRAS 18079-1756 with an estimated spectral type of $\rm B1-B0.5$. Shock-excited lines of {\h2} and [FeII], as seen in the near-infrared spectra obtained with UKIRT-UIST, lend support to this picture. Cold dust emission shows a massive clump of mass 1375~M{\sun} enveloping {\g12}. Study of the molecular gas kinematics using the MALT90 and JCMT archival data unravels the presence of both infall activity and large-scale outflow suggesting an early stage of massive star formation in {\g12}. A network of filamentary features are also revealed merging with the massive clump mimicking a hub-filament layout. Velocity structure along these indicate bulk inflow motion. 
\end{abstract}


\section{INTRODUCTION}
Massive stars ($\rm M \gtrsim 8 M_{\odot}$) play a vital role in the evolution of the universe given their radiative, mechanical and chemical feedback. 
They dictate the energy budget of the galaxies through powerful radiation, strong winds and supernovae events. Despite this, most aspects of the processes 
involved in their formation are far less understood in contrast to the low-mass regime. A universal theory elucidating the formation mechanism across 
the mass range, though much sought after, is still not well established. Tremendous efforts have been going on, since the last decade or so, to investigate whether 
high-mass star formation can be understood as a `scaled-up' version of the processes involved in the low-mas domain via the {\it Core Accretion} hypothesis. 
This advocates for formation of high-mass stars from pre-stellar cores to form single or binary protostars with enhanced accretion via a rotationally
supported disk that also launches protostellar outflows. This model adequately circumvents the `radiation pressure problem', while leaving many 
questions open regarding the timescales of collapse and fragmentation in massive cores. Alternate theories, like {\it Competitive Accretion} and 
{\it Protostellar Merger}, have also been proposed as viable mechanisms. The debate is still not sealed on the preferred mechanism and the influence of prevailing conditions on each. 
On the observational front, probing the early stages of massive star formation, in particular, remains a challenge. Rarity of sources (owing to fast evolutionary time scales), 
large distances, complex, embedded and influenced environment and high extinction are factors which hinder the building up of a proper observational 
database crucial for validating proposed theories. The current status on the theoretical and observational scenarios of high-mass star formation can be found in the excellent reviews by 
\citet{2014prpl.conf..149T} and \citet{2018NatAs...2..478M} which also give an  update on the literature in this field.

%%%%%%%%%%%%%%%%%%%%%%%% irac 2mass rgb %%%%%%%%%%%%%5
\begin{figure*}
\hspace*{-0.6cm}
\centering 
\includegraphics[scale=0.47]{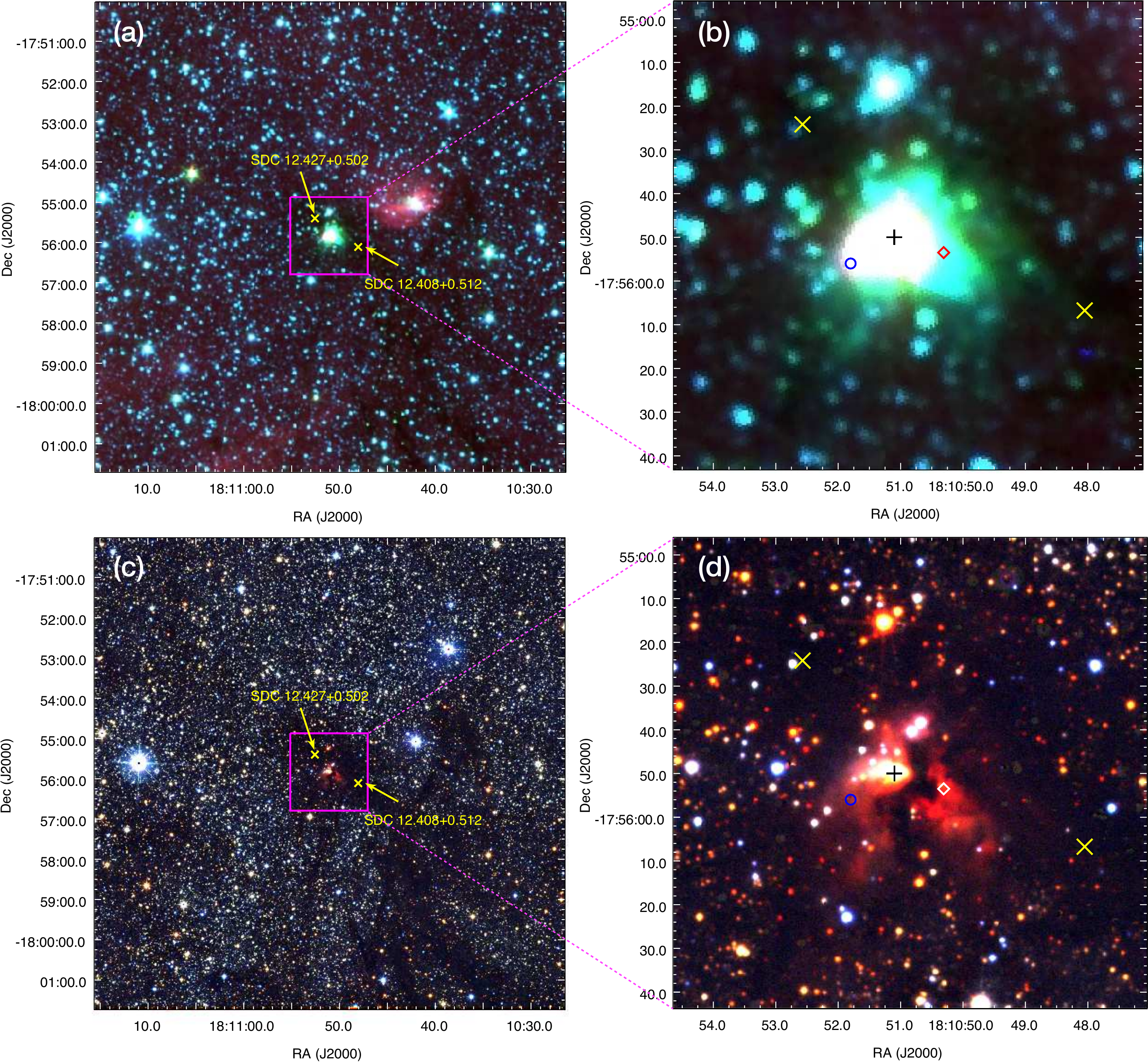} 
\caption{(a) Colour composite image of the region around {\g12} using IRAC 3.6~{\um} (blue), 4.5~{\um} (green) and 8.0~{\um} (red) bands. (b) A zoom-in showing the EGO {\g12}. IRDCs are shown with the `$\times$' symbol and the position of IRAS~18079-1756 associated with {\g12} is indicated with a diamond mark. The cross marks the position of the 2MASS point source, J18105109-1755496. The location of the $\rm H_2O$ maser is shown as a blue circle. (c) and (d) are colour composites created from the UKIDSS $J$ (1.25~{\um}), $H$ (1.63~{\um}) and $K$ (2.20~{\um}) band data, covering the same area as in (a) and (b), respectively.}
\label{irac_ukidss_rgb}
\end{figure*}
%%%%%%%%%%%%%%%%%%%%%%%%%%%%%%%%%%%%%%%%%%%%5

A step towards strengthening the observational domain was taken when the large-scale {\it Spitzer} Galactic Legacy Infrared Mid-Plane Survey Extraordinaire (GLIMPSE) \citep{2003PASP..115..953B} unfolded the presence of a 
significant population of objects displaying enhanced and extended emission in the IRAC 4.5~{\um} band. Following the conventional colour coding of 
GLIMPSE colour-composite images, these objects were christened as `green fuzzies' or `extended green objects' (EGOs) by \citet{2009ApJS..181..360C} 
and \citet{2008AJ....136.2391C}. Post detection, several studies focussed on ascertaining the nature of these objects and the research towards identification of the 
spectral carriers responsible for the enhanced 4.5~{\um} emission were initiated \citep{{2004ApJS..154..333M},{2004ApJS..154..352N},{2005ApJ...630L.181R},{2005MNRAS.357.1370S},
{2006AJ....131.1479R},{2007MNRAS.374...29D},{2009ApJS..181..360C},{2009ApJ...702.1615C},{2010AJ....140..196D},{2011ApJ...743...56C},{2011ApJ...729..124C},
{2012ApJ...748....8T},{2012ApJS..200....2L},{2013ApJS..206...22C}}.
Some of the above studies have associated EGOs with shock-excited $\rm H_2$ line and/or CO bandhead emission in protostellar outflows. In addition, 
observations show that majority of EGOs are
co-located with infrared dark clouds (IRDCs) and with Class II Methanol masers which are distinct signposts of massive star formation \citep{{2006ApJ...641..389R},
{2007ApJ...662.1082R},{2005A&A...434..613S},{2006ApJ...638..241E}}. Studies till date support a picture wherein EGOs can be regarded as candidates for outflows from massive YSOs (MYSOs) and hence, they offer a unique sample of sources to investigate the early phases of massive star formation.

\par In this paper, we focus on the EGO, {\g12} catalogued as a ``possible" outflow candidate and associated with the luminous infrared source, IRAS~18079-1756
\citep{2008AJ....136.2391C}. The kinematic distance ambiguity towards {\g12} has been resolved by \citet{2012ApJS..202....1H}. Following them and \citet{2010ApJ...710..150C}, we adopt a distance of 2.4~kpc in our study. \citet{2013ApJ...765..129V} estimate the far-infrared (FIR) luminosity, from the IRAS fluxes, to be $\rm \sim 10^{4} L_{\odot}$. From literature, {\g12} is designated as an ultracompact (UC) {\hii} \citep{{1984ApJ...281..225J},{2007ApJ...669L..37W}}. 
{\g12} has been observed as part of several surveys such as the Millimeter Astronomy Legacy Team 90~GHz (MALT90) survey 
\citep{{2013PASA...30...57J},{2011ApJS..197...25F}}  
and the 6~cm Red MSX Source survey by \citet{2009A&A...501..539U}. The latter was aimed at identifying candidate MYSOs. 
Apart from this, a millimeter study of southern IRAS sources by 
\citet{1997ApJS..110...71O} reports IRAS~18079-1756 as an outflow candidate from the red- and blueshifted molecular outflow features observed 
in the $\rm CO~(2-1)$ transition and a redshifted line dip in the $\rm CS~(2-1)$ transition. $\rm H_2O$ maser emission is detected towards {\g12} \citep{{1981ApJ...250..621J},{2013ApJ...764...61C}}.
\citet{2011ApJS..196....9C}, in their study, have identified a 95~GHz Class I methanol maser towards {\g12}. In addition, a few molecular line surveys also include {\g12} \citep{{2003ApJS..149..375S},{2013ApJ...764...61C}}.

\par In {\fig}\ref{irac_ukidss_rgb}, we present the near-infrared (NIR) and the mid-infrared (MIR) colour-composite images of the field of {\g12} developed from the UKIDSS (Section \ref{ukidss}) and {\it Spitzer}-IRAC (Section \ref{mir}) data, respectively.
The images not only reveal the characteristic, extended and enhanced 4.5~{\um} emission defining the EGOs, but also show extended 
$K$-band nebulosity associated with {\g12}. The morphology of the $K$-band emission is more confined to a narrower north-east and south-west
stretch with a distinct dark lane in-between. A network of filamentary structures are seen towards the south-west and west, being more prominent 
in the NIR colour composite image. These filaments seem to converge towards {\g12} suggesting a `hub-filament' scenario. Such systems have been detected in other star forming complexes and discussed in various studies \citep{{2013A&A...555A.112P},{2018ApJ...852...12Y}}. Two IRDCs, SDC 12.427+0.502 and SDC 12.408+0.512 from the catalogue of {\it Spitzer} 
dark clouds by \citet{2009A&A...505..405P}, are seen to lie on either side of {\g12} and marked on the images in {\fig} \ref{irac_ukidss_rgb}. 

\par In presenting the multiwavelength study towards the {\g12}, we have organized the paper in the following way. Section \ref{obs} outlines the observations and data reduction details, along with the archival databases used for this study. Section \ref{results} deals with the various results obtained. In Section \ref{discussion} we discuss the results, where we explore different scenarios to explain the nature of the radio continuum emission and elaborate on the gas kinematics. The summary of this comprehensive study is compiled in Section \ref{summary}.

\section{OBSERVATIONS AND ARCHIVAL DATA} {\label{obs}}

\subsection{Radio continuum observation}

The ionized emission associated with {\g12} is probed at low radio frequencies of 610 and 1390~MHz using the Giant Metrewave Radio Telescope (GMRT)
located at Pune, India. GMRT consists of an array of 30 antennas, each of diameter 45~m and arranged in a Y-shaped configuration. 
The central square consists of 12 antennas spread randomly over an area of 1~km$^{\text 2}$ with the shortest baseline being $\sim$100~m. 
The remaining 18 antennas are uniformly stretched along the three arms ($\sim$14~km each) providing the longest baseline of $\sim$25~km. 
This hybrid configuration enables radio mapping of small-scale structures at high-resolution, along with large-scale, diffuse emission at low-resolution.  

\par Our GMRT observations were carried out on 2017 August 22 and 2017 July 21 at 1390 and 610~MHz, respectively, 
with a bandwidth of 32~MHz over 256 channels. We selected the radio sources 3C286 and 3C48 as primary flux calibrators. The phase calibrators, 1911-201 (at 1390~MHz) and 1822-096 (at 610~MHz) were observed after each 40-mins scan of the target, to calibrate the phase and amplitude variations over the full observing run.
The details of the observations are given in {\tab}\ref{radio_obs}. Data reduction is performed using the NRAO Astronomical Image Processing Software (AIPS).
%%%%%%%%%%%%%%%%%%%%%%%%%%%%%%%%%%%
\begin{table}
\caption{Details of GMRT observations towards {\g12}}
\begin{center}
\centering
\begin{tabular}{l l l} \hline \hline \
Details 				 & 610~MHz				& 1390~MHz \\
\hline \
Date of Obs.		& 21 July 2017 	& 22 August 2017 \\
Flux calibrators  & 3C286  	&	3C286, 3C48 \\
Phase calibrators 	& 1822-096 &1911-201 \\
Integration time & $\sim 5$~hrs & $\sim 5$~hrs \\
Synthesized beam &  $\sim7.6'' \times 4.8''$  &   $\sim 3.0'' \times 2.4''$ \\ 
{\it rms} noise ($\mu $Jy/beam) &  94    & 29.7   \\
\hline \
\end{tabular}
\label{radio_obs}
\end{center}
\end{table}
%%%%%%%%%%%%%%%%%%%%%%%%%%%%%%%%%%%%
The data sets are carefully examined to identify bad data (non-working antennas, bad baselines, RFI, etc) using the tasks {\tt UVPLT}, and {\tt TVFLG}. 
Subsequent flagging of the bad data is performed using the tasks {\tt UVFLG}, {\tt TVFLG} and {\tt SPFLG}. 
After flagging, the gain and bandpass calibration is carried out following standard procedure. Channel averaging was restricted to keep the bandwidth smearing negligible. The calibrated and channel averaged data are cleaned and deconvolved using the task {\tt IMAGR} by adopting the wide-field imaging procedure (``3D" imaging) to account for the w-term effect. Primary beam correction is done using the task {\tt PBCOR}. 

\par Galactic diffuse  emission contributes to the system temperature, which becomes relevant at low frequencies (especially at 610~MHz). 
Since our target source is close to the Galactic plane and the flux calibrators are located away from this plane, rescaling of the final images becomes essential.
The scaling factor is estimated under the assumption that the Galactic diffuse emission follows a power-law spectrum. The sky temperature, ${T_{sky}}$ at frequency,
$\nu$ was determined using the equation
\begin{equation}
T_{sky} = T_{sky}^{408}\bigg(\frac{\nu}{408~\textrm{MHz}}\bigg)^\gamma
\end{equation}
\noindent
where $\gamma$ is the spectral index of the Galactic diffuse emission and is taken to be -2.55 \citep{1999A&AS..137....7R} and $\it{T_{sky}^{\rm 408}}$ 
is the sky temperature at 408~MHz obtained from the all-sky 408~MHz survey of \citet{1982A&AS...47....1H}. We estimate the scaling factors to be 1.25 and 2.46 at 1390 and 610~MHz, respectively. These values are used to rescale our images.

\subsection{Infrared observations}

\subsubsection{Spectroscopy} \label{spectroscopy_steps}
NIR spectroscopic observations towards {\g12} were carried out with the 3.8-m United Kingdom Infrared Telescope (UKIRT), Hawaii.
Observations were taken with the UKIRT 1-5~{\um} Imager Spectrometer (UIST, \citealt{2004SPIE.5492.1160R}). UIST consists of a 1024$\times$1024 InSb array. 
In spectroscopy mode, the camera with a plate scale of 0.12{\arc}~pixel$^{-1}$ was used. The observations were made using the 4 ($\sim$0.48\arc)-pixel-wide 
and 120{\arc} long slit. Spectra were obtained in two grism set-ups, namely $HK$ and $KL$ that cover the spectral range of $1.395-2.506$~{\um} and $2.229-2.987$~{\um}, with spectral resolution of 500 and 700, respectively. Flat-field and Argon arc lamp observations were made ahead of the of the target observations on each night. 
The slit was oriented at an angle of 55$^{\circ}$ east of north centred on {\g12} 
($\rm \alpha_{J2000}= 18^{h}10^{m}51.1^s, \delta_{J2000} = -17\degree 55\arcmin 50\arcsec$) so as to also sample the outflow-like, extended feature towards 
the south-west of the central bright source. The telluric standard, SA0 160915, an A0V type star, was observed for telluric and instrumental corrections. Since the target has an extended morphology, nodding along the slit would result in overlapping features. Thus, the science target observation was performed by nodding between the target and a blank sky position. However,
for the standard star, the source was nodded along the slit in an ABBA pattern between two positions A and B along the slit \citep{{2009MNRAS.397..849P},{2009MNRAS.399.2165R}}. Details of the observation are given in {\tab}\ref{UKIRT_obs_spec}. 

%%%%%%%%%%%%%%%%%%%%%%%%%%%%%%%%%%%
\begin{table}
\caption{Details of UKIRT-UIST spectroscopic observations towards {\g12}}
\begin{center}
\centering
\begin{tabular}{c c c c c} \hline \hline \
Date~&Grism~& Exposure  & Integration  &Standard star \\
(yyyymmdd)	&		&time (s)			&time (s)      \\
\hline \	\
20150402		&HK		&120 &720	& SAO 160915 \\
~~20150405		&KL		&50	&300	& SAO 160915 \\
\hline \
\end{tabular}
\label{UKIRT_obs_spec}
\end{center}
\end{table}
%%%%%%%%%%%%%%%%%%%%%%%%%%%%%%%%%%%%
\par The initial data reduction is carried out by the {\tt ORAC-DR} pipeline at UKIRT. Subsequent reductions are carried out using suitable tasks from the Starlink packages, {\tt FIGARO} and {\tt KAPPA} \citep{2008ASPC..394..650C}. 
The spectra from the two nodded beams of the reduced spectral image of the standard star are extracted and averaged after bad-pixel masking and flat-fielding. 
The averaged spectrum is then wavelength calibrated using the observed Argon arc spectrum. Photospheric lines are removed from the standard star spectrum after a careful interpolation of the continuum across these lines. The standard star spectrum is then divided by a blackbody spectrum of the temperature similar to the photospheric temperature of the standard star. Since our target involves extended emission, sky subtraction is done by subtracting the off-slit sky from the on-target spectral image followed by the FIGARO task {\tt POLYSKY} that subtracts the residual sky. Subsequent to this, correction for telluric lines is achieved by dividing the bad-pixel masked, flat-fielded, sky-subtracted and wavelength calibrated target spectral image by the standard star spectral image.  As the sky conditions were not photometric, we have not performed the flux calibration. 

\subsubsection{Imaging} \label{cont_sub}

We imaged {\g12} in the broad-band {\it H} filter and the narrow-band filter centred on the {\feii} line at 1.644~{\um} using the UKIRT Wide-Field Camera (WFCAM, \citealt{2007A&A...467..777C}). The WFCAM consists of four 2048 $\times$ 2048 HgCdTe Rockwell Hawaii-II arrays each with a field of view of 13.65$\arcmin$ $ \times $ 13.65$\arcmin $ and a pixel scale of 0.4{\arc} pixel$^{-1}$. Details of the imaging observations are included in Table \ref{UKIRT_obs_img}.
The data reduction was carried out by the Cambridge Astronomical Survey Unit (CASU). 
%%%%%%%%%%%%%%%%%%%%%%%%%%%%%%%%%%%
\begin{table}
\caption{Details of UKIRT-WFCAM imaging observations made towards {\g12}}
\begin{center}
%\hspace*{-0.5cm}
\centering
\begin{tabular}{c c c c } \hline \hline \
Date~& Filter~& Exposure  & Integration  \\
(yyyymmdd)	&		&time (s)			&time (s)    \\
\hline \	\
20170705		&{\it H}	& 5	&	180 \\
~~20170705		&{\feii}		& 40		&	1440 \\

\hline \
\end{tabular}
\label{UKIRT_obs_img}
\end{center}
\end{table}
%%%%%%%%%%%%%%%%%%%%%%%%%%%%%%%%%%%%

\par Continuum subtraction of the narrow-band {\feii} is performed following the steps described in \citet{2005MNRAS.359....2V} employing multiple Starlink packages. The sky background was fitted and removed from the images using the KAPPA tasks {\tt SURFIT} and {\tt SUB}. The {\feii} and {\it H}-band images are aligned using the task {\tt WCSALIGN}. Since the seeing conditions were different for the {\feii} and {\it H}-band observations, the image with lower point spread function (PSF) was smoothed to the full width at half maximum (FWHM) of the image with larger PSF. For scaling the broad-band image, sky-subtracted flux counts of discrete point sources in both the narrow-band and broad-band images were measured. The average value of the ratio of the fluxes ($H/${\feii}) was computed and used to scale the {\it H}-band image. Subsequently, the scaled {\it H}-band image was subtracted from the {\feii} image to construct the continuum subtracted image.

\subsection{NIR data from UWISH2 and UKIDSS survey} \label{ukidss}
The UKIRT Widefield Infrared Survey for {\h2} (UWISH2) is a 180 square degree survey of the Galactic Plane to probe the 1-0 S(1) ro-vibrational line of {\h2} 
($\lambda$ = 2.122~{\um}) \citep{2011MNRAS.413..480F}. This survey used the WFCAM at UKIRT. CASU processed data of the region associated with {\g12} was retrieved. We have also used 
the {\it K}-band image obtained as a part of the UKIRT Infrared Deep Sky Survey Galactic Plane Survey (UKIDSS-GPS, \citealt{2008MNRAS.391..136L}) 
from the WFCAM Science Archive. Continuum subtraction of the {\h2} image is carried out following the same procedures detailed in Section \ref{cont_sub}. 

\subsection{MIR data from the Spitzer Space Telescope and the Midcourse Space Experiment} \label{mir}

In order to probe the emission at the MIR bands, we made use of the images of the region around {\g12} from the archives of the {\it Spitzer Space Telescope} and the images from the Midcourse Space Experiment (MSX) survey \citep{2001AJ....121.2819P}. The Infrared Array Camera 
(IRAC) is one of the instruments on the {\it Spitzer Space Telescope} that has simultaneous broadband imaging capability at 3.6, 4.5, 5.8 and 8.0~{\um} with angular resolutions $\sim$ 2.0{\arc} \citep{2004ApJS..154...10F}. The MSX survey mapped the Galactic plane in four mid-infrared spectral bands, 8.28, 12.13, 14.65, and 
21.3~{\um} at a spatial resolution of $ \sim $ 18.3{\arc}.
In order to investigate the physical properties of the dust core associated with {\g12}, we use the 12.13 and 14.65~{\um} MSX band images, and the level-2 PBCD 8.0~{\um} image of the GLIMPSE survey.

\subsection{FIR data from Hi-GAL survey}
FIR data used to study the nature of the cold dust emission was retrieved from the archives of the {\it Herschel Space Observatory}. This is a 3.4-m telescope that covers the spectral regime of $55-671$~{\um} \citep{2010A&A...518L...1P}. We use the level 2 processed 
images from the Photodetector Array Camera and Spectrometer (PACS, \citealt{2010A&A...518L...2P}) and Spectral and Photometric Imaging Receiver 
(SPIRE, \citealt{2010A&A...518L...3G}) observed as a part of the Herschel infrared Galactic plane survey (Hi-GAL, \citealt{2010PASP..122..314M}). 
The Hi-GAL observations were carried out in `parallel mode’ covering 70, 160~{\um} (PACS) as well as 250, 350 and 500~{\um} (SPIRE). 
The images have resolutions of 5,13, 18.1, 24.9 and 36.4 {\arc} at 70, 160, 250, 300, and 500~{\um}, respectively. 

\subsection{APEX+Planck data}
The APEX+Planck image is a combination of the APEX Telescope Large Area Survey of the Galaxy (ATLASGAL) \citep{2009A&A...504..415S} at 870~{\um}, which used the LABOCA bolometer array and the 850~{\um} map from the Planck/HFI instrument. The combined data covers emission at larger angular scales, thus revealing the structure of the cold Galactic dust in more detail \citep{2016A&A...585A.104C}.  The combined image has an angular resolution of 21{\arc}.

\subsection{SMA observation}
{\g12} was observed using the Submillimeter Array (SMA) on 2008 July 1 and 8 in its extended configuration. The phase reference center was $\rm \alpha_{J2000}= 18^{h}10^{m}51.8^s, \delta_{J2000} = -17\degree 55\arcmin 56\arcsec$).  In both observations, QSO 1924-292 was observed for gain correction and Callisto was used for flux-density calibration. The absolute flux level is accurate to about 15\%. Bandpass was corrected by observing QSO 3C454.3. The 345 GHz receivers were tuned to 267~GHz for the lower sideband and 277~GHz for the upper sideband. The frequency spacing across the spectral band is 0.812~MHz or $\rm \sim0.9~km~s^{-1}$. The 1.1 mm continuum data were acquired by averaging all the line-free channels over both the upper and lower spectral bands in the two datasets. The visibility data are calibrated with the IDL superset MIR package and imaged with the MIRIAD\footnote{\url{http://admit.astro.umd.edu/miriad/}} package.  The MIRIAD task {\tt SELFCAL} is employed to perform self-calibration on the continuum data.  The synthesized beam size and {\it rms} noise of the continuum emission from combining both compact and extended configuration data are $\rm  \sim 1.5'' \times 1.0''$ and $\rm \sim 3~mJy/beam$, respectively. The lines are not imaged due to low signal-to-noise levels.

%%%%%%%%%%%%%%%%%%%%%%%%%%%%% radio map %%%%%%%%%%%%%%%%%%%%%%%%%%%%%%
\begin{figure*}
%\hspace*{-0.6cm}
\centering
\includegraphics[scale=0.17]{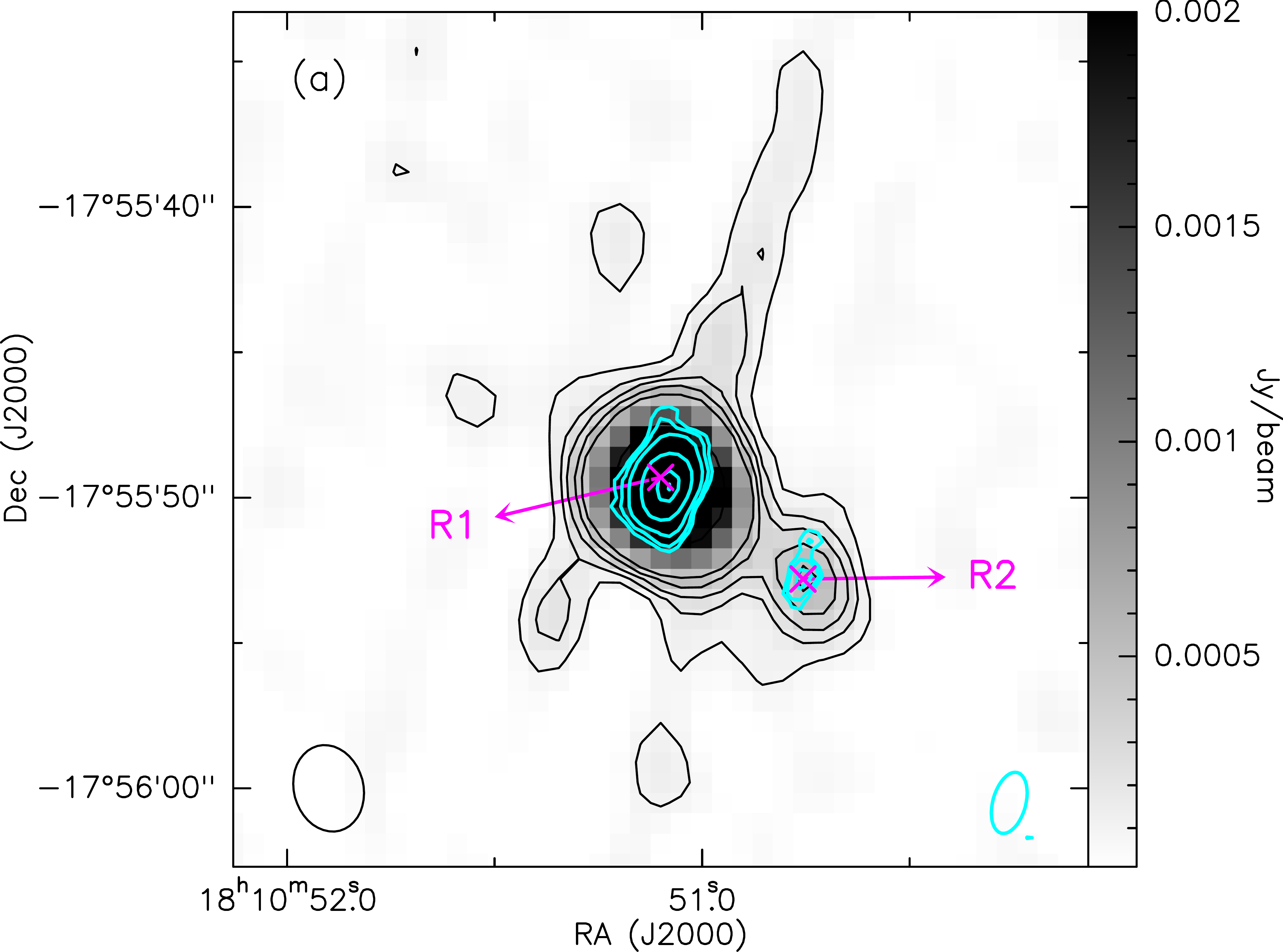} \quad\includegraphics[scale=0.17]{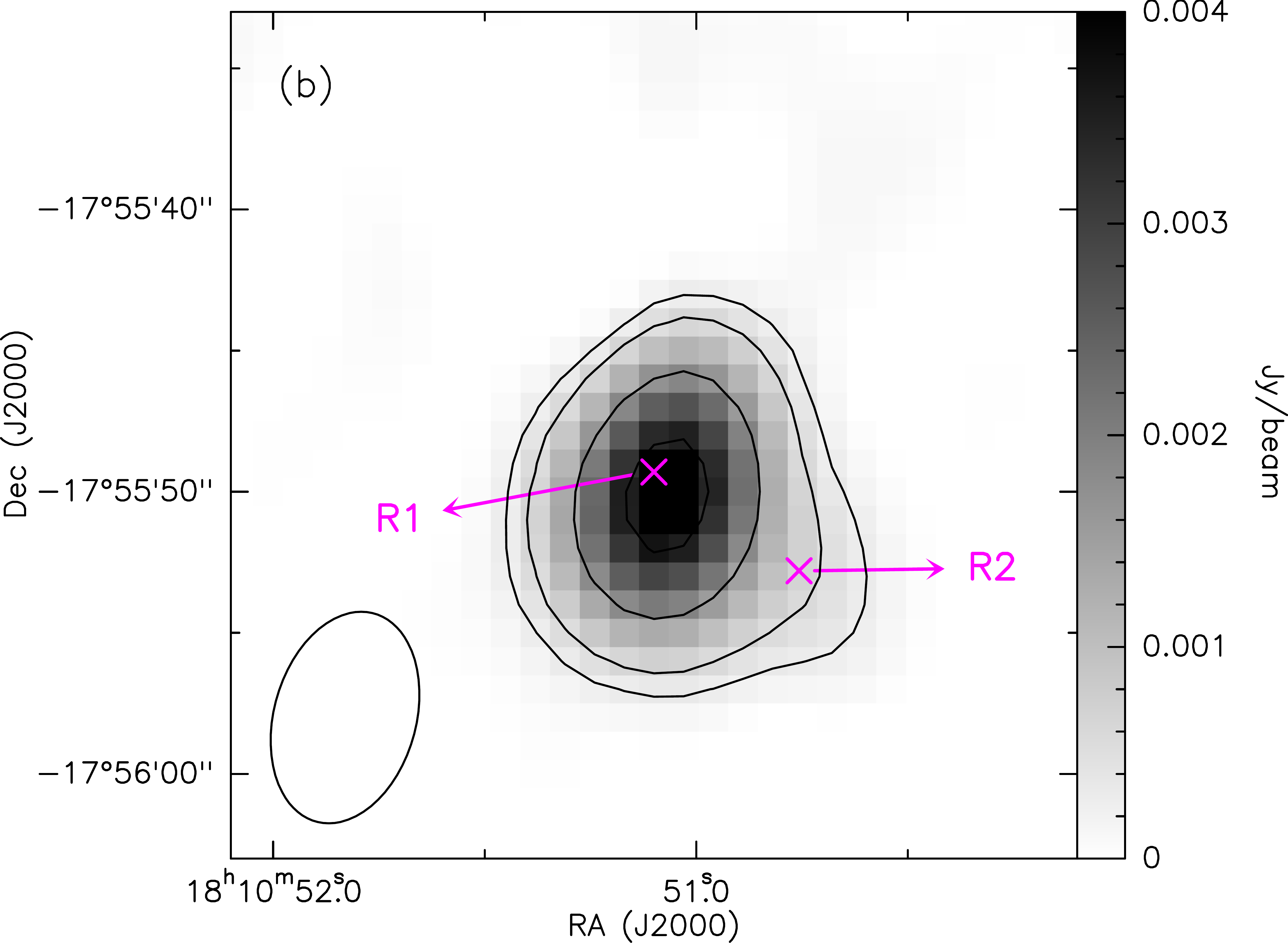}
\caption{(a) The grey scale shows the high resolution radio continuum map of {\g12} at 1390~MHz with the contour levels 3, 6, 9, 18, 63, 150 and 
172 times $\sigma$ ($\sigma \sim 29.7~\mu$Jy/beam).  The beam size is  $\sim 3.0'' \times 2.4''$. Positions of R1 and R2 are also labelled.
The contours of the 6~cm radio map are overlaid in cyan with the contour levels 3, 4, 6, 12, 21 and 24 $\sigma$ ($\sigma \sim 0.15$ mJy/beam) and
the beam size is $\sim 2.2'' \times 1.1''$. (b) The radio continuum map of {\g12} at 610~MHz with contour levels 
3, 6, 18, 38 and 60 times $\sigma$ ($\sigma \sim 94~\mu$Jy/beam). The beam size is $\sim7.6'' \times 4.8''$. 
The positions of the two radio peaks detected in the 1390~MHz map is indicated by `x'. The restoring beams in the 1390 and 610~MHz 
bands are represented as open ellipses towards the bottom-left of each image and of the 6~cm map is represented as an open cyan ellipse
towards the bottom-right in (a). }
\label{radio}
\end{figure*}

%%%%%%%%%%%%%%%%%%%%%%%%%%%%%%%%%%%%%%%%%%%%%%%%%%%%%%%%%%%%%%%%%%%%

\subsection{Molecular line data from MALT90 survey}
To understand the gas kinematics in our region of interest, molecular line data were obtained from the MALT90 survey \citep{{2013PASA...30...57J},{2011ApJS..197...25F}}. The survey, carried out using the ATNF Mopra 22-m telescope, has simultaneously 
mapped the transitions of 16 molecules near 90~GHz with a spectral resolution of $\rm 0.11~km~s^{-1}$. The Mopra Telescope is a 22-m single-dish radio
telescope operated by The Commonwealth Scientific and Industrial Research Organisation's Astronomy and Space Science division. The data reduction 
was performed using CLASS90 (Continuum and Line Analysis Single-dish Software), a GILDAS\footnote{\url{http://www.iram.fr/IRAMFR/GILDAS}} software 
(Grenoble Image and Line Data Analysis  Software).

\subsection{JCMT archival data}
The molecular line data for the $J=3-2$ transition of $\rm ^{12}CO$, $\rm ^{13}CO$ and $\rm C^{18}O$ were downloaded from the archives of the Heterodyne
Array Receiver Program (HARP) mounted on the {\it James Clerk Maxwell Telescope} (JCMT) operated by the East Asian Observatory. JCMT is a 15~m telescope
and is the largest single-dish astronomical telescope which operates in the submillimetre wavelength region of the spectrum. HARP is a Single Sideband
array receiver that can be tuned between 325 and 375 GHz and has an instantaneous bandwidth of $ \sim $2 GHz and an Intermediate Frequency of 5 GHz.
It comprises of 16 detectors laid out on a 4$\times$4 grid, with an on-sky projected beam separation of 30{\arc}. At 345 GHz the beam size is
14{\arc} \citep{2009MNRAS.399.1026B}.

\subsection{TRAO observation}

The molecular line data for the $J=1-0$ transition of $\rm ^{13}CO$ was obtained from the Taeduk Radio Astronomy Observatory (TRAO). TRAO is a 14~m radio telescope with a single-horn receiver system operating in the frequency range of 86 to 115 GHz and is located on the campus of the Korea Astronomy and Space Science Institute (KASI) in Daejeon, South Korea. The main FWHM beam sizes for the $\rm ^{12}CO~(1-0)$ and $\rm ^{13}CO~(1-0)$ lines are 45{\arc} and 47{\arc} respectively. The system temperature ranges from 150 K, for $\rm 86-110$~GHz to 450 K, for 115 GHz and $\rm ^{12}CO$  \citep{2018ApJS..234...28L}.

\section{RESULTS} \label{results}
\subsection{Emission from Ionized Gas} \label{radio_text}
%%%%%%%%%%%%%%%%% spectral index %%%%%%%%%%%5
\begin{figure}
%\hspace{-0.3cm}
\centering
\includegraphics[scale=0.20]{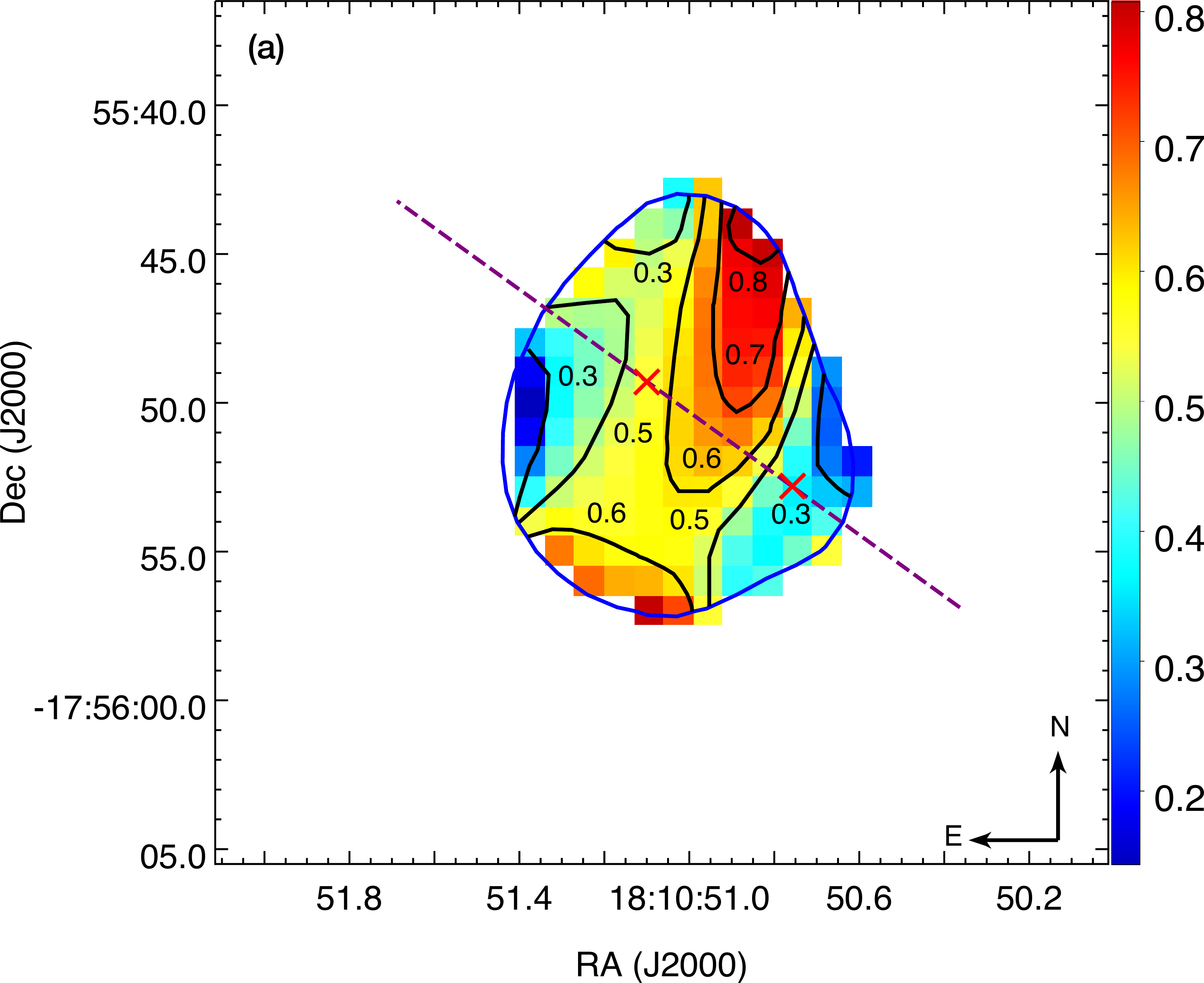} \quad\includegraphics[scale=0.205]{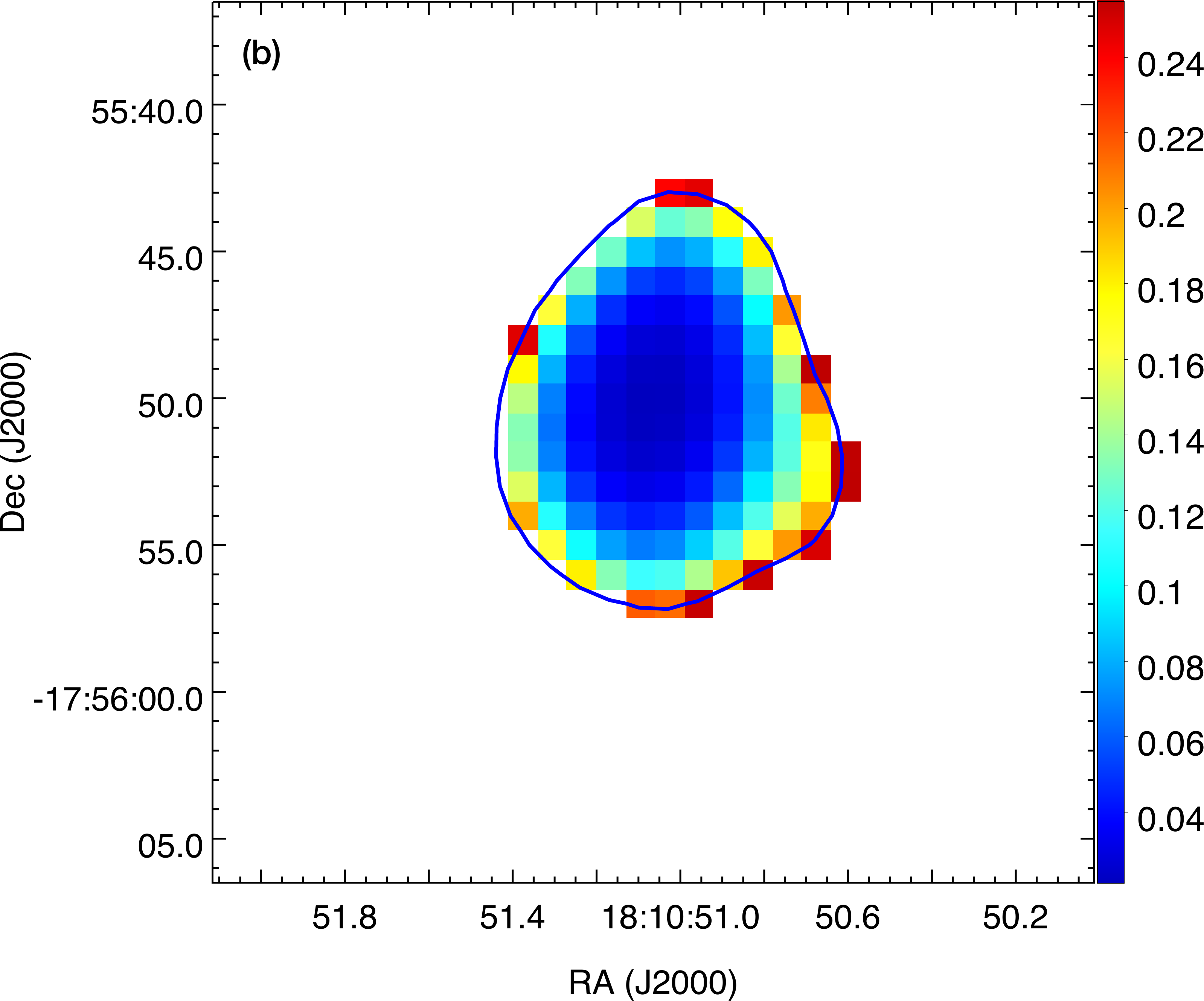}
\caption{(a) Spectral index map of {\g12} between 1390 and 610~MHz. Black curves represent the spectral index levels. The blue contour shows
the 5$\sigma$ ($\sigma \sim 0.4 \times 10^{-4}$ Jy/beam) level of the 610~MHz map used to construct the spectral index map.
The red `x's mark the positions of the radio components, R1 and R2. The dashed purple line indicates the possible direction of the ionized jet. The spectral index varies from 0.3 to 0.7 along the possible jet axis. The error map is shown in (b). The errors involved are $\lesssim$ 0.15, barring a few pixels at the edges.}
\label{specind}
\end{figure}
%%%%%%%%%%%%%%%%%%%%%%%%%%%%%%%%%% 

%%%%%%%%%%%%%%%%%%%%%%%%%%%radio table %%%%%%%%%%%%%%5 
\begin{table*}
\centering 
\small
\caption{Peak coordinates, peak and integrated flux densities, and deconvolved sizes of the components R1 and R2 associated with {\g12}.}
\begin{center}
\hspace*{-0.8cm}
\begin{tabular}{c c c c c c c c c c c c} \hline \hline \

& \multicolumn{2}{c}{~~~~~~~~~~Peak Coordinates~~~~} & \multicolumn{3}{c}{~~~Deconvolved size (\arc$\times$\arc)}~~~ &\multicolumn {3}{c}{ ~~~~~~~~~~~Peak flux (mJy/beam)~~~~~~~~~~~} & \multicolumn{3}{c}{~~~~~~~~~~~~Integrated flux (mJy)~~~~~~~~~~~}   \\

%&						& Peak coordinates			&& Peak flux density (mJy/beam)		&&& Integrated flux density (mJy) \\
\hline \
Component 		& RA (J2000)   	& Dec (J2000)	 	&610 MHz  & 1390 MHz		 & 5 GHz $^{\ast}$ & 610 MHz  & 1390 MHz		 & 5 GHz $^{\ast}$     & 610 MHz  & 1390 MHz		 & 5 GHz $^{\ast}$ \\
\hline \
R1					&18 10 51.10	&-17 55 49.30		& 2.6 $\times$ 0.6   & 1.9 $\times$ 1.7	 & 1.4 $\times$ 1.2 & 4.4			& 5.3				& 3.5			&		4.7		& 7.9	  	& 6.2  \\  
R2					&18 10 50.76	& -17 55 52.80	& -	&	1.5 $\times$ 1.2 $^{\dagger}$  & 1.1 $\times$ 0.6 $^{\dagger}$	&-		& 0.6				& 1.1			&-			& 0.7          & 1.1   \\ 
\hline 
\end{tabular}
$^{\dagger}$ Upper limits which is half the FWHM of the restoring beam.\\ 
$^{\ast}$ Values for R1 are from \citet{2009A&A...501..539U} and for R2 they are estimated from the available map.
 
\label{peak_position}

\end{center}
\end{table*}
%%%%%%%%%%%%%%%%%%%%%%%%%%%%%%%%%%%%%%%%

The radio continuum maps at 1390 and 610~MHz, probing the ionized gas emission associated with {\g12}, are shown in {\fig} \ref{radio}. 
The 1390~MHz map reveals the presence of a linear structure in the north-east and south-west direction comprising of an extended emission with
two distinct and compact components, labelled R1 and R2 in the figure. The component R1 is well resolved, whereas R2 seems to be barely resolved. 
In this figure, we also plot the contours of the high-resolution 6~cm (5~GHz) map obtained using VLA by \citet{2009A&A...501..539U} as part of the RMS survey
towards candidate massive YSOs. Both components are also visible in the 6~cm map. In comparison, the lower-resolution 610~MHz shows a single, almost spherical emission region with the peak position coinciding with R1. However, a discernible elongation is evident towards R2. In addition, the 1390~MHz map shows a narrow extension in the north-west and south-east direction. Given that the maps (especially 1390~MHz) have low level stripes in the said direction, 
it becomes difficult to comment on the genuineness of this feature. 

Table \ref{peak_position} compiles the coordinates, peak and integrated flux densities of R1 and R2. The deconvolved sizes and integrated flux densities are estimated by fitting 2D Gaussians using the task {\tt IMFIT} from Common Astronomy Software Application (CASA)\footnote{\url{https://casa.nrao.edu}} \citep{2007ASPC..376..127M}). At 610~MHz, the components are not resolved so the values obtained are assigned to R1 and hence should be treated as upper limits. For the component R1, the 5~GHz values are quoted from \citet{2009A&A...501..539U}.  As for the component R2, that is barely resolved in both 1390~MHz and 5~GHz maps, we have set an upper limit to its size at both frequencies. This is taken to be the FWHM of the respective restoring beams \citep{2009A&A...501..539U}. Further, at 5~GHz, we take the peak flux density to be the same as the integrated flux density.

In order to get an in-depth knowledge of the nature of the observed radio emission, we generate the spectral index map using our 1390 and 610~MHz maps.
The spectral index, $\alpha$, is defined as $S_\nu \propto \nu^\alpha$, where, $S_\nu$ is the flux density at frequency $\nu$. GMRT is not a scaled array, 
hence, each frequency is sensitive to different spatial scales. To circumvent this, we generate new maps in the {\it uv} range ($\rm 0.7 - 47 K\lambda$) 
common to both frequencies. Keeping in mind the requirement of same pixel size and resolution, pixel and beam-matching is taken into account while 
generating the new maps. The spectral index map is then constructed using the task {\tt COMB} in AIPS. Further, to ensure reliable estimates of
the spectral index, we retain only those pixels with flux density greater than 5$\sigma$ ($\sigma$ being the {\it rms} noise of the map) in both maps.
The generated spectral index map and the corresponding error map, which has the same resolution as that of the 610~MHz map ($\sim7.6'' \times 4.8''$), are presented in {\fig}\ref{specind}.
As seen from the figure, the spectral index values vary between 0.3 and 0.9 with the estimated errors involved being less 
than $\sim$ 0.15, barring a few pixels at the edges. These values indicate that the region is dominated by thermal bremsstrahlung emission of varying optical depth \citep{{1993RMxAA..25...23R},{1993ApJ...415..191C},{1999ApJ...527..154K},{2016ApJS..227...25R}}. Moreover, spectral index values in the range of $0.4-0.9$ are also typically seen in regions associated with thermal jets \citep[e.g.][]{{1975A&A....39....1P},{1986ApJ...304..713R},{2016MNRAS.460.1039P},{2016A&A...596L...2S}}. We will
revisit these results obtained in a later section where we explore various scenarios to adequately explain the nature of the radio emission.

\subsection{Emission from shock indicators} {\label{NIR_spectroscopy}}

As discussed in the introduction, there is growing evidence in literature associating EGOs with MYSOs, notwithstanding the ongoing debate regarding their exact nature. Several mechanisms, like shocked emission in outflows, fluorescent emission or scattered continuum from MYSOs \citep{{2004ApJS..154..352N},{2010AJ....140..196D},{2012ApJ...748....8T}}, are invoked to identify the spectral carriers of the enhanced 4.5~{\um} emission.
The picture of shocked emission from outflows suggests the spectral carriers to be molecular and atomic shock indicators like {\h2} and {\feii} as well as the broad CO bandhead. All of these have distinct features within the 4.5~{\um} IRAC band. However, \citet{2012MNRAS.419..211S}, while investigating the population of MYSOs in the G333.2-0.4 region, opine that the excess 4.5~{\um} could not be attributed to the {\h2} lines as these would be too faint to be detected at this wavelength. Instead, they support a scattered continuum or the CO bandhead origin. From the $L$- and $M$-band spectra of two EGOs, \citet{2010AJ....140..196D} show the {\h2} line hypothesis to be consistent with one of them (G19.88-0.53), while in the other target (G49.27-0.34), the spectra shows only continuum emission. So far, spectroscopic studies of EGOs in the 4.5~{\um} and the NIR are few  \citep{{2010AJ....140..196D},{2015A&A...573A..82C},{2016ApJ...829..106O}}, thus keeping the debate on their nature ongoing. In the NIR domain, a few studies have focussed towards narrow-band imaging \citep{{2012ApJS..200....2L},{2013ApJS..208...23L}}. Based on the UWISH2 survey images, \citet{{2012ApJS..200....2L},{2013ApJS..208...23L}} present a complete {\h2} line emission census of EGOs in the Northern Galactic Plane. 
 
\subsubsection{Narrow-band Imaging}\label{NIR_imaging}
%%%%%%%%%%%%%%%%%%%%%%%%%%%%% Fe-H, H2-K %%%%%%%%%%%%%%%%%%%%%%%%%%%%%%
\begin{figure}
%\hspace*{-0.6cm}
%\vspace*{0.7cm}
\centering
\includegraphics[scale=0.16]{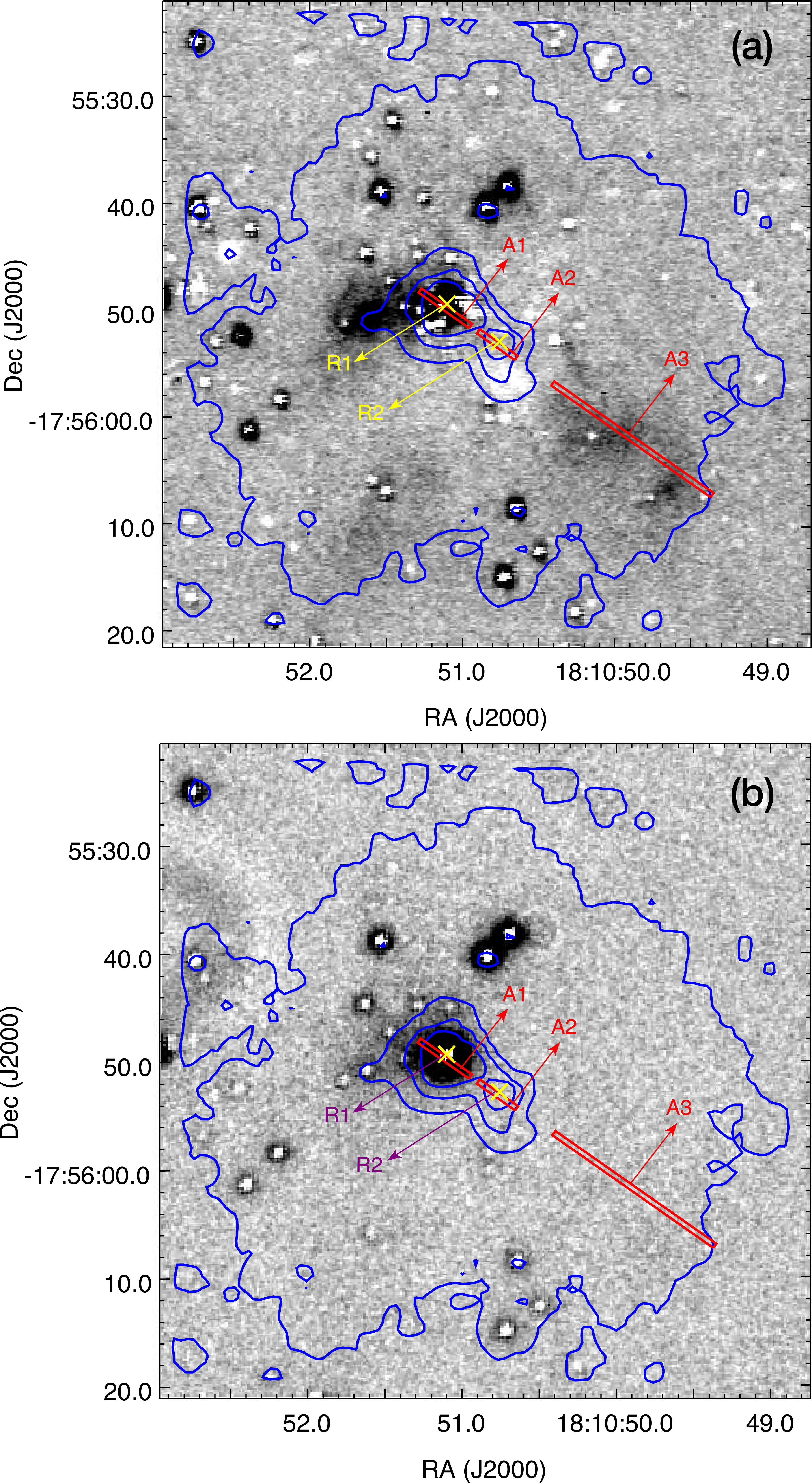}
\caption{(a) Continuum subtracted {\h2} image  made using the UWISH2 survey data. (b) Continuum subtracted {\feii} image made using UKIRT-WFCAM observations towards {\g12}. The positions of the identified radio components R1 and R2 are indicated. The blue contours represent the 4.5~{\um} emission with the levels 3, 60, 120, and 220$\sigma$ ($\sigma \sim 1.5$~MJy/sr).
The red rectangles show the orientation of the slit and denote the apertures used for spectra extraction (see text in Section \ref{NIR_spectra})}.
\label{NIR_narrowband}
\end{figure}
%%%%%%%%%%%%%%%%%%%%%%%%%%%%%%%%%%%%%%%%%%%%%%%%%%%%%%%%%%%%%%%%%%%%

\par {\h2} line emission towards {\g12} has been investigated in \citet{2012ApJS..200....2L}. They ascribe the extended emission seen in the continuum-subtracted image to be the result of residuals of continuum subtraction rather than real {\h2} line emission. In order to carefully scrutinize the NIR picture of {\g12}, we revisit the {\h2} line emission from images retrieved from the UWISH2 survey. In addition, we also probe the {\feii} line image which is a robust indicator of shocks as compared to the {\h2} lines \citep{2014ApJS..214...11S}.
\par Following the procedure outlined in Section \ref{cont_sub} we construct the continuum subtracted {\h2} and {\feii} line images which are presented in {\fig}\ref{NIR_narrowband}.
In the continuum-subtracted {\h2} image, the morphology is similar to that obtained by \citet{2012ApJS..200....2L}. An extended emission is seen towards the peak of the 4.5~{\um} emission coinciding with the location of the the radio component R1.
Ideally a narrow-band continuum filter should enable a better continuum subtraction but in the absence of the same, we have ensured PSF matching and proper scaling of the broad $K$-band image. Contrary to the suggestion by \citet{2012ApJS..200....2L}, we believe that the extended {\h2} line emission detected in the continuum-subtracted image is genuine. This finds strength in the spectra obtained and discussed in the next section. In addition, diffuse line emission is seen towards the north-east and east of R1 as well towards the south-west. The continuum-subtracted {\feii} image shows a weak, extended emission coinciding with the brighter part of the {\h2} line emission.

\subsubsection{NIR spectroscopy}\label{NIR_spectra}
%%%%%%%%%%%%%%%%%%%%%%%%%%%%% HK spectrum %%%%%%%%%%%%%%%%%%%%%%%%%%%%%%
\begin{figure*}
%\hspace{-0.6cm}
\centering
\includegraphics[scale=0.60]{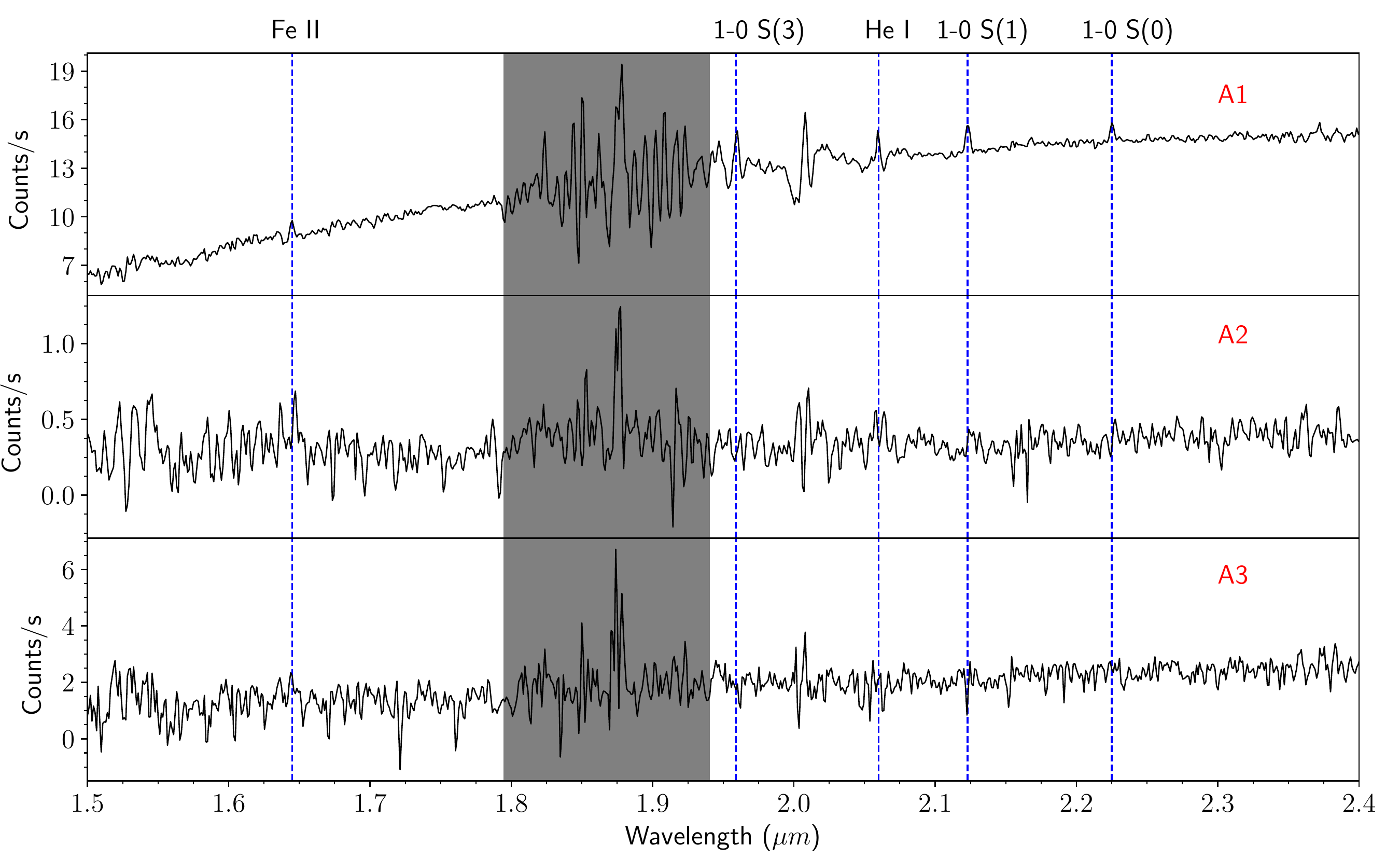} 
\caption{The $HK$ spectrum of {\g12} extracted over the apertures A1, A2 and A3. The aperture A1 covers the radio component R1 and the extended {\h2} emission seen towards the north-east, A2 covers the second radio component R2 and A3 samples the detached, extended emission seen towards the south-west. The shaded area marks the region of poor sky transparency. The identified spectral
lines along aperture A1 are marked over the spectrum with the details given in Table \ref{spectral_lines}. No emission lines above the noise level are detected in the spectra extracted over A2 and A3.}
\label{spechk_A2A3}
\end{figure*}
%%%%%%%%%%%%%%%%%%%%%%%%%%%%%%%%%%%%%%%%%%%%%%%%%%%%%%%%%%%%%%%%%%%%
%%%%%%%%%%%%%%%%%%%%%%%%%%%%% L spectrum %%%%%%%%%%%%%%%%%%%%%%%%%%%%%%
\begin{figure*}
\hspace{0.2cm}
%\centering
\includegraphics[scale=0.56]{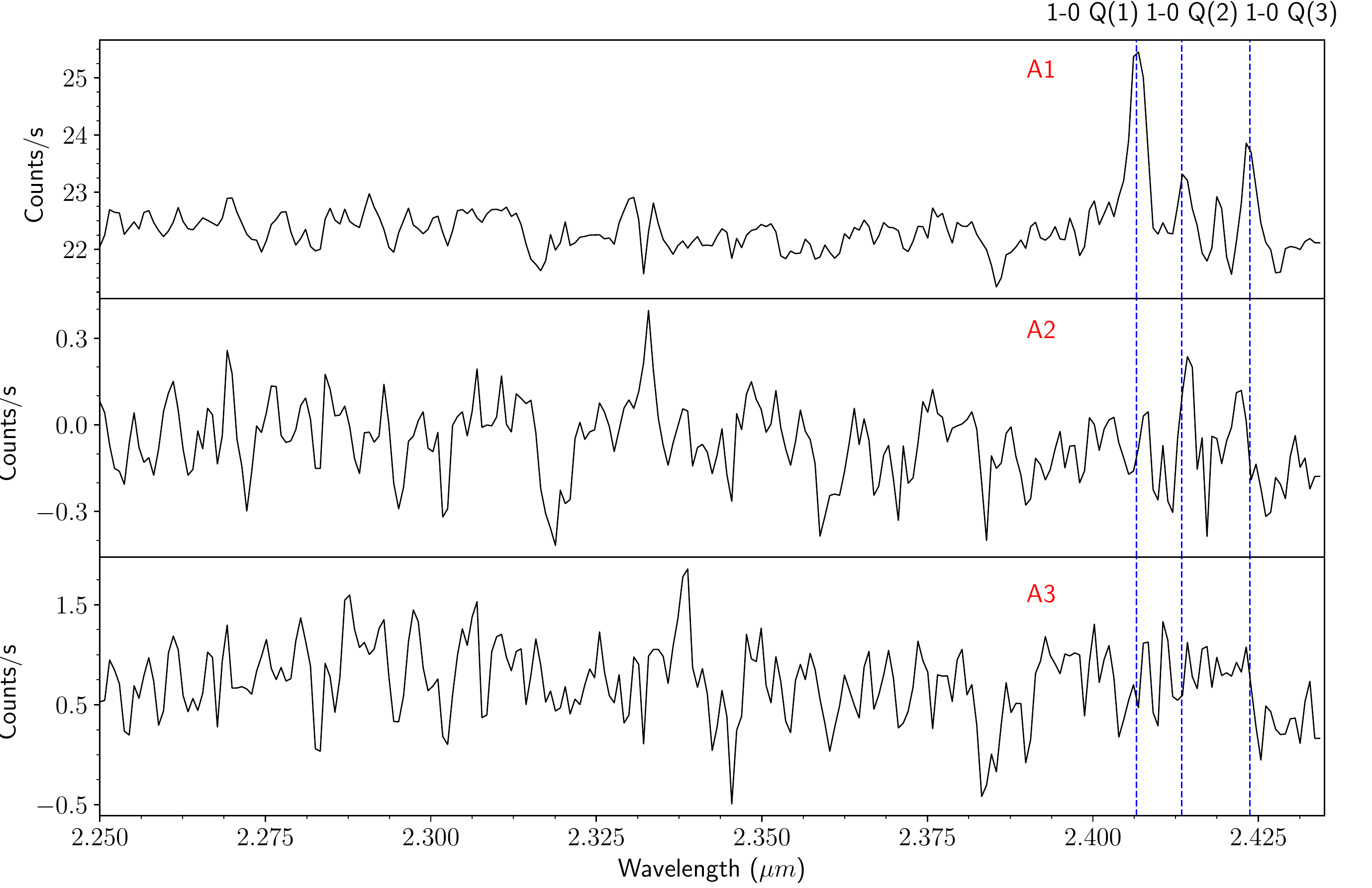} 
\caption{The $KL$ spectrum of {\g12} extracted over the apertures A1, A2 and A3. The regions covered by all the three apertures are the same as given in {\fig}\ref{spechk_A2A3}. The identified spectral
lines along aperture A1 are marked over the spectrum with the details given in Table \ref{spectral_lines}. No emission lines above the noise level are detected in the spectra extracted over A2 and A3.}
\label{spec-l}
\end{figure*}
%%%%%%%%%%%%%%%%%%%%%%%%%%%%%%%%%%%%%%%%%%%%%%%%%%%%%%%%%%%%%%%%%%%%
%%%%%%%%%%%%%%%%%%%%%%%%%%%%%%%%%%%%%%%%%%%%%%%%%%%%%%%%%%%%%
\begin{table}
\caption{Lines detected in the spectra extracted from aperture A1 towards {\g12}.}
\begin{center}
\begin{tabular}{c c} \hline \hline \
Line		& Wavelength ($\mu$m) 	\\
\hline \
{\feii}						&1.644 \\
{\h2} 1-0 S(3)		&1.958 \\
He I						&2.059 \\
{\h2} 1-0 S(1)		&2.122 \\
{\h2} 1-0 S(0)		&2.224 \\
{\h2} 1-0 Q(1)		&2.407 \\
{\h2} 1-0 Q(2)		&2.413 \\
{\h2} 1-0 Q(3)		&2.424 \\
\hline \
\end{tabular}
\label{spectral_lines}
\end{center}
\end{table}
%%%%%%%%%%%%%%%%%%%%%%%%%%%%%%%%%%%%%%%%%%%%%%%%%%%%%%%%%%%%%%%%%%% 
As is clear from earlier discussions, studies towards identifying the spectral carriers of the 4.5~{\um} emission are crucial in understanding the nature of EGOs and confirming their association with MYSOs. Given the lack of sensitive spectrometers in the 4.5~{\um} region, spectroscopy in the NIR becomes indispensable. We probe {\g12} with NIR spectroscopy to understand further the results obtained from narrow-band imaging. From the continuum-subtracted line images shown in {\fig}\ref{NIR_narrowband} and the UKIDSS $K$-band image shown in {\fig}\ref{irac_ukidss_rgb}, presence of faint nebulosity around the peak position (that coincides with the 4.5~{\um} peak) and towards the south-west is clearly visible. The slit orientation shown in {\fig}\ref{NIR_narrowband} ensures that the regions harbouring the radio components and the extended {\h2} line emission towards the north-east of the peak and the detached elongated nebulosity towards the south-west are probed. 

\par The $HK$ spectra extracted over the three identified apertures (marked in {\fig}\ref{NIR_narrowband}) are shown in {\fig}\ref{spechk_A2A3}. The top-panel of {\fig}\ref{spechk_A2A3} shows the spectrum over aperture A1 with the line details listed in Table \ref{spectral_lines}. This aperture covers the radio component R1 and portions of the extended {\h2} emission seen towards the north-east of the 4.5~{\um} peak. The spectrum shows clear detection of three emission lines of molecular {\h2} with the most prominent feature being the $\rm 1-0 S(1)$ line at 2.122~{\um}. No {\h2} line is detected in the blue part ($\rm 1.5 - 1.8~\mu m$) of the spectrum but there is a weak {\feii} line detected at 1.644~{\um}  These lines of {\h2} and {\feii} are commonly observed in outflows/jets. In addition, He I at 2.059~{\um} is also seen in the extracted spectrum. Apart from the emission lines, the continuum slope is seen rising towards the red thus, indicating a highly reddened source. 
{\fig}\ref{spechk_A2A3} also plots the extracted spectra over the apertures A2 and A3 in the middle and lower panels, respectively. Aperture A2 covers the second radio component R2 and aperture A3 samples the detached, extended emission seen towards the south-west. No emission lines above the noise level are detected in these and the spectra displayed are flat.

\par In {\fig}\ref{spec-l}, we present the extracted spectra in the $KL$ band. The displayed spectra has been truncated at 2.45~{\um} due to poor signal-to-noise ratio owing to less than optimal sky transparency. In aperture A1, three additional emission lines of molecular {\h2} are prominent. The other two apertures do not show the presence of any spectral feature. The detected lines are listed in {\tab}\ref{spectral_lines}.

\par The observed {\h2} line emissions seen in the spectra of {\g12} can be attributed to either thermal or non-thermal excitation. The thermal emission  mostly originates from shocked neutral gas in outflows/jets that are heated up to a few 1000 K, whereas, the non-thermal emission is understood to be due to UV fluorescence by
non-ionizing UV photons. These two competing mechanisms populate different energy levels thus yielding different line ratios \citep{{2003MNRAS.344..262D},{2015A&A...573A..82C},{2016MNRAS.456.2425V}}. UV fluorescence excites higher vibrational levels.   
The {\h2} lines detected in {\g12} originate from the upper vibrational level, $\nu = 1$ suggesting a low level of excitation. The absence of high vibrational state transitions supports the shock-excited origin of the detected lines.  
Lack of fluorescent {\h2} line emission in {\g12} may also be due to veiling of UV photons from the central star due to high extinction. Nevertheless, given the association with an outflow source, shock-excited origin is most likely the case.

\subsection{Emission from the dust component} \label{dust}

%%%%%%%%%%%%%%%%%%%%%%% Dust emission %%%%%%%%%%%%%%%%%%%%%%%%%%%%%%%%
\begin{figure*}
\hspace*{0.3cm}
%\centering
\includegraphics[scale=0.45]{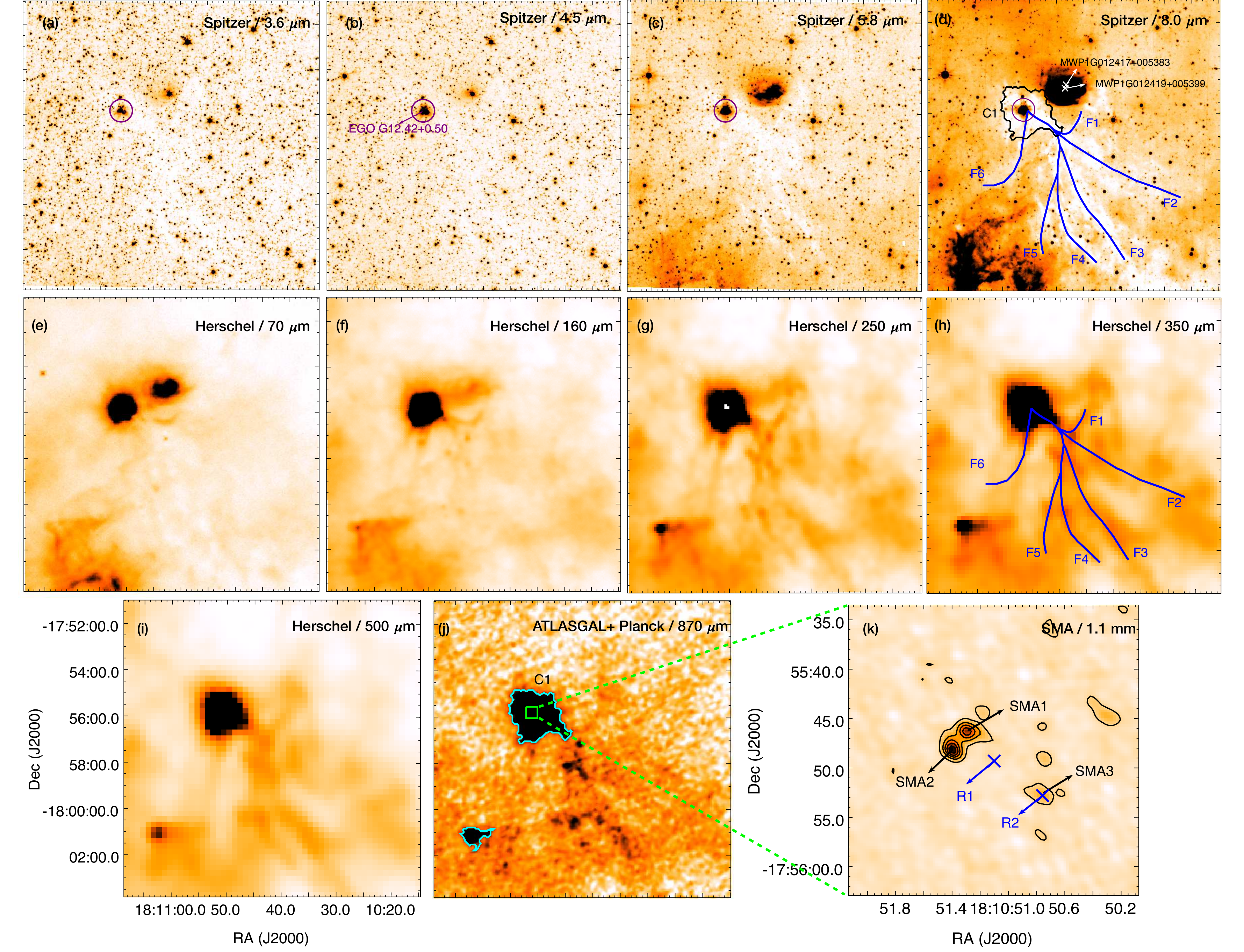}
\caption{Dust emission in the region associated with {\g12} at the mid- and far-infrared wavelengths ($ \rm 3.6~\mu m-1.1~mm$). All the images from (a)$-$(j) have the same field of view. Skeletons of six clearly identified filaments are overlaid on the 8.0 and 350~{\um} maps. The position of the EGO, {\g12} is shown within a purple circle on the 3.6, 4.5, 5.8 and 8.0~{\um} IRAC maps. The location of the two infrared dust bubbles (MWP1G012417+005383 and MWP1G012419+005399) are indicated by white `x's on the 8.0~{\um} map, (d). (d) and (j) show the retrieved aperture of clump C1. Another clump detected towards south-east of {\g12} is shown in the 870~{\um} map, (j). (k) shows the SMA 1.1~mm map with the contour levels 3, 12, 21, 30 and 39 times $ \sigma $ ($ \sigma \sim 3 $~mJy/beam). The blue `x's on the 1.1~mm map mark the positions of the radio components R1 and R2.}
\label{FIR}
\end{figure*}
%%%%%%%%%%%%%%%%%%%%%%%%%%%%%%%%%%%%%%%%%%%%%%%%%%%%%%%%%%%%%

The dust emission at MIR and FIR wavelengths sampled in the IRAC, Hi-Gal, ATLASGAL-Planck and SMA wavelengths (3.6~{\um}$-$1.1~mm) in the region associated with {\g12} is shown in {\fig}\ref{FIR}. 
In the IRAC bands, various emission mechanisms come into play and contribute towards the warm dust component \citep{2008ApJ...681.1341W}. Thermal emission from the circumstellar dust heated by the stellar radiation and emission from the UV excited polycyclic aromatic hydrocarbons in the Photo Dissociation Regions are known to be the dominant contributors. In the shorter IRAC wavelengths (3.6, 4.5~{\um}), where mostly the stellar sources are sampled, emission from the stellar photosphere would also be appreciable. Apart from this, shock-excited {\h2} line emission and diffuse emission in the Br$ \alpha $ and Pf$ \beta $ lines would also exist. Further, in case of {\hii} regions, one expects significant contribution from the Ly$ \alpha $ heated dust \citep{1991MNRAS.251..584H}.  
The morphology in the IRAC bands is similar and the emission becomes more prominent at 8.0~{\um}. Dark filamentary features (bright in the negative images shown) are seen in silhouette towards the south-west in the 8.0~{\um} map. The skeletons of the six clearly identified filamentary features are overlaid on the 8.0~{\um} map. In addition, an extended emission feature is seen towards the north-west of {\g12}, being prominent in the 5.8, 8.0, and 70~{\um} images. 
Two infrared dust bubbles (MWP1G012417+005383 and MWP1G012419+005399) are found to be associated with this feature and are marked in {\fig}\ref{FIR}(d). No further literature is available on these bubbles so we drop them in further discussion. 

\par As we move towards the longer wavelengths, cold dust emission associated with {\g12} becomes enhanced and more extended. 
From the ATLASGAL-Planck combined 870~{\um} map, we identify two clumps using the 2D {\it Clumpfind} algorithm \citep{1994ApJ...428..693W} with 2$\sigma$ ($\sigma$ = 0.3 Jy/beam) threshold and optimum contour levels. The apertures of the identified clumps are overlaid on the 870~{\um} map in {\fig}\ref{FIR}(j). While one of the clumps, hereafter C1, is associated with {\g12}, another clump lies towards the south-east of {\g12} at an angular distance of $\sim 6'$. From the $\rm H^{13}CO^+$ molecular line data (Section \ref{molecular-line}), we estimate the LSR velocity of this clump to be $\rm 31.5~km~s^{-1}$. Comparing this with the estimated LSR velocity of {\g12} ($\rm 18.3~km~s^{-1}$), it is unlikely that the clump has any association with {\g12}. 
The identified filaments now appear in emission and are shown on the 350~{\um} map. Interestingly, these filaments seem to converge towards clump C1. As mentioned in the introduction, the morphology has an uncanny resemblance to a hub-filament structure, detailed discussion of which is presented in Section \ref{hub_filament}. 
Furthermore, in the high-resolution 1.1~mm SMA map, the inner region of the cold dust clump, C1 associated with {\g12} is seen to harbour two, dense and bright compact cores labelled on the map as SMA1 and SMA2. Additionally, a few bright emission knots are detected in the SMA map including the one highlighted as SMA3 which coincides with the radio component R2.

\subsubsection{Properties of SMA cores}
From the SMA 1.1~mm map shown in {\fig}\ref{FIR}(k), SMA1 and SMA2 show up as dense, compact cores possibly in a binary system. SMA3, on the other hand, looks more like a clumpy region of density enhancement. Following the method described by \citet{2008A&A...487..993K} the masses of the SMA components are computed using the equation
\begin{eqnarray}
  M & = &
  \displaystyle 0.12 \, M_{\odot}
  \left( {\rm e}^{1.439 (\lambda / {\rm mm})^{-1}
      (T / {\rm 10 ~ K})^{-1}} - 1 \right) \nonumber \\
  & & \displaystyle
 \left( \frac{\kappa_{\nu}}{0.01 \rm ~ cm^2 ~ g^{-1}} \right)^{-1}
  \left( \frac{F_{\nu}}{\rm Jy} \right)
  \left( \frac{d}{\rm 100 ~ pc} \right)^2
  \left( \frac{\lambda}{\rm mm} \right)^{3}
  \label{sma_mass_eqn}
\end{eqnarray}
\noindent
where the opacity is 
\begin{equation}
\kappa_{\nu}=0.1({\nu}/\textrm{1000~GHz})^{\beta}~{\rm cm}^{2} {\rm g}^{-1} 
\end{equation}
$\beta$ is the dust emissivity spectral index which is fixed at 2.0 \citep{{1983QJRAS..24..267H},{1990AJ.....99..924B},{2010A&A...518L.102A}}. $F_\nu$ is the integrated flux density of each component, $d$ is the distance to the source and $\lambda$ is the wavelength taken as 1.1~mm. The temperature, $T$ is taken to be 26.8~K for SMA1 and SMA2 and 22.7~K for SMA3, from their positions in the dust temperature map (Section \ref{cold_dust}).
The peak positions and flux densities, integrated flux densities, the deconvolved sizes  and the masses of the of the 1.1~mm SMA cores are presented in {\tab}\ref{sma_params}. The deconvolved sizes and integrated flux densities of the cores are evaluated by fitting 2D Gaussians to each component using the 2D fitting tool of CASA viewer. From the mass and size estimates, SMA1 and SMA2 qualify as potential high-mass star forming cores satisfying the criterion, $ m(r) > 870{\rm M_\odot}(r/\rm pc)^{1.33} $ \citep{2010ApJ...716..433K}.

%%%%%%%%%%%%%%%%%%%%%%% Dust SED %%%%%%%%%%%%%%%%%%%%%%%%%%%%%%%%
\begin{figure}
%\hspace*{-0.3cm}
\centering
\includegraphics[scale=0.40]{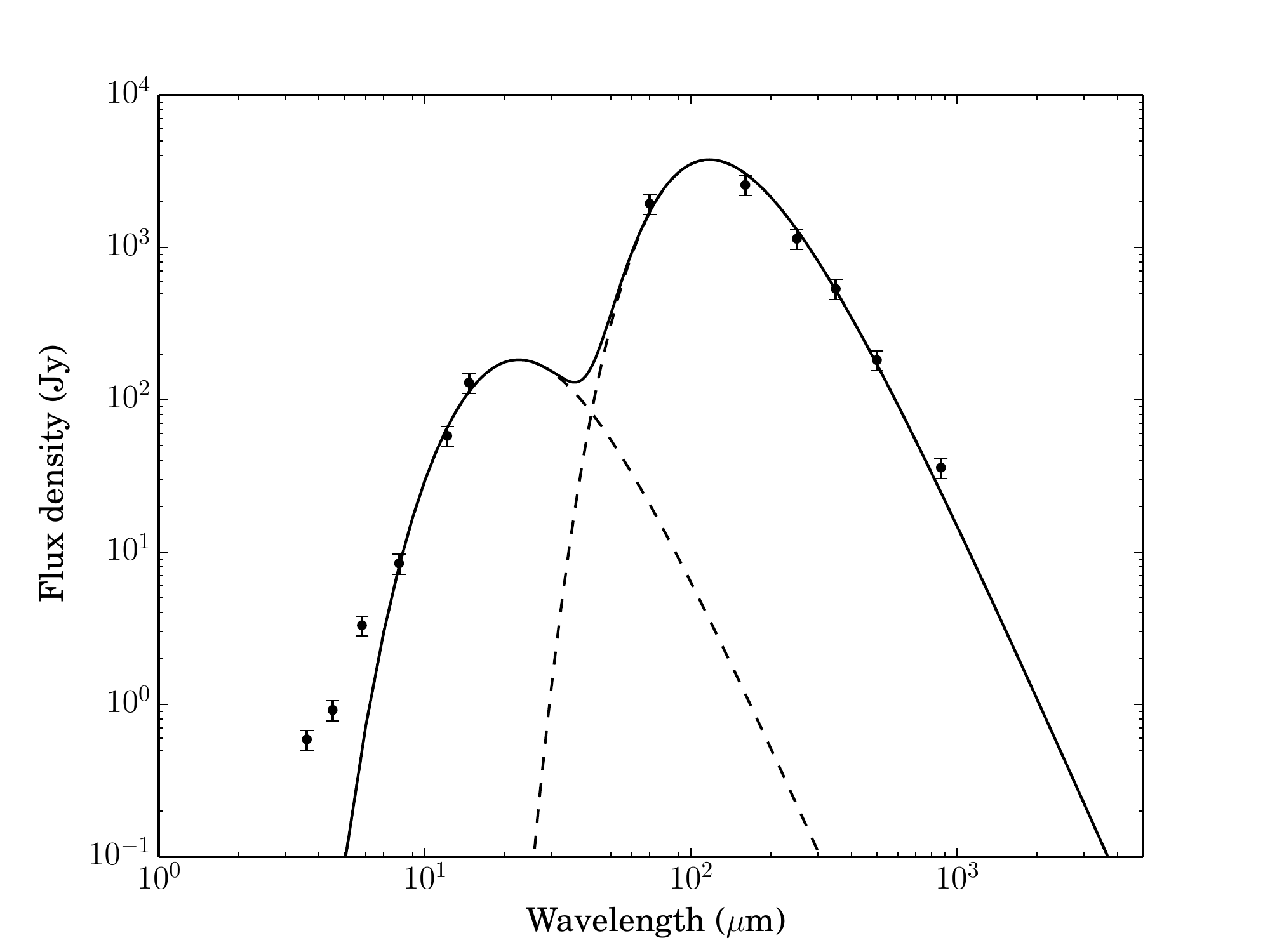}
\caption{Spectral energy distribution of the dust core associated with {\g12} in the wavelength range of 3.6 to 870~{\um}. Assumed 15\% errors are indicated. The solid curve represents the best fit two-component model with a warm component at 183~K and a cold envelope at 25~K.}
\label{Dust_SED}
\end{figure}

%%%%%%%%%%%%%%%%%%%%%%%%%%%%%%%%%%%%%%%%%%%%%%%%%%%%%%%%%%%%%
%%%%%%%%%%%%%%%%%%%%%%%%%%%%%%%%%%%%%%%%%%%%%%%%%5
\begin{table*}
\caption{Physical parameters of the 1.1~mm continuum emission near {\g12}.}
\begin{center}

%\hspace*{-1.2cm}
\begin{tabular}{c c c c c c c } \hline \hline 

Component 	 &\multicolumn{2}{c}{Peak position} & Deconvolved size  &Integrated flux	&Peak flux 	&Mass \\
			 &RA (J2000) $({^h}~{^m}~{^s})$			&Dec (J2000)	$(\degree~\arcmin~\arcsec)$	 & (\arc$\times$\arc)			& (mJy)			& (mJy/beam) 	&(M{\sun}) \\			
\hline \
SMA1		&18 10 51.3 		&-17 55 46.3 		&1.4$ \times $0.5		&190		&109		&	14.8 \\	
SMA2 		&18 10 51.4		&-17 55 48.1		&1.3$ \times $0.4		&221		&136  		&	17.2 \\
SMA3		&18 10 50.8		&-17 55 52.8		&1.5$ \times $0.7		&57			&34			&	5.5 \\

\hline \
\end{tabular}
\label{sma_params}

\end{center}
\end{table*}

%%%%%%%%%%%%%%%%%%%%%%%%%%%%%%%%%%%%%%%%%%%%%%%%%%
%%%%%%%%%%%%%%5
\begin{table*}
\caption{Integrated flux densities of the dust core associated with {\g12}.}
\begin{center}

%\hspace*{-1.2cm}
\begin{tabular}{l c c c c c c c c c c c c} \hline \hline \
Wavelength ({\um})		&3.6		&4.5		&5.8		&8.0		&12.13	&14.65	&70			&160		&250		&350		&500		&870	\\	
\hline \
Flux density (Jy)			&0.6		&0.9		&3.3		&8.4		&57.7		&129.8	&1942.6	&2575.5	&1140.7	&535.3	&182.5	&35.9	\\
\hline \
\end{tabular}
\label{SED_flux}

\end{center}
\end{table*}
%%%%%%%%%%%%%%%%%%%%%%%%%%%%%%%%%%%%%%%%%%%%%%%%%%%%%%%%%%%

%%%%%%%%%%%%%%%%%%%%%%%%%%%%%%%%%%%%%%%%%%%%%%%%%%%%%%%%%%%%%%%%%%%%%5
\begin{table*}
\tiny
\caption{Derived physical parameters of identified clumps associated with {\g12}. The peak position, radius, mean temperature and column density, total column density, mass, and volume number density of the identified clumps are listed.}
\begin{center}

%\hspace*{-1.1cm}
\centering  
\begin{tabular}{c c c c c c c c c} \hline \hline 

Clump 		&\multicolumn{2}{c}{Peak position} 			&Radius		&Mean $T_d$		&Mean $N(H_2)$		&$\Sigma N(H_2)$		& Mass			& Number density, $n(H_2)$		\\

& $\rm \alpha(J2000)~({^h}~{^m}~{^s})$		& $\rm \delta(J2000)~(\degree~\arcmin~\arcsec)$		&(pc)		&(K)			& (${\sc 10^{22}}$ cm$^{-2}$)		&($10^{23}$ cm$^{-2}$)		& (M\sun)		&($10^{3}$ cm$^{-3}$) 	\\

\hline \
C1	&18 10 49.64		& -17 55 59.40 		& 0.8	&	19.9$\pm$1.9		& 3.3$\pm$0.9	& 23.2		&1375		& 10.4	\\
C2	& 18 10 42.75		& -17 57 08.92 		& 0.3	& 16.1$\pm$0.4		& 1.2$\pm$0.1	& 1.0		& 59			& 10.7	\\
C3 	& 18 10 36.91		& -17 55 02.54 		& 0.4	& 17.1$\pm$0.8		& 0.7$\pm$0.1	& 1.5		& 92			& 4.2	\\
C4 	& 18 10 37.83 	& -17 58 04.57		& 0.4	& 16.7$\pm$0.7		& 0.9$\pm$0.1	& 2.1		& 127		& 4.9	\\
C5	& 18 10 27.03		& -17 58 17.79		& 0.4	& 17.7$\pm$0.9		& 0.7$\pm$0.1	& 1.1		& 66			& 4.4	\\
C6 	& 18 10 23.08 	& -17 59 55.48		& 0.4	& 18.1$\pm$0.9		& 0.7$\pm$0.1	& 1.2		& 70			& 4.3	\\
C7	& 18 10 19.16		& -17 59 27.19		& 0.2	& 18.2$\pm$0.8		& 0.7$\pm$0.1	& 0.4		& 24			& 7.2	\\
C8	& 18 10 43.71		& -17 58 04.98		& 0.5	& 15.8$\pm$0.6		& 1.2$\pm$0.2	& 3.6		& 214		& 6.0	\\
C9	& 18 10 38.76		& -18 00 24.61		& 0.2	& 15.6$\pm$0.7		& 1.2$\pm$0.2	& 0.7		& 41			& 11.1 \\
C10	& 18 10 35.81		& -18 00 52.40		& 0.3	& 16.0$\pm$0.8		& 1.1$\pm$0.2	& 0.9		&55			& 9.5	\\
C11	& 18 10 43.67		& -18 00 24.95		& 0.7	& 16.3$\pm$0.8		& 1.0$\pm$0.2	& 6.0		& 359		& 3.5	\\
C12	& 18 10 38.73		& -18 02 02.59		& 0.2	& 16.3$\pm$1.1		& 0.8$\pm$0.2	& 0.5		& 29			& 9.7	\\
\hline \\	
\end{tabular}
\label{clump}

\end{center}
\end{table*}
%%%%%%%%%%%%%%%%%%%%%%%%%%%%%%%%%%%%%%%%%%%%%%%%%%%%%%%%

%%%%%%%%%%%%%%%%%%%%%%% Temp-CD-Chi^2 maps %%%%%%%%%%%%%%%%%%%%%%%%%%%%%%%%
\begin{figure}
%\hspace*{0.2cm}
\centering
\includegraphics[scale=0.30]{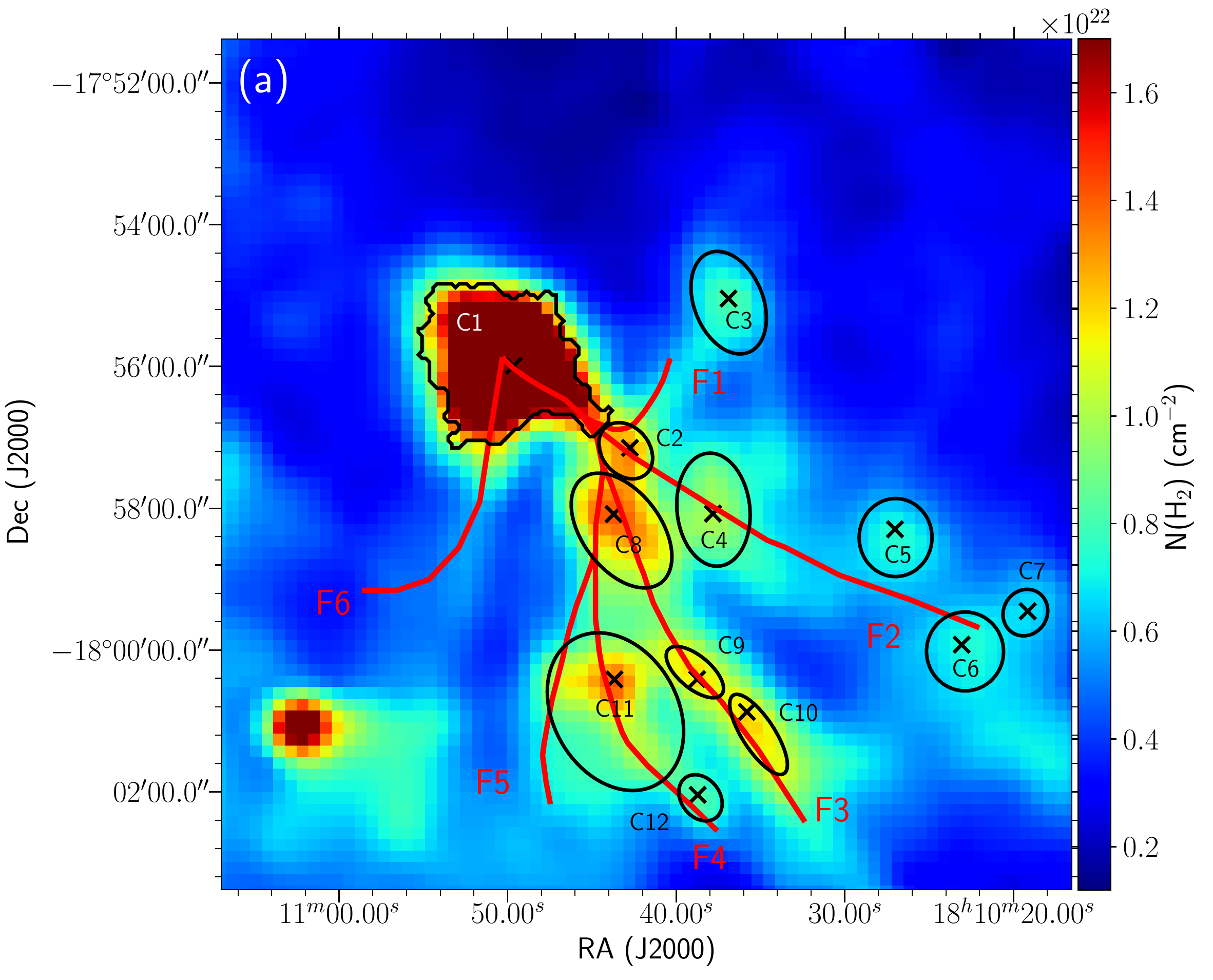} \quad\includegraphics[scale=0.30]{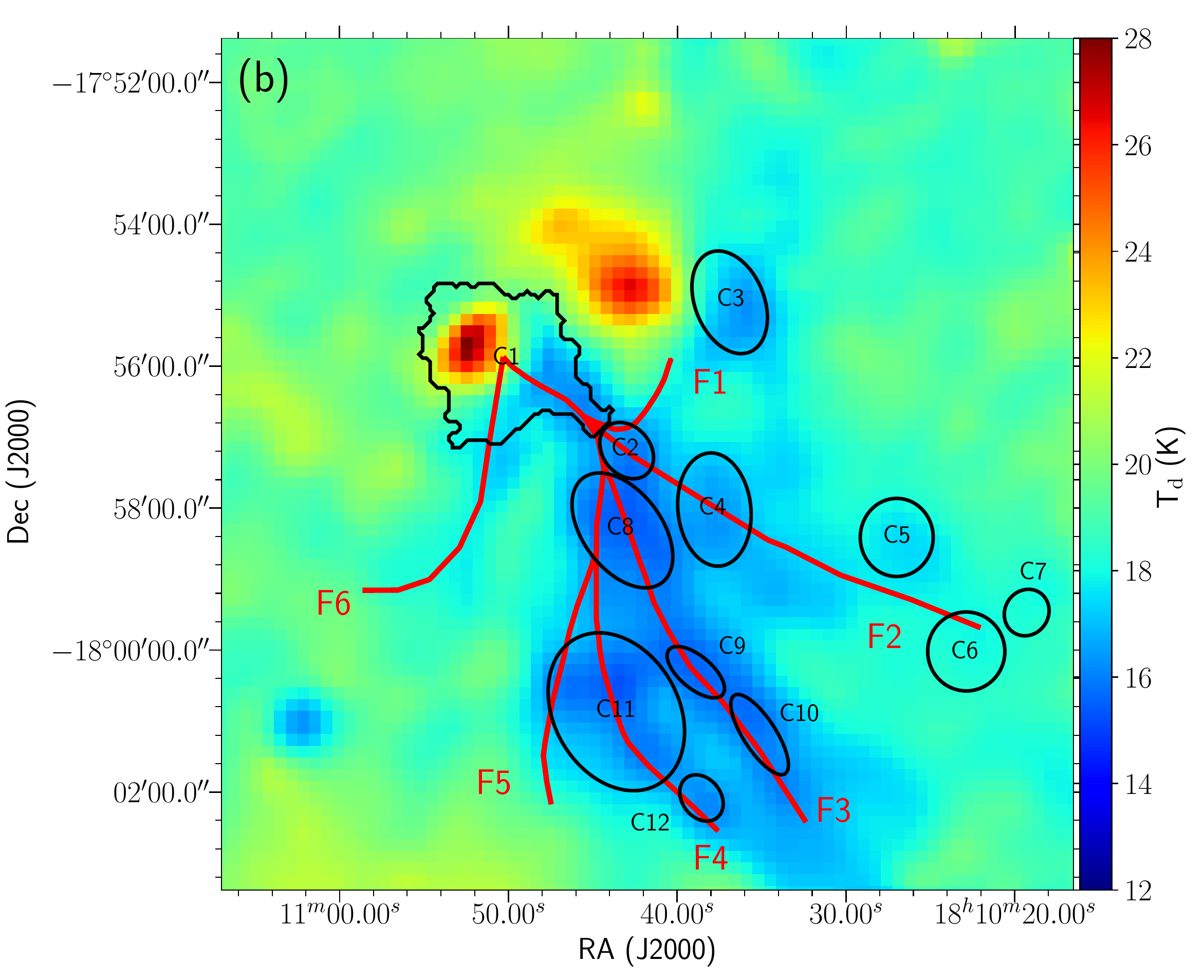} \quad\includegraphics[scale=0.30]{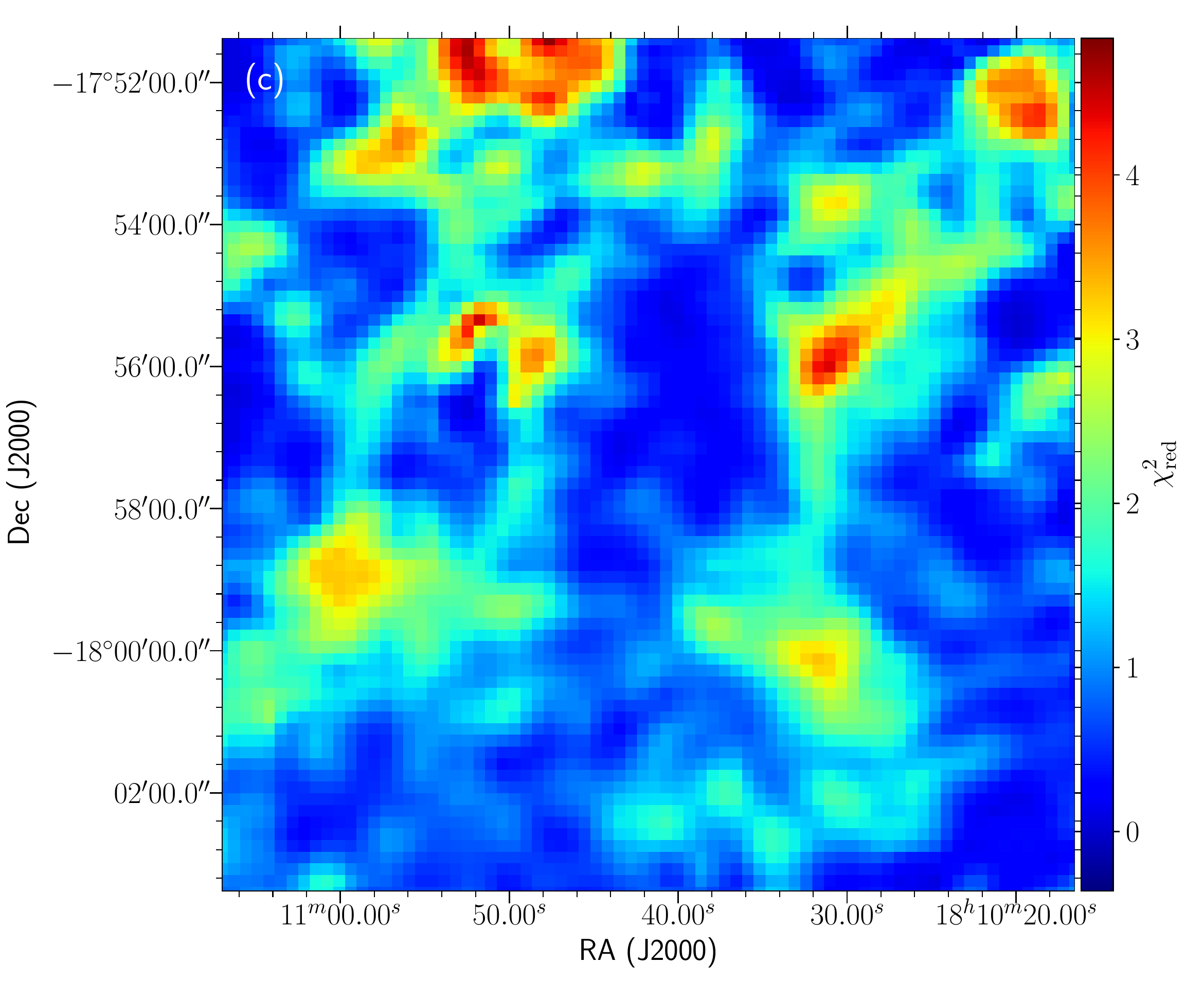}
\caption {(a) Column density, (b) Temperature and (c) Reduced ${\chi^{2}}$ maps towards {\g12} generated using the {\it Herschel} FIR data and the ATLASGAL-Planck data. The {\it Clumpfind} retrieved clump,C1 and the visually identified clumps from the column density map are marked on the maps. The `x's mark the positions of the peak column densities of each clump in the column density map. Skeletons of the filaments identified from the 8.0~{\um} map is overlaid on the column density and temperature maps.}
\label{temp_cd}
\end{figure}

%%%%%%%%%%%%%%%%%%%%%%%%%%%%%%%%%%%%%%%%%%%%%%%%%%%%%%%%%%%%%

\subsubsection{SED modelling of C1} \label{2_comp}
\par In an attempt to understand the properties of the dust clump (C1) associated with {\g12}, we model the infrared flux densities with a two-component modified blackbody using the following functional form \citep{1998ApJ...507..794L} 
\begin{equation}
S_{\nu}=[\Omega_1\;{\it a}\;B_{\nu}(T_1)  + \Omega_2\;(1-{\it a})\;B_{\nu}(T_2)] (1-e^{-\tau_\nu})
\end{equation}
where
\begin{equation}
\tau_{\nu} = \mu_{H_2}m_{H}\kappa_{\nu}N(H_2) 
\label{tau}
\end{equation}
\noindent
where $S_\nu$ is the integrated flux density of C1, $\Omega_1$ and $\Omega_2$ are the solid angles subtended by the apertures used for estimating the flux densities at the FIR and MIR wavelengths, respectively. $a$ is the ratio of optical depth in the warmer component to the total optical depth, $B_{\nu}(T_1)$ and $B_{\nu}(T_2)$ are the blackbody functions at dust temperatures $T_1$ and $T_2$, respectively, $\mu_{H_2}$ is the mean molecular weight \citep[taken as 2.8;][]{2008A&A...487..993K}, $m_{H}$ is the mass of hydrogen atom, $\kappa_{\nu}$ is the dust opacity and $N(H_2)$ is the hydrogen column density. For opacity, we assume the function $\kappa_{\nu}=0.1({\nu}/\rm 1000~GHz)^{\beta}~cm^{2}~g^{-1}$, where $\beta$ is the dust emissivity spectral index for which, a value of 2.0 is adopted as in the previous section.

\par In addition to the {\it Spitzer}-IRAC, {\it Herschel} and ATLASGAL wavebands, we have also included flux densities from the MSX survey\footnote{\url{https://irsa.ipac.caltech.edu/applications/MSX/MSX/}} at 12.13 and 14.65~{\um} to constrain the model in the MIR wavelength.  
The integrated flux densities of the dust clump at the MIR wavelengths are measured within the area defined by the 4$\sigma$ contour level of the 8.0~{\um} image (274 arcsec$^2$) and longward of 70~{\um}, the integration is done over the area defined by the {\it Clumpfind} aperture for C1 (13720 arcsec$^2$). Background emission is estimated using the same apertures on nearby sky region (visually scrutinized to be smooth) and subtracted.
Estimated flux densities are listed in Table \ref{SED_flux}.
\citet{{2009A&A...504..415S},{2013A&A...551A..98L}} use a conservative 15\% uncertainty in the flux densities of the {\it Herschel} bands. We adopt the same value here for all the bands. 
Model fitting is carried out using the non-linear least square Levenberg-Marquardt algorithm with $ T_1$, $T_2$, $N(H_2)$ and $a$ taken as free parameters. The best fit temperature values are 25.0$\pm$1.0~K (cold) and 183.2$\pm$12.0~K (warm), respectively. The model fit also gives an estimate of the hydrogen column density, $ N(H_2) = \rm 2.1 \times 10^{22}~cm^{-2} $. This result shows that the dust clump in {\g12} consists of an inner warm component surrounded by an extended outer, cold envelope traced mostly by the FIR wavelengths. It should be noted here that we have excluded the data points below 8.0~{\um} while fitting the model. This is because the emission at 4.5 and 5.8~{\um} may largely be dominated by shock excitation and the 3.6~{\um} emission may arise from even hotter components. The SED and the best fit modified blackbody are shown in {\fig}\ref{Dust_SED}. 
The bolometric luminosity estimated from the two-component SED model over $8.0-870$~{\um} is $ 2.8 \times 10^4 $~L{\sun}. It is a factor of 1.6 higher to that obtained by \citet{2013ApJ...765..129V}, who use the IRAS band flux densities. However, our values are in fair agreement to the estimate of 
$\rm 3.2 \times 10^4$~L{\sun} \citep{1997ApJS..110...71O} where flux densities between $\rm 2.1-1.3~mm$ are included.  

\subsubsection{Nature and distribution of cold dust emission} \label{cold_dust}

\par We probe the nature of the cold dust associated with {\g12}, using the {\it Herschel} FIR bands which cover the wavelength range ($ \rm 160-500 $~{\um}) and the combined ATLASGAL-Planck data at 870~{\um}. The dust temperature and the line-of-sight average molecular hydrogen column density maps are generated by a pixel-by-pixel modified single-temperature blackbody model fitting. While fitting the model, we assume the emission at these wavelengths to be optically thin. Following the discussion in several papers \citep{{2010A&A...518L..98P},{2010A&A...518L..99A},{2011A&A...535A.128B},{2018A&A...612A..36D}}, we exclude the 70~{\um} data point as the optically thin assumption would not hold. In addition, the emission here would have significant contribution from the warm dust component thus modelling with a single-temperature blackbody would over-estimate the derived temperatures. Given this, the model fitting is done with only five points which lie on the Rayleigh-Jeans tail. 

\par The first step towards the generation of the temperature and column density maps is to have the maps from SPIRE, PACS and ATLASGAL-Planck in the same units. The units of the SPIRE map which is in MJy sr$^{-1}$ is converted to Jy pixel$^{-1}$ which is the unit for the 160~{\um} PACS map. Similarly, the ATLASGAL-Planck map that has the unit of Jy beam$^{-1}$ is also converted to Jy pixel$^{-1}$. The maps are at different resolutions and pixel sizes. The pixel-by-pixel routine makes it mandatory to convolve and regrid the maps to a common resolution and pixel size of   
36{\arc} and 14{\arc}, respectively which are the parameters of the 500~{\um} map (as it has the lowest resolution).
Convolution kernels are taken from \citet{2011PASP..123.1218A} for the {\it Herschel} maps. Since no pre-made convolution kernel is available for the ATLASGAL-Planck map, we use a Gaussian kernel. These preliminary steps are carried out using the software package, HIPE\footnote{The software package for {\it Herschel} Interactive Processing Environment (HIPE) is the application that allows users to work with the Herschel data, including finding the data products, interactive analysis, plotting of data, and data manipulation.}

\par The maps include sky/background emission which is a result of the cosmic microwave background and the diffuse Galactic emission. In order to correct for the flux offsets due to this background contribution, we select a relatively uniform and dark region (free of bright, diffuse or filamentary emission) at a distance of $\sim$0.25{\degree} from {\g12}. The same region is used for background subtraction in all the five bands.  
Using the method described in several papers \citep{{2011A&A...535A.128B},{2013A&A...551A..98L},{2017MNRAS.465.4753R},{2017MNRAS.472.4750D},{2018A&A...612A..36D}}
the background values, ${I_{bg}}$ are estimated to be -2.31, 2.15, 1.03, 0.37 and 0.08 Jy pixel$^{-1}$ at 160, 250, 350, 500 and 870~{\um}, respectively. The negative flux value at 160~{\um} is due to the arbitrary scaling of the PACS images.
 
\par To probe an extended area encompassing {\g12} and the related filaments, we select  
a 12.8{\arcmin}$\times$12.8{\arcmin} region centred at $\rm \alpha_{J2000}=18^{h}10^{m}41.8^s, \delta_{J2000}=-17\degree 57\arcmin 23\arcsec$. The model fitting algorithm was based on the following formulation \citep{{1990MNRAS.244..458W},{2011A&A...535A.128B},{2013A&A...551A..98L},{2015MNRAS.447.2307M}}:

\begin{equation}
S_{\nu}(\nu)-I_{bg}(\nu) = B_{\nu}(\nu,T_d)\; \Omega\; (1-e^{-\tau_\nu})
\end{equation}
where $\tau_{\nu}$ is given by Eqn. \ref{tau}, $S_{\nu}$ is the observed flux density, $B_{\nu}(\nu,T_d)$ is the Planck function, $T_d$ is the dust temperature, $\Omega$ is the solid angle in steradians, from where the flux is measured (solid angle subtended by a $ \rm 14''\times 14'' $ pixel) and the rest of the parameters are the same as used in the previous section. 
Following the same procedure discussed in Section \ref{2_comp}, SED modelling for each pixel is carried out keeping the dust temperature, $T_d$ and column density, $N(H_2)$ as free parameters. The dust temperature and column density maps generated are displayed in {\fig}\ref{temp_cd} along with the reduced ${\chi^{2}}$ map. The reduced ${\chi^{2}}$ map indicates that the fitting uncertainties are small with a maximum value of 4 towards the bright central emission where 
the 250~{\um} image ({\fig}\ref{FIR}(e)) has a few bad pixels. The column density map reveals a dense, bright region towards clump, C1 that envelopes {\g12}. Also clear is increased density along the filamentary structures identified in Section \ref{dust}. The apertures of the clump C1 identified from the 870~{\um} is overlaid on the maps. Using $3 \times 3$ pixel grids, local column density peaks are identified above  3$ \sigma$ threshold ($\rm \sigma = 2.3\times 10^{21}~cm^{-2}$). 11 additional clumps were thus identified located within the 3$ \sigma$ contour.
Subsequent to this, a careful visual inspection is done and ellipses are marked to encompass most of the clump emission. 

\par Two high-temperature regions are seen in the dust temperature map coinciding with {\g12} and the two bubbles discussed earlier. The warmest temperature in the map is found to be 28.6~K and is located a pixel to the north-east of SMA1, SMA2 and peak position of R1. 
The mean dust temperature and column density of C1 is found to be 19.9~K and 3.3$\rm \times 10^{22}~cm^{-2}$, respectively. It has to be noted here that the mean temperature we obtain here is less than the temperature of the cold component we estimate from the two-component model by $\rm \sim 5~K$. This is because, unlike the two-component modelling, here we do not include the emission at 70~{\um}. Similarly the column density we obtain here is greater than the column density estimated using the two-component fit by a factor of $\sim 1.6$. A striking feature noticed is the distinct low dust temperatures along the filaments.    

\subsubsection{Properties of cold dust clumps} \label{clump_text}

Several physical parameters of the identified clumps are derived. The enclosed area within the {\it Clumpfind} retrieved aperture of C1 is used to determine the effective radius, $ \rm r=(A/\pi)^{0.5} $ \citep{2010ApJ...712.1137K}, where A is the area. For the visually identified clumps ($ \rm C2-C12 $), the effective radius is taken to be the geometric mean of the semi-major and semi-minor axes of the ellipses bounding the clumps. From the derived column density values, we estimate the mass of the dust clumps using the following expression
\begin{equation}
M_{\rm C} = \mu_{H_2} m_H A_{\rm pixel} \Sigma N (H_2)
\end{equation}
where $A_{\rm pixel}$ is the area of a pixel in $\rm cm^2$,  $\mu_{H_2}$ is the mean molecular weight (2.8), $m_{H}$ is the mass of hydrogen atom. The volume number density of the clump is estimated using the expression,
\begin{equation}
n_{(H_2)} =  \frac{3~M_{\rm C}}{4\pi R^3 \mu m_H}
\end{equation}
The peak position, radius, mean temperature and column density, integrated column density, mass, and volume number density of the identified clumps are listed in {\tab}\ref{clump}. The clump enclosing {\g12}, C1 is the largest and most massive clump having a radius 0.8~pc, column density $\rm 3.3\times 10^{22}~cm^{-2}$ and mass 1375~M{\sun}. \citet{2015MNRAS.450.1926H} derives the radius, column density and mass of the clump associated with {\g12} to be 0.57~pc, $\rm 1.3\times 10^{23}~cm^{-2}$ and 724~M{\sun}, respectively. Apart from a larger size estimated by us, the other factors contributing to this difference in the estimated values of mass and column density are the different opacity and dust temperature values adopted by \citet{2015MNRAS.450.1926H}.

\subsection{Molecular line emission from G12.42+0.50} \label{molecular-line}
%%%%%%%%%%%%%%%%%%%%%%%%%%%%%%%%%%%%%%%%%%%%%%%%%%%%%%%%%%
\begin{table*}
\caption{Details of the detected molecular line transitions towards the clump, C1 enveloping {\g12}. The details are extracted from Table 2 of \citet{2014A&A...562A...3M} and Table 2 of \citet{2011ApJS..197...25F}.}
\begin{center}

%\hspace*{-1.2cm}
\begin{tabular}{l l} \hline \hline 

Transition 		& Comments	\\

\hline \

H$^{13}$CO$^+~(1-0) $ 			& six hyperfine (hf) components; high-density and ionization tracer  \\
$\rm C_2H~(1-0)~3/2-1/2$		& three hf components; photodissociation region tracer\\										
HCN 	$ (1-0) $								& three hf components; high-density and infall tracer \\
$\rm HCO^+~(1-0)$ 					& high-density, infall, kinematics and ionization tracer\\	
HNC $ (1-0) $								& three hf components; high-density and cold gas tracer \\
$\rm HC_3N~(10-9) $				& six hf components; High-density and hot-core tracer\\
$\rm N_2H^+~(1-0)$					& 15 hf components, seven have a different frequency; high density and CO-depleted gas tracer\\

\hline \
\end{tabular}
\label{molecule_details}

\end{center}
\end{table*}
%%%%%%%%%%%%%%%%%%%%%%%%%%%%%%%%%%%%%%%%%%%%%%%%%%%%%%%%%%%%%%%%%%%%5
%%%%%%%%%%%%%%%%%%%%%%%%%%%%%%%%%%%%%%%%%%%%%%%%%%
\begin{figure*}
\vspace*{0.6cm}
\centering
\includegraphics[scale=0.45]{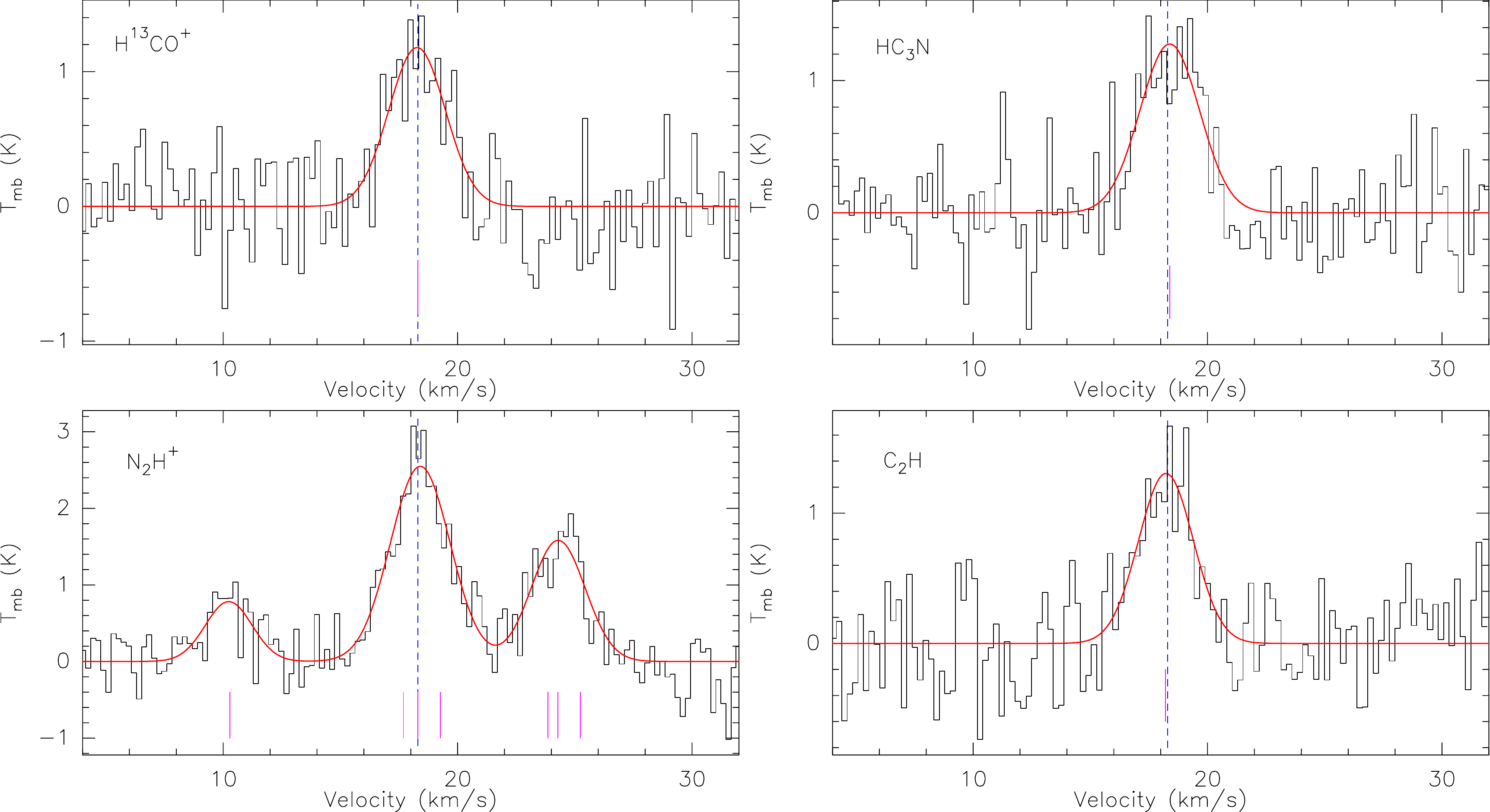} 
\caption{Spectra of the optically thin molecular lines ($\rm H^{13}CO^+$, $\rm HC_3N$, $\rm C_2H$ and $\rm N_2H^+$) associated with {\g12} obtained from the MALT90 survey. The spectra are extracted towards the peak of the 870~{\um} ATLASGAL emission. The dashed blue line indicates the LSR velocity, $\rm 18.3~km~s^{-1} $, estimated from the the optically thin $\rm H^{13}CO^+$ line. The magenta lines indicate the location of the hyperfine components for each transition.}
\label{molecule_thin}
\end{figure*}

%%%%%%%%%%%%%%%%%%%%%%%%%%%%%%%%%%%%%%%%%%%%%%%%%%%%%%%%%%%%%
 %%%%%%%%%%%%%%%%%%%%%%% optically thick %%%%%%%%%%%%%%%%%%%%%%%%%%%
\begin{figure*}
\vspace*{0.6cm}
%\centering
\includegraphics[scale=0.45]{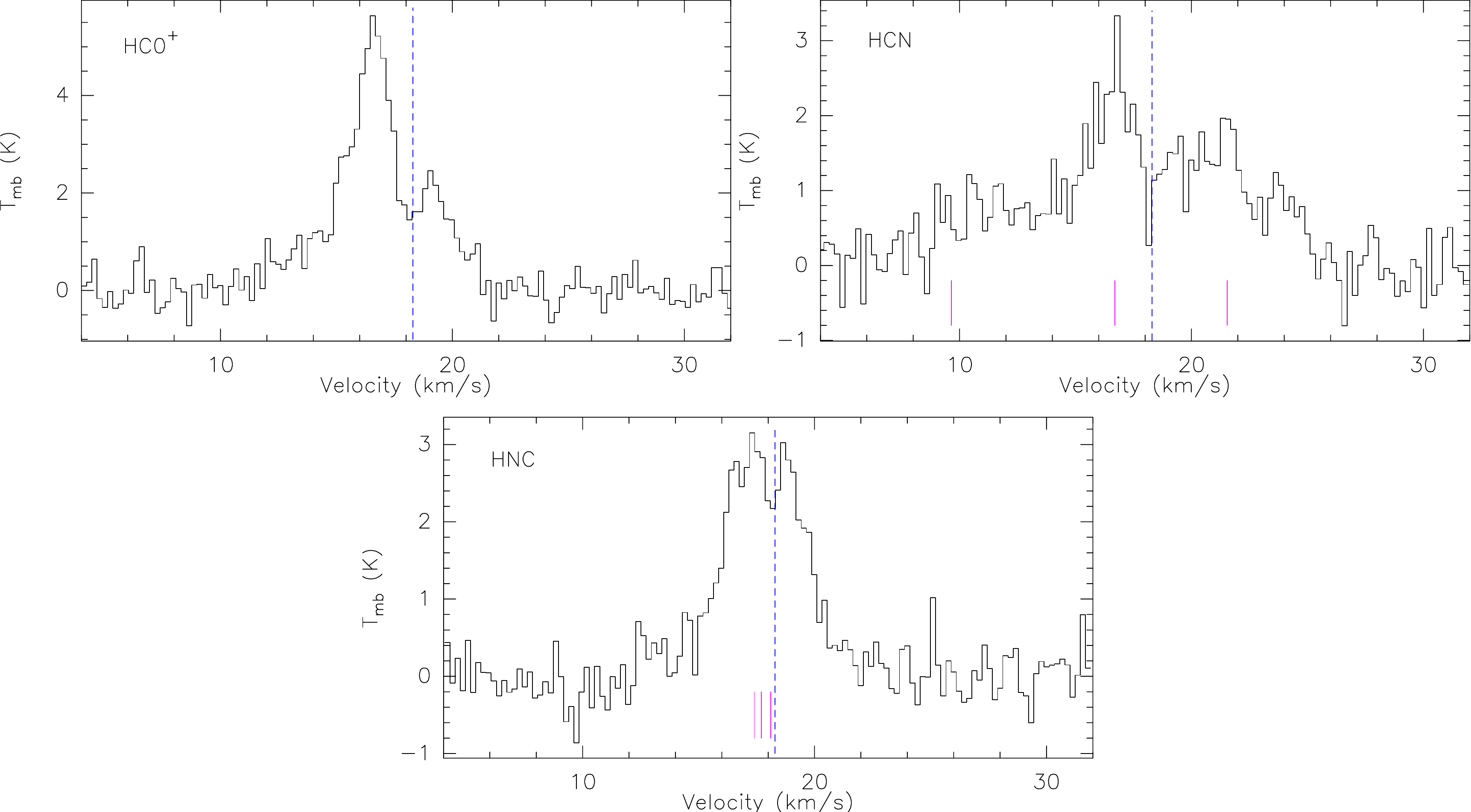}
\caption{Same as {\fig}\ref{molecule_thin} but for the optically thick transitions of $\rm HCO^+$, HCN, and HNC.}
\label{molecule_thick}
\end{figure*}

%%%%%%%%%%%%%%%%%%%%%%%%%%%%%%%%%%%%%%%%%%%%%%%%%%%%%%%%%%%%%
 
 %%%%%%%%%%%%%%%%%%%%%%%%%%%%%%%%%%%%%%%%%%%%%%%%%%%%%%%%%%%
\begin{table*}
\caption{Parameters of the optically thin molecular transitions detected towards {\g12}. The line width ($\Delta V $), main beam temperature ($ \rm T_{mb} $) and velocity integrated intensity ($\rm \int T_{mb}$) are obtained from the {\tt hfs} fitting method of CLASS90 for all the molecules except for $\rm H^{13}CO^+$, for which a single Gaussian profile is used to fit the spectrum. The column densities ($N$) of molecules are estimated using RADEX, and their fractional abundances ($x$) are determined using the mean {\h2} column density of the clump, C1.}
\begin{center}

%\hspace*{-1.2cm}
\begin{tabular}{l c c c c c} \hline \hline 

Transition 		& $\Delta V$ 	& $\rm T_{mb}$ 	& $\rm \int T_{mb}$		& $N$		&$x$\\

&($\rm km~s^{-1}$)		&(K)		&($\rm K~km~s^{-1}$)			&($\rm 10^{14}~cm^{-2}$)	&	 ($10^{-9}$)\\

\hline \

$\rm H^{13}CO^+$ 	&2.9		&1.2 	&3.6		& 0.1 	& 0.3\\
$\rm N_2H^+$		&3.2		&2.5		&8.5		& 4.1		& 12.4\\
$\rm HC_3N$		&3.0		&1.3		&4.1		&1.4 		& 4.2\\
$\rm C_2H$		&2.7		&1.3		&3.8		& 5.5		& 16.7\\

\hline \
\end{tabular}
\label{mol_table}

\end{center}
\end{table*}
%%%%%%%%%%%%%%%%%%%%%%%%%%%%%%%%%%%%%%%%%%%%%%%%%%%%%%%%%%%%%%%%%%%%5
%%%%%%%%%%%%%%%%%%%%%%%%%%%% CO %%%%%%%%%%%%%%%%%%%%%%%%%%
\begin{figure*}
\centering 
\includegraphics[scale=0.41]{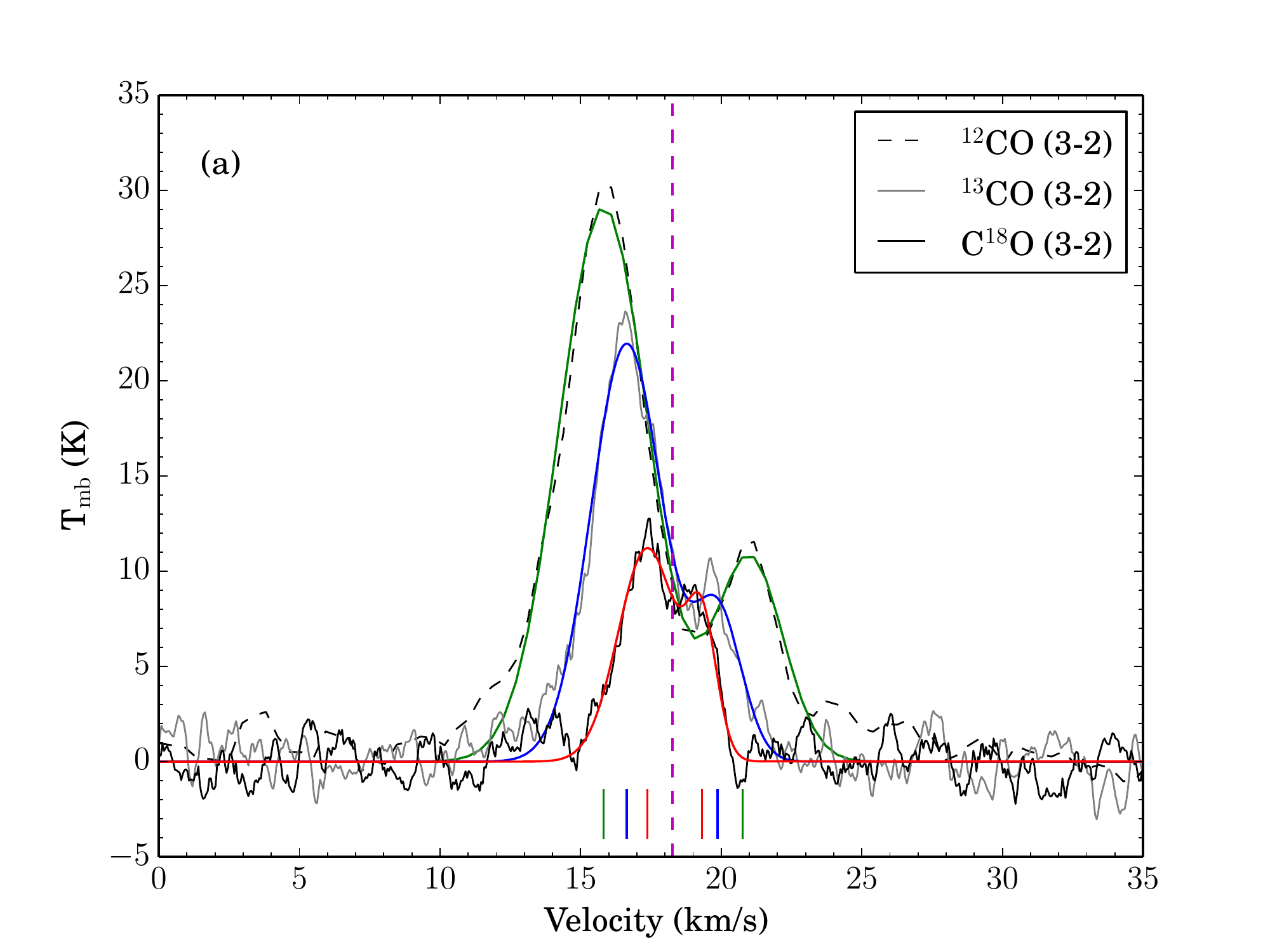} \quad\includegraphics[scale=0.41]{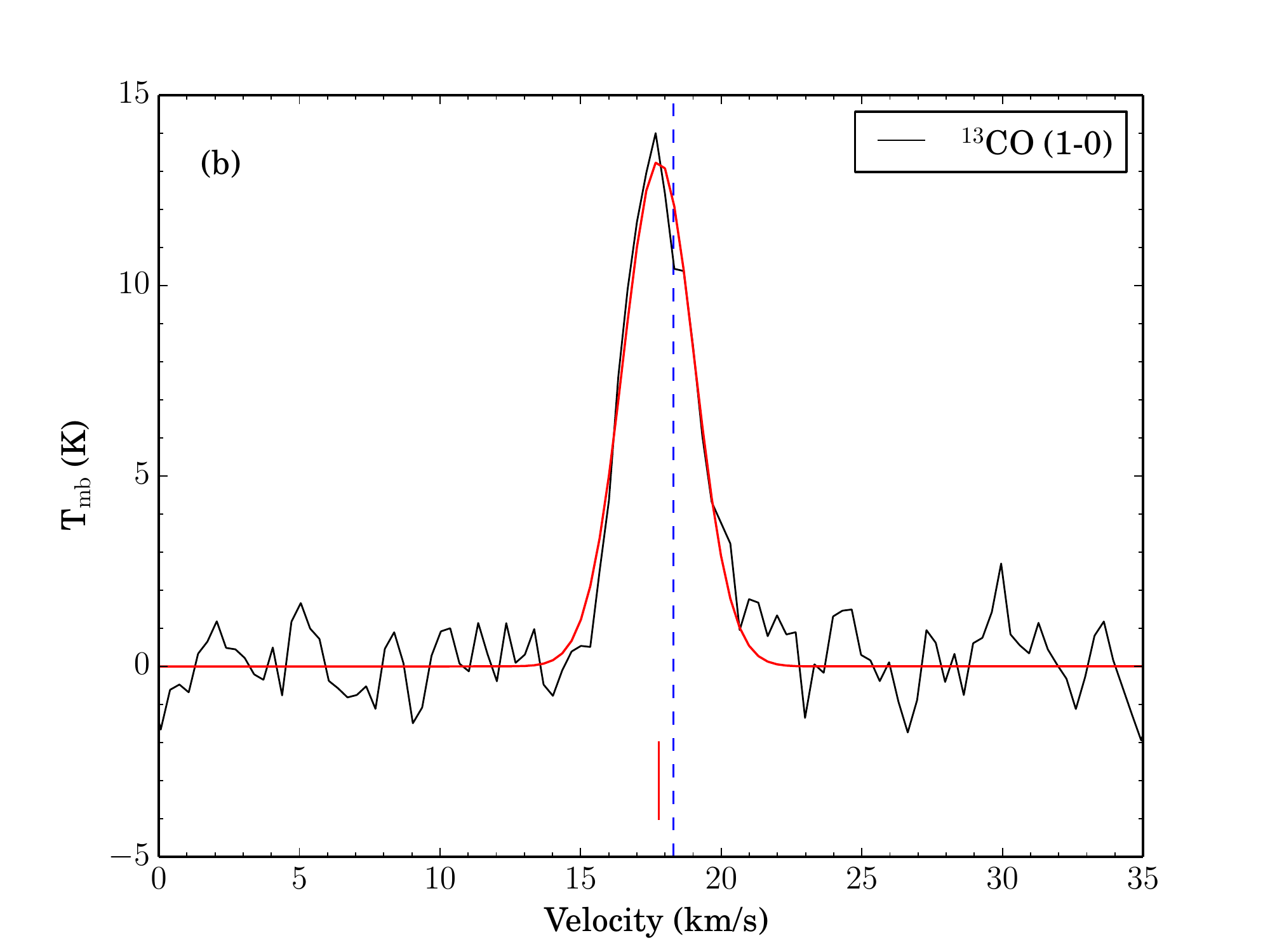}
\caption{(a) Rotational transition lines of isotopologues of the $\rm CO~(3-2)$ observed towards {\g12} fitted with double Gaussians. The spectra of  $\rm ^{12}CO$, $\rm ^{13}CO$ and $\rm C^{18}O$ are boxcar-smoothed by three, eight and eleven channels which correspond to velocity smoothing of 1.2, 3.4 and $\rm 4.6~km~s^{-1}$, respectively. The fit to the $\rm ^{12}CO$ spectrum is depicted in green, $\rm ^{13}CO$ in blue and $\rm C^{18}O$ in red. The dashed magenta line corresponds to the LSR velocity, $\rm 18.3~km~s^{-1}$. The positions of the red- and blueshifted components are indicated in green, blue and red lines for $\rm ^{12}CO$, $\rm ^{13}CO$ and $\rm C^{18}O$ lines respectively. (b) Spectrum of the $J=1-0$ transition of $\rm ^{13}CO$ obtained from TRAO. The Gaussian fit to the spectrum is sketched in red. The dashed blue line corresponds to the LSR velocity. The spectrum shows a blue shifted single peak, indicated by a red line. The red and blue lobes are not resolved, probably due to larger beam size of TRAO compared to JCMT. }
\label{CO_plot}
\end{figure*}
%%%%%%%%%%%%%%%%%%%%%%%%%%%%%%%%%%%%%%%%%%%%%%%%%%%%%%%%
%%%%%%%%%%%%%%%%%%%%%%%%%%%%%%%%%%%%%%%%%%%%%%%%%%%%%%%%%%%
\begin{table*}
\caption{The retrieved parameters, peak velocities, velocity widths and peak fluxes of the molecular transitions $\rm ^{12}CO~(3-2)$, $\rm ^{13}CO~(3-2)$, $\rm C^{18}O~(3-2)$ and $\rm ^{13}CO~(1-0)$ towards {\g12}. R and B in parentheses denote the red and blueshifted components.}
\begin{center}

%\hspace*{-1.2cm}
\begin{tabular}{l c c c c c c } \hline \hline 

Transition 		&\multicolumn{2}{c}{$V$} & \multicolumn{2}{c}{ $\Delta V$} 	& \multicolumn{2}{c}{$\rm T_{mb}$} 	\\
						&\multicolumn{2}{c}{($\rm km~s^{-1}$)}				&\multicolumn{2}{c}{($\rm km~s^{-1}$)}		&\multicolumn{2}{c}{(K)}\\

\hline \
$\rm ^{12}CO~(3-2)$		&21.0 (R)	 & 15.8 (B)		&	2.8 (R)		&3.7 (B)		&10.7 (R)		&29.1(B) \\
$\rm ^{13}CO~(3-2)$		&19.9 (R)		&	16.6 (B)		&	2.0 (R)		&3.9 (B)		&7.7 (R)			&21.9  (B)\\
$\rm C^{18}O~(3-2)$ 	&19.3 (R)		&	17.4 (B)	&	1.2 (R)		&2.4 (B)		&6.6 (R)			&11.2 (B) \\
$\rm ^{13}CO~(1-0)$		&\multicolumn{2}{c}{17.8}		&\multicolumn{2}{c}{3.0}	&\multicolumn{2}{c}{13.3} \\

\hline \
\end{tabular}
\label{CO_table}

\end{center}
\end{table*}
%%%%%%%%%%%%%%%%%%%%%%%%%%%%%%%%%%%%%%%%%%%%%%%%%%%%%%%%%%%%%%%%%%%%
The molecular line emission provides information on the kinematics and chemical structure of a molecular cloud in addition to throwing light on its evolutionary stage. Data from the MALT90 survey, JCMT archives and observation from TRAO are used to probe these aspects in the star forming region associated with {\g12}.

\par Of the 16 molecules covered by the MALT90 survey, 7 molecular species, namely $\rm HCO^+$, $\rm H^{13}CO^+$, HCN, HNC, $\rm C_2H$, $\rm N_2H^+$ and 
$\rm HC_3N$ are detected towards the clump C1 enveloping {\g12}. The details of the detected transitions taken from \citet{2014A&A...562A...3M} and \citet{2011ApJS..197...25F} are listed in {\tab}\ref{molecule_details}. \citet{2014A&A...562A...3M} also gives an excellent review on the physical conditions and environment required for the formation of these species. The spectrum of each molecule is extracted towards the 870 {\um}, ATLASGAL emission peak. The spectra of the optically thin molecular species,
$\rm H^{13}CO^+$, $\rm C_2H$, $\rm N_2H^+$ and $\rm HC_3N$, are shown in {\fig}\ref{molecule_thin} and the spectra of the optically thick molecular species, $\rm HCO^+$, HCN and HNC, are plotted in {\fig}\ref{molecule_thick}. 
We use the hyperfine structure ({\tt hfs}) method of CLASS90 to fit the observed spectra for the optically thin transitions of $\rm C_2H$, $\rm N_2H^+$ and $\rm HC_3N$. Since the molecule $\rm H^{13}CO^+$ has no hyperfine components, a single Gaussian profile is used to fit the spectrum. The Gaussian fit yields a LSR velocity of $\rm 18.3~km~s^{-1}$, which is consistent with the value estimated using the $\rm N_2H^+$ line of the same survey ($\rm 18.3~km~s^{-1}$; \citealt{2015MNRAS.451.2507Y}). The fit to the spectra are indicated by solid red line, and the LSR velocity and the location of the hyperfine components are indicated by the dashed blue and solid magenta lines, respectively in {\fig}\ref{molecule_thin}.
The retrieved line parameters that include the peak velocity ($V_{\rm LSR}$), line width ($\Delta V$), main beam temperature ($\rm T_{mb}$) and the velocity integrated intensity ($\rm \int T_{mb}$) are tabulated in {\tab}\ref{mol_table}. Beam correction is applied to the antenna temperature to obtain the main beam temperature using the equation,  
 $\rm T_{mb} = T_A/\eta_{mb}$ \citep{2014ApJ...786..140R}, where $\rm \eta_{mb}$ is assumed to be 0.49 \citep{2005PASA...22...62L} for the MALT90 data.

\par To estimate the column density of these transitions, we use RADEX, a one dimensional non-local thermodynamic equilibrium radiative
 transfer code \citep{2007A&A...468..627V}. The input parameters to RADEX include the peak main beam temperature,  background temperature assumed to be 2.73 K \citep{{2006MNRAS.367..553P},{2015MNRAS.451.2507Y}},  kinetic temperature, which is assumed to be same as the dust temperature \citep{{2012ApJ...756...60S},{2016ApJ...833..248Y}, {2016ApJ...818...95L}}, line width, and {\h2} number density. The dust temperature and {\h2} number density towards the clump, C1 of {\g12} are taken from {\tab}\ref{clump} presented in Section \ref{clump_text}. The column densities of the optically thin transitions are also tabulated in {\tab}\ref{mol_table}. From the mean hydrogen column density of the clump, we also calculate the fractional abundances of the detected molecules. These estimates are in good agreement with typical values obtained for IR dark clumps and IRDCs \citep{{2014A&A...562A...3M},{2011A&A...527A..88V}}. 

 \par From {\fig}\ref{molecule_thick}, it is evident that the $ J=1-0 $ transitions of the molecules, $\rm HCO^+$, HCN and its metastable geometrical isomer, HNC, display distinct double-peaked line profiles with self-absorption dips coincident with the LSR velocity. The blue-skewed profile seen in $\rm HCO^+$ is very prominent with the blueshifted emission peak being much stronger than the redshifted one. In case of the HCN transition, the central hyperfine component shows a blue-skewed double profile where the redshifted component is rather muted in the noise. Such blue asymmetry is usually indicative of infalling  gas \citep{{2003ApJ...592L..79W}, {2016A&A...585A.149W}}. In Section \ref{infall}, we discuss in detail the $\rm HCO^+$ line profile. In comparison, in the HNC transition, the blueshifted and redshifted peaks have similar intensities. Similar line profiles are detected towards the star forming region AFGL 5142 \citep{2016ApJ...824...31L}. These authors have attributed it to low-velocity expanding materials entrained by high-velocity jets. An alternate reason could be of a collapsing envelope. In case of {\g12}, however, no conclusive explanation can be proposed given the resolution of the data. Higher resolution observations are hence required to resolve the kinematics and explain the double peaked profile of HNC.

\par The rotational transition line data of the isotopologues of the CO molecule,
 $\rm ^{12}CO~(3-2)$, $\rm ^{13}CO~(3-2)$ and $\rm C^{18}O~(3-2)$ taken from  archives of JCMT and $\rm ^{13}CO~(1-0)$ observed with TRAO are used to understand the large-scale outflows associated with {\g12}. The rotational transitions of the CO molecule is an excellent tracer of outflow activity in star forming regions \citep{{2001ApJ...552L.167Z},{2002A&A...383..892B}}. Different transitions trace different conditions of the ISM and probe different parts of the cloud. While the CO $J=3-2$ transition has a distinct upper energy level temperature and critical density of 33.2~K and $ \rm 5 \times 10^4~cm^{-3} $, respectively \citep{1999ApJ...527..795K}, the lower $J$ CO transitions effectively trace the kinematics of low density material of the cloud \citep{2013A&A...549A...5R}. Typically, the $\rm ^{12}CO$ line is optically thick and the $\rm ^{13}CO$ and $\rm C^{18}O$ lines are optically thin and are high density tracers. While $\rm ^{12}CO$ can effectively map the spatial and kinematic extent of the outflows and $\rm ^{13}CO$ can map them to some extent, the $\rm C^{18}O$ can trace the cloud cores under the optically thin assumption \citep{2015MNRAS.453.3245L}.
The spectra of these molecular species are extracted towards the peak of the 870~{\um}, ATLASGAL emission and shown in {\fig}\ref{CO_plot}(a) and (b). The spectra of the isotopologues of $\rm CO~(3-2)$ transition show red and blueshifted profiles. However, the $\rm ^{13}CO~(1-0)$ transition shows a single component, blueshifted profile. This is due to the large beam size of TRAO where the blue and the red components are unresolved. A double Gaussian is used to fit the 
 spectra of  $\rm ^{12}CO~(3-2)$, $\rm ^{13}CO~(3-2)$ and $\rm C^{18}O~(3-2)$, and a single Gaussian profile is fitted to the $\rm ^{13}CO~(1-0)$ line. The fitted profiles are also shown in the Figures. 
The retrieved parameters are peak velocities, velocity widths, and peak fluxes 
which are listed in {\tab}\ref{CO_table}.
Beam correction is applied to the antenna temperature, taking $\rm \eta_{mb}$ to be 0.64 for the JCMT \citep{2009MNRAS.399.1026B} and 0.54  for TRAO \citep{2018ApJS..234...28L}.  
Detailed discussion on the outflow feature will be presented in Section \ref{outflow}.

\section{DISCUSSION} \label{discussion}
\subsection{Nature of radio emission}\label{nat_rad}
Based on the GMRT maps and the radio spectral index estimation, two scenarios unfold in understanding the nature of the radio emission. The thermal radio emission could be explained as due to individual
ultracompat (UC) {\hii} regions or given the association with an EGO, one can 
explore the case of an ionized jet. We discuss the possibilities of these two scenarios in the following sections. 

\subsubsection{UC {\hii} region} \label{UCHii}
%%%%%%%%%%%%%%%%%%%%%%%%%%%%%%%%%%%%%%
\begin{table*}
\caption{Physical parameters of the radio continuum emission from the UC {\hii} region associated with component R1 of {\g12}.}
\begin{center}

%\hspace*{-1.2cm}
\begin{tabular}{c c c c c c c c c c} \hline \hline 
& \

Source 		& $\theta_{src}$  &Radius 	& $T_e$ 	& $N_{Ly}$  & log ($N_{Ly}$)		&$EM$ 	& $n_e$	 & $t_{dyn}$ \\

&			&(arcsec)		& (pc)	& (K	)    	& ($10^{45}$ s$^{-1}$)   && (pc~cm$^{-6}$ )	&(cm$^{-3}$)		& ($10^{-3}$~yr)	\\

\hline \
 & R1 			& 1.8	& 0.01		&7416$\pm$437	&4.1 	&45.6		& 1.8 $\times$ 10$^6$		& 9.4$\times$ 10$^3$	& 0.4    \\

\hline \
\end{tabular}
\label{radio_physical_param_tab}

\end{center}
\end{table*}
%%%%%%%%%%%%%%%%%%%%%%%%%%%%%%%%%%%%%%%%%%%
We first investigate under the UC {\hii} region framework.   
Morphologically, R1 appears to be a compact, spherical radio source. The association of R1 with a hot molecular core ($\sim $183 K; Section \ref{2_comp}) supports the interpretation of the emission as being due to photoionization, since hot cores are often associated with UC {\hii} regions \citep{{2000prpl.conf..299K},{2002ARA&A..40...27C},{2016A&A...593A..49B}}.  
Assuming the continuum emission at 1390~MHz to be optically thin and arising from a homogeneous, isothermal medium, we derive the 
Lyman continuum photon flux ($N_{Ly}$), the emission measure (EM) and the electron number density ($n_e$). These physical parameters are 
estimated using the following formulation \citep{2016A&A...588A.143S}
\begin{equation}
\bigg[\frac{N_{Ly}}{\textrm s^{-1}}\bigg] = 4.771 \times 10^{42}  \bigg[\frac{S_\nu}{\textrm {Jy}}\bigg] \bigg[\frac{T_e}{ \textrm {K}}\bigg]^{-0.45} \bigg[\frac{\nu}{\textrm {GHz}}\bigg]^{0.1} \bigg[\frac{d}{\textrm {pc}}\bigg]^2
\end{equation}
\begin{multline}
\bigg[\frac{EM}{\textrm{pc}\; \textrm{cm}^{-6}}\bigg] = 3.217 \times 10^{7}  \bigg[\frac{S_\nu}{\textrm{Jy}}\bigg] \bigg[\frac{\nu}{\textrm{GHz}}\bigg]^{0.1} \bigg[\frac{T_e}{\textrm{K}}\bigg]^{0.35} \\ \bigg[\frac{\theta_{src}}{\textrm{arcsec}}\bigg]^{-2}
\end{multline}
\begin{multline}
\bigg[\frac{n_e}{\textrm{cm}^{-3}}\bigg] = 2.576 \times 10^{6}  \bigg[\frac{S_\nu}{\textrm{Jy}}\bigg]^{0.5} \bigg[\frac{\nu}{\textrm{GHz}}\bigg]^{0.05} \bigg[\frac{T_e}{\textrm K}\bigg]^{0.175} \\ \bigg[\frac{\theta_{src}}{\textrm {arcsec}}\bigg]^{-1.5} \bigg[\frac{d}{\textrm {pc}}\bigg]^{-0.5}
\end{multline}
\noindent
where $S_\nu$ is the integrated flux density of the ionized region, $T_e$ is the electron temperature, $\nu$ is the frequency, $\theta_{src}$ 
is the deconvolved size of the ionized region, and $d$ is the distance to the source. We estimate $T_e$ from the derived electron temperature gradient in the Galactic disk by  \citet{2006ApJ...653.1226Q}.
We use their empirical relation,  
$ T_e = (5780 \pm 350) + (287 \pm 46) R_G$, where $R_G$ is the Galactocentric distance. $R_G$ is estimated to be 5.7~kpc following \citet{2008ApJ...684.1143X}.
This yields an electron temperature of 7416$\pm$437~K.
The derived physical parameters of the UC {\hii} region are listed in 
Table \ref{radio_physical_param_tab}. 

\par If a single ZAMS star is responsible for the ionization of this UC {\hii} region, then from \citet{1973AJ.....78..929P}, the estimated Lyman continuum 
photon flux corresponds to a spectral type of $\rm B1-B0.5$.  Following \citet{2011MNRAS.416..972D}, the Lyman
continuum flux from the UC {\hii} region is suggestive of a massive star of mass 
$\sim 9-12$~M\sun. As discussed earlier, the estimate is made under the assumption of optically thin emission. 
Hence, this result only provides a lower limit, since the emission at 1390~MHz could be partially optically thick as is evident from our radio spectral index estimations. In addition to this, several studies show that there could be appreciable absorption of Lyman continuum photons by dust \citep{{2001ApJ...555..613I},{2004ApJ...608..282A},{2011A&A...525A.132P}}. It is further noticed that if the total infrared luminosity of {\g12} ($2.8 \times 10^4$, Section \ref{2_comp}) were to be produced by a ZAMS star, it would correspond to a star with spectral type between $\rm B0-O9.5$ \citep{1973AJ.....78..929P}. Taking a B0 star, the Lyman continuum photon flux is expected to be $ \rm 2.3 \times 10^{47}~s^{-1}$. At optically thin radio frequencies such a star could generate an {\hii} region with a flux density of $\sim 400$~mJy, which is higher than observed flux density value of 7.9~mJy observed. This could be suggestive of the central source going though a strong accretion phase, with it still being in a pre-UC {\hii} or very early UC {\hii} region phase \citep{2010ApJ...725..734G}. An intense accretion activity could stall the expansion of the {\hii} region which results in weaker radio emission. 
The above picture is congruous with the infall scenario associated with R1 and 
the evidence of global collapse of the molecular cloud associated with {\g12}. Detailed discussion on molecular gas kinematics are presented in Section \ref{infall}. 

%%%%%%%%%%%%%%%%%%%%%%%%%%%%% kendall %%%%%%%%%%%%%%%%%%%%%%%%%%%%%%
\begin{figure}
%\hspace{-0.5cm}
\centering
\includegraphics[scale=0.38]{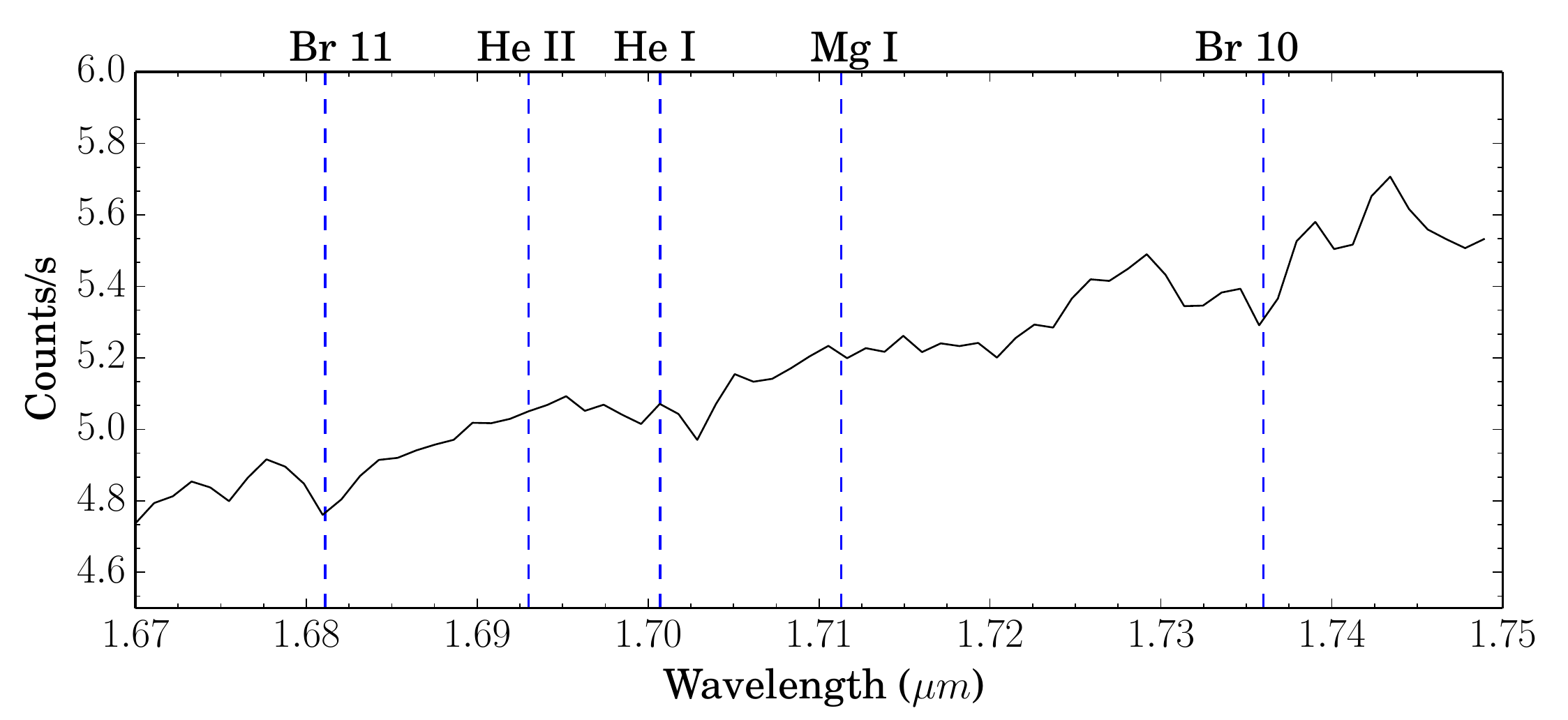} 
\caption{The UKIRT-UIST spectrum extracted over a 6 pixel wide aperture centered on the radio component, R1 is shown here. The spectral range is chosen to be the same as the VLT-ISAAC spectrum towards the infrared source, IRAS~18079-1756, associated with {\g12}, studied by \citet{2003A&A...408..313K}. The absorption lines identified by these authors are indicated in the plot.}
\label{kendall}
\end{figure}
%%%%%%%%%%%%%%%%%%%%%%%%%%%%%%%%%%%%%%%%%%%%%%%%%%%%%%%%%%%%%%%%%%%%

\par From the Lyman continuum photon flux and the electron density estimates, we compute the radius of the Str\"omgren sphere, 
that is defined as the radius at which the rate of ionization equals the rate of recombination, assuming that the {\hii} region is expanding 
in a homogeneous and spherically, symmetric medium. The radius of the Str\"omgren sphere, $R_s$ is given by the expression,
 
\begin{equation}
R_s = \bigg(\frac{3 N_{Ly}}{4 \pi n_{0}^2 \alpha_B}\bigg)^{1/3}
\end{equation}

\noindent
where $\alpha_B$ is the radiative recombination coefficient taken to be $\rm 2.6 \times 10^{-13}~cm^3~s^{-1}$ \citep{1997ApJ...489..284K} and $n_0$ is 
the mean number density of atomic hydrogen which is estimated to be $\rm 2.1 \times 10^4~cm^{-3}$ from the clump detected in the column density map (Section \ref{cold_dust}). Thus, the radius of the Str\"omgren sphere, $R_s$ for the resolved component, R1 is calculated to be 0.007~pc. 
Using this, the dynamical age the {\hii} region is determined from the expression
\begin{equation}
t_{dyn} = \bigg[\frac{4\, R_s}{7\, c_i}\bigg] \bigg[\bigg(\frac{R_{\hii}}{R_s}\bigg)^{7/4} - 1\bigg]
\end{equation}
\noindent
where $R_{\hii}$, is the radius of the {\hii} region, $c_i$ is the isothermal sound speed in the ionized medium, which is typically assumed to be $\rm 10~km~s^{-1}$. 
$R_{\hii}$ is estimated to be 0.01~pc by taking the geometric mean of the deconvolved size given in {\tab}\ref{peak_position}. The dynamical age of the
UC {\hii} region associated with component R1 is determined to be $ 0.4 \times 10^3$~yr. Since this estimation is made under a not so realistic assumption that the medium in which the {\hii} region expands is homogeneous, the results obtained may be considered representative at best. 
The derived physical parameters of the UC {\hii} region are tabulated in {\tab} \ref{radio_physical_param_tab}. 
The estimated values of electron density and emission measure are in the range found for UC {\hii} regions around stars of spectral type $\rm B1-B0.5 $ \citep{1994ApJS...91..659K}. Furthermore, the size estimates for UC {\hii} regions are proposed to be $\rm \lesssim 0.1~pc$ \citep{{1989ApJS...69..831W},{2002ASPC..267...81K}}, in agreement with that derived for the component R1. The dynamical timescales obtained indicate a very early phase of the UC {\hii} region \citep{{1989ApJS...69..831W},{2002ARA&A..40...27C}}. \citet{1989ApJS...69..831W} estimate that it would take  $\sim 10^4$~yr for an UC {\hii} region to expand against the gravitational pressure of the confining dense molecular cloud. 
\par On a careful scrutiny of the point sources in the region, it is seen that a red  2MASS\footnote{This publication makes use of data products from the Two Micron All Sky Survey, which is a joint project of the University 
of Massachusetts and the Infrared Processing and Analysis Center/California Institute of Technology, funded by the NASA and the NSF} source (J18105109-1755496; $J$ = 13.727, $H$ = 11.011, $K$ = 9.351) is located at the peak position of R1 (within $\sim 0.3''$). 
Investigating its location in $JHK$ colour-colour diagrams (e.g {\fig} 6(d) of \citet{2017MNRAS.472.4750D}) suggests a highly embedded Class II YSO  which in all likelihood could be the ionizing source. Detailed spectroscopic observations of this source is presented in \citet{2003A&A...408..313K}. 
In the observed wavelength range of $1.67-1.75$~{\um}, the VLT/ISAAC $H$-band spectra, presented by these authors, show the presence of broad absorption features of He I ($\sim$ 1.7~{\um}) and hydrogen. We did a careful examination of our UKIRT spectroscopic observations. We extracted the spectrum over a 6 pixel wide aperture (estimated from other stellar sources along the slit) centred on R1, a zoom in of which is shown in {\fig}\ref{kendall}. The spectral range is chosen such that it matches the VLT spectrum of \citet{2003A&A...408..313K} (refer {\fig}5 of their paper). Inspite of the poor signal-to-noise, the spectrum does show hint of 
the Br 11 line and possibly the Br 10 line as well as detected by \citet{2003A&A...408..313K}. Based on these absorption lines and the absence of the 1.693~{\um} He II absorption line, \citet{2003A&A...408..313K} suggest this source to be a main-sequence star of spectral type B3 ($\pm 3$ subclasses). This is consistent with the spectral type derived from our measured radio flux densities. However, the absence of emission lines in their observed spectra prompted the authors to speculate a late evolutionary stage. This contradicts the results obtained from our $HK$ and $KL$ spectra which show the presence of several emission lines that are listed in Table \ref{spectral_lines}, indicating an early evolutionary phase. The results from the molecular line analysis discussed in Section \ref{kinematic_signature} is also in agreement with this picture.
The compact component R2 can either be an independent UC {\hii} region or an externally ionized density clump. If we consider it as an UC {\hii} region then
the observed Lyman continuum flux translates to an ionizing source of spectral type $\rm B3-B2$ \citep{1973AJ.....78..929P} and a mass of $6-9$~M\sun \citep{2011MNRAS.416..972D}. 

\subsubsection{A possible thermal jet?} \label{jet}

\par Even with the compelling possibility of R1 being an UC {\hii} region, we explore an alternate scenario along the lines of a possible thermal jet. This is motivated by the very nature of {\g12} which is identified as an EGO and hence likely to be associated with jets/outflows. Further, several observational manifestations are consistent with the characteristics of thermal radio jets listed in \citet{1996ASPC...93....3A} and \citet{1997IAUS..182...83R}.

{\g12} is a weak radio source (integrated flux density $<$ 10~mJy) displaying a linear morphology, including components R1 and R2, in the north-east and south-west direction. It is also seen to be associated with a large scale molecular outflow ({\fig}\ref{outflow_moment}(a), Section \ref{outflow}) with the candidate jet located at its centroid position and the observed elongation aligned with the outflow axis. From the radio spectral index map shown in {\fig}\ref{specind}, we see that along the direction of the radio components R1 and R2, the spectral index varies between $\sim 0.3-0.7$. These values of spectral index are consistent with the radio continuum emission originating due to the thermal free-free emission from an ionized collimated stellar wind \citep{{1975A&A....39....1P},{1986ApJ...304..713R},{1998AJ....116.2953A}}. Similar range of spectral index values are also cited in literature for systems harbouring thermal radio jets \citep{{1994ApJ...420L..91A}, {2010ApJ...725..734G},{2016A&A...596L...2S}}. Additional support for the thermal jet hypothesis comes from the angular size spectrum. \citet{2016ApJ...826..208G} and \citet{2017ApJ...843...99H} have discussed the trend of the angular size spectrum where the jet features show a decrease in size with frequency as expected from the variation of electron density with frequency \citep{{1975A&A....39....1P},{1986ApJ...304..713R}}. In case of {\g12}, the 1390~MHz and 5~GHz sizes show this trend with the upper limit from 610~MHz being consistent. It should be noted here that in the 5~GHz map, all structures upto $\sim 20\arcsec$ would be well-imaged \citep{2009A&A...501..539U}. However, the size dependence is not conclusive given the resolution of the two maps.
Presence of shock-excited emission lines in the NIR (Section \ref{NIR_spectroscopy}) further corroborates with this ionized jet scenario. Additionally, a $\rm H_2O$ maser is seen to be associated with {\g12} \citep{2013ApJ...764...61C}, located at an angular distance of $ \sim $12{\arc} from the radio peak.
The position of this is indicated in {\fig}\ref{irac_ukidss_rgb}(b) and (c) and {\fig}\ref{outflow_moment}(b). $\rm H_2O$ masers have often been found in the vicinity of thermal radio jets, and in some cases both the thermal jet and 
$\rm H_2O$ masers are powered by the same star \citep{1995ApJ...453..268G}. 

\par The two competing schemes deliberated above are in good agreement with our observation making it difficult to be biased towards any. However, recent studies speculate about the co-existence of UC/HC {\hii} regions and ionized jets. From the investigation of the nature of the observed centimetre radio emission in G35.20-0.74N, \citet{2016A&A...593A..49B} discuss the possibility of it being a UC {\hii} region as well as a radio jet being powered by the same YSO suggesting an interesting transitional phase where the UC {\hii} region has started to form while infall and outflow processes of the main accretion phase is still ongoing. Similar scenario is also invoked for the MYSO, G345.4938+01.4677 by \citet{2016ApJ...826..208G}. Both these examples conform well with our results. 
\citet{2010ApJ...725..734G} discuss about a 
string of radio sources which are likely to be the ionized emission due to shocks from fast jets wherein the separation of the inner 
lobe from the central object is $\sim 0.03$~pc. \citet{2003ApJ...587..739G} also examines a radio triple source in which case the central
source harbours a high-mass star in an early evolutionary phase and ejects collimated stellar wind which ionizes the surrounding medium
giving rise to the observed radio emission. In this case, the separation between the central source and the radio lobe is $\sim$ 0.14~pc. 
For {\g12}, component R2, at a distance of $\sim 0.07$~pc from R1, can also be conjectured to be a clumpy, enhanced density region (SMA3) ionized by the emanating jet. The star forming region of {\g12} has also been speculated to be harbouring a cluster \citep{{1984ApJ...281..225J},{2003A&A...408..313K}}. With the detection of R1, R2, SMA1 and SMA2, it reveals itself as a potentially active star forming complex.

\subsection{Kinematic Signatures of gas motion} \label{kinematic_signature}

\subsubsection{Infall activity} \label{infall}
%%%%%%%%%%%%%%%%%%%%%%%%%%%%%%%%%%%%%%%%%%%%%%%%%%%%%%
\begin{table}
\caption{The infall velocity, $V_{\rm inf}$ and mass infall rate, $\dot{M}_{\rm inf}$ of the clump, C1 associated with {\g12}, estimated using the blue-skewed optically thick $\rm HCO^+$ line }
\begin{center}

%\hspace*{-1.2cm}
\begin{tabular}{c c c c} \hline \hline 

$V_{\rm LSR}$ 		&$V_{\rm inf}$  	& $ \delta V$ 	& $\dot{M}_{\rm inf}$ 	\\

($\rm km~s^{-1}$)		&($\rm km~s^{-1}$)		&		&($10^{-3}$ M\sun yr$^{-1}$)			\\

\hline \
18.3			& 1.8		&-0.6		& 9.9 \\

\hline \
\end{tabular}
\label{infall-tab}

\end{center}
\end{table}
%%%%%%%%%%%%%%%%%%%%%%%%%%%%%%%%%%%%%%%%%%%%%%%%%%%%%%%%%%%%%%%%
%%%%%%%%%%%%%%%%%%%%%%%%%%%% infall_grid %%%%%%%%%%%%%%%%%%%%%%%%%%
\begin{figure}
%\centering
%\hspace{0.3cm}
\includegraphics[scale=0.30]{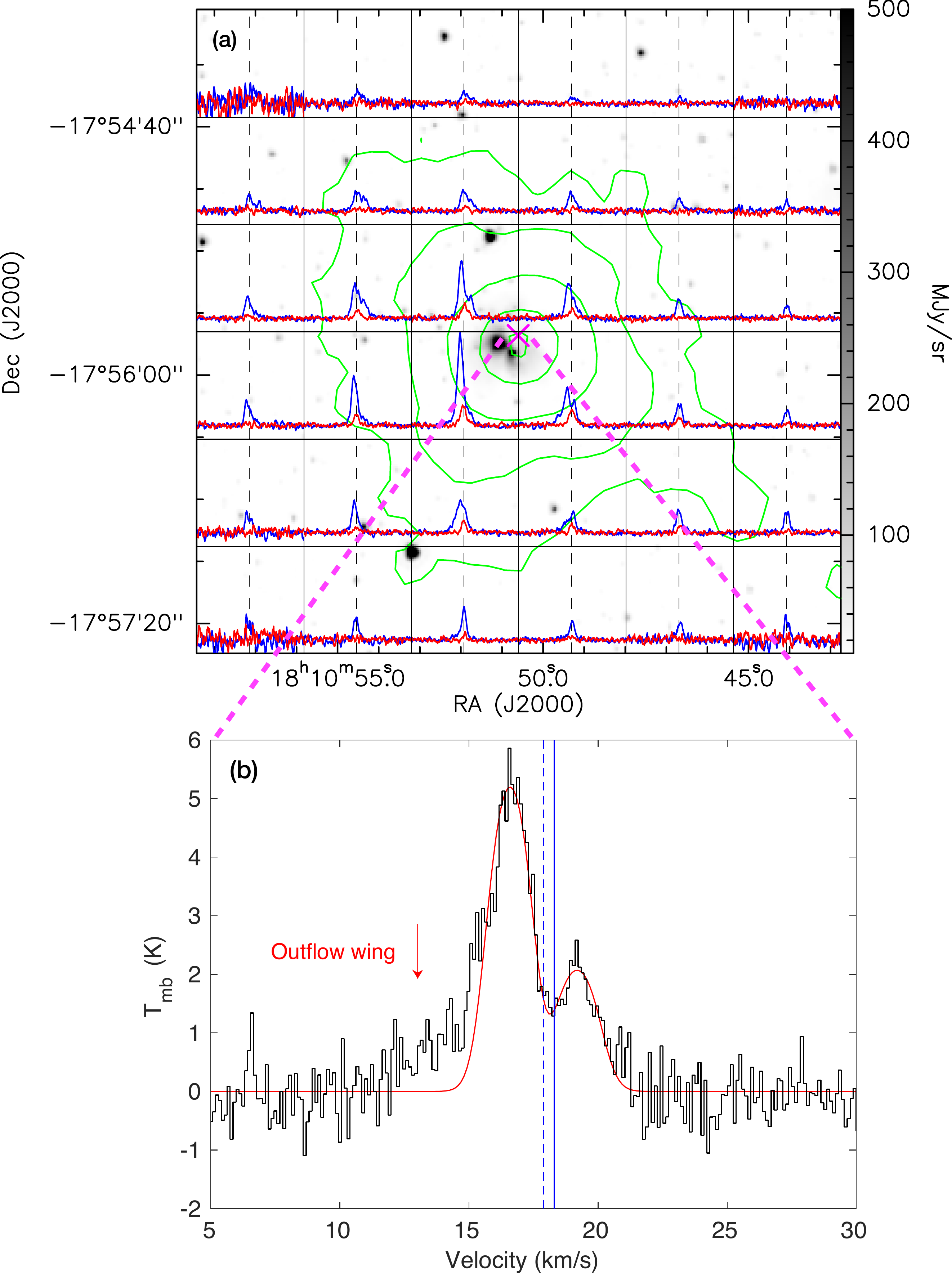}
\caption{(a) The grey scale shows the 4.5~{\um} map. The green contours represent the ATLASGAL 870~{\um} emission. The contour levels are 3, 9, 27, 72 and 108 times $\sigma$ ($ \sigma\sim 0.06$~Jy/beam). The $\rm HCO^+$ spectrum shown in blue and the $\rm H^{13}CO^+$ spectrum shown in red are overlaid. The spectra are boxcar-smoothed by five channels that corresponds to a velocity smoothing of $\rm 0.6~km~s^{-1}$. The dashed vertical lines indicate the LSR velocity estimated by averaging the peak positions of the $\rm H^{13}CO^+$ line in all the regions where the line is detected. The peak position of the 870~{\um} emission is marked by a magenta `x'. (b) The $\rm HCO^+$ spectrum extracted towards the ATLASGAL 870~{\um} peak . The best fit obtained using the `two-layer' model is shown in red. The solid blue line represents the LSR velocity, $\rm 18.3~km~s^{-1} $ derived from the optically thin $\rm H^{13}CO^+$ line and the dashed blue line represents the LSR velocity obtained from the model fit. The red arrow points to a blue-wing which could indicate a possible molecular outflow. }
\label{infall_grid}
\end{figure}
%%%%%%%%%%%%%%%%%%%%%%%%%%%%%%%%%%%%%%%%%%%%%%%%%%%%%%%5
%%%%%%%%%%%%%%%%%%%%%%%%%%%%%%%%%%%%%%%%%%%%%%%%%%%%%%%%%%%
\begin{table*}
\caption{Best fit parameters retrieved from the 
 model for the self-absorbed $\rm HCO^+$ line observed towards {\g12}.}
\begin{center}

%\hspace*{-1.2cm}
\begin{tabular}{c c c c c c c c} \hline \hline 

$\tau_0$ 		& $\Phi$ 	& $J_c$ 	& $J_f$		& $J_r$		& $ V_{\rm cont} $ 	& $\sigma$		& $V_{\rm rel}$ \\

											&&(K)		&(K)			&(K)				&($\rm km~s^{-1}$)	&($\rm km~s^{-1}$)	&($\rm km~s^{-1}$) \\

\hline \
1.3		& 0.3	&13.9		&7.4		&10.6		& 17.9		&0.7		&1.3 \\
\hline \
\end{tabular}
\label{two_layer}

\end{center}
\end{table*}
%%%%%%%%%%%%%%%%%%%%%%%%%%%%%%%%%%%%%%%%%%%%%%%%%%%%%%%%%%%%%%%%%%%%5

The double-peaked, blue-asymmetric $\rm HCO^+$ line profile with a self-absorption dip shown in {\fig}\ref{molecule_thick} is a characteristic signature of infall activity \citep{{2005ApJ...620..800D},{2006A&A...445..979P},{2013A&A...555A.112P},{2018ApJ...852...12Y}}. In order to probe the gas motion in the entire clump associated with {\g12}, we generate a grid map of the $\rm HCO^+$ line profile which is presented in  {\fig}\ref{infall_grid}(a). $\rm HCO^+$ line profiles are displayed in blue. For comparison, we also plot the optically thin transition, $\rm H^{13}CO^+$, in red. The grey scale map shows the 4.5~{\um} map emission with the ATLASGAL contours (in green) overlaid. The spectra shown here are averaged over regions gridded to an area given by the square of the beam size (36{\arc}) of the Mopra radio telescope. The $\rm HCO^+$ spectrum displays blue-skewed line profiles in all the grids within ATLASGAL contour revealing a strong indication of the clump in global collapse. 
For the molecular cloud to be collapsing, the gravitational energy of the cloud has to overcome the kinetic energy that supports it from collapsing. The gravitational stability of the cloud can be inspected using the virial parameter, $ \alpha_{\rm vir} = 5\sigma^2R/(GM_{\rm C}) $ which 
needs to be lower than unity \citep{1992ApJ...395..140B} for a collapsing cloud. $\sigma$ is the velocity dispersion which is taken from the FWHM of the optically thin $\rm H^{13}CO^+$ line and is estimated to be $\rm 1.2~km~s^{-1}$. Taking $R$ and $ M_{\rm C} $ as the radius and mass of the clump, C1, the virial parameter, $\alpha_{\rm vir}$ is calculated to be $\sim 0.9$. In comparison, \citet{2018ApJ...852...12Y} obtain a value of 0.58 for the EGO G022.04+0.22 and \citet{2011A&A...530A.118P} in their study of massive cores obtain values in the range $0.1-0.8$. Given the presence of infall and outflow activity, that could significantly increase the velocity dispersion, the derived estimate towards {\g12} is likely to be an overestimate.

\par To support the picture of protostellar infall, we estimate the infall velocity and mass infall rate. First, to quantify the blue-skewness of the $\rm HCO^+$ profile, we calculate the asymmetry parameter, $\delta V$, using the following expression \citep{2013RAA....13...28Y},
\begin{equation}
\delta V = \frac{(V_{\rm thick}-V_{\rm thin})}{\Delta V_{\rm thin}}
\end{equation}
Here, $\delta V$ is defined as the ratio of the difference between the peak velocities of the optically thick line, $V_{\rm thick}$ and the optically thin line, $V_{\rm thin}$, and the FWHM of the optically thin line denoted by $\Delta V_{\rm thin}$.
Using values of $ V_{\rm thin} = 18.3{\rm ~km~s^{-1}}$ and $\Delta V_{\rm thin}=2.9{\rm ~km~s^{-1}}$ from the Gaussian fit to the $\rm H^{13}CO^+$  line and $ V_{\rm thick} = 16.5{\rm ~km~s^{-1}}$, the peak of the blue component of the $\rm HCO^+$ line, $\delta V$ is estimated to be $-0.6$.
According to \citet{1997ApJ...489..719M}, the criteria for a bona fide blue-skewed profile is $\delta V < -0.25$. 
Furthermore, we estimate the mass infall rate ($\dot{M}_{\rm inf}$) of the envelope using the equation, $\dot{M}_{\rm inf}=4\pi R^2V_{\rm inf}\rho$ \citep{2010A&A...517A..66L}, where $V_{\rm inf}$ = $V_{\rm thin} - V_{\rm thick}$ = $V_{\rm H^{13}{CO}^+} - V_{\rm HCO^+}$ is the infall velocity and $\rho$ is the average volume density of the clump given by $\rho = M_{\rm C}/\frac{4}{3} \pi R^3$. The clump mass, $M_ {\rm C}$ and radius, $R$ are taken from Section \ref{clump_text}. The infall velocity, $V_{\rm inf}$ and the mass infall rate are estimated to be $\rm 1.8~km~s^{-1}$ and $\rm 9.9 \times 10^{-3}$ M\sun yr$^{-1}$, respectively. The mass infall rate estimate is higher compared to the value of $6.4 \times 10^{-3}$ M\sun yr$^{-1}$ derived by \citet{2015MNRAS.450.1926H}. As discussed in Section \ref{clump_text}, our clump mass and radius estimates are higher.  
Nevertheless, both the estimates fall in the range seen in other high mass star forming regions \citep{{2010ApJ...710..150C},{2010A&A...517A..66L},{2013MNRAS.436.1335L}}. 

\par To further understand the properties of the infalling gas, we extend our analysis and fit the  $\rm HCO^+$ line  with a `two-layer' model following the discussion in \citet{2013MNRAS.436.1335L}. Here, we briefly repeat the salient features of the model with a description of the equations and the terms. In this model, a continuum source is located in between the two layers, with each layer having an optical depth, $ \tau_0 $ and velocity dispersion, $ \sigma $, and an expanding speed, $ V_{\rm rel} $ with respect to the continuum source. This is the infall velocity introduced earlier.  $ V_{\rm rel} $ is negative if the gas is moving away and positive when there is inward motion. The brightness temperature at velocity, $ V $ is given by

\begin{multline}
\Delta T_{B}=(J_{f}-J_{cr})[1-exp(-\tau_{f})]\\
+(1-\Phi)(J_{r}-J_{b})\times[1-exp(-\tau_{r}-\tau_{f})]
\end{multline}
where
\begin{equation}
J_{cr}=\Phi J_{c}+(1-\Phi)J_{r}
\end{equation}
and
\begin{equation}
\tau_{f}=\tau_{0}exp\bigg[\frac{-(V-V_{\rm rel}-V_{\rm cont})^{2}}{2\sigma^{2}}\bigg]
\end{equation}
\begin{equation}
\tau_{r}=\tau_{0}exp\bigg[\frac{-(V+V_{\rm rel}-V_{\rm cont})^{2}}{2\sigma^{2}}\bigg]
\end{equation}

\noindent 
Here $J_{c}$, $J_{f}$, $J_{r}$, $J_{b}$ are the Planck temperatures of the continuum source, the ``front" layer, the ``rear" layer
and the cosmic background radiation, respectively. $J$ is the blackbody function at temperature, $T$ and frequency, $\nu$ and is expressed as
\begin{equation}
J=\frac{h\nu}{k}\frac{1}{exp (T_{0}/T)-1}
\end{equation}
\noindent
where $T_0=h\nu/k$, $h$ is Planck's constant, and $k$ is Boltzmann's constant. $\Phi$ and $V_{\rm cont}$ are the filling factor and systemic velocity (or the LSR velocity) of the continuum source, respectively. 
The $\rm HCO^+$ profile and the fitted spectrum (in red) are displayed in {\fig}\ref{infall_grid}(b). The LSR velocities determined from the model fit (dashed blue) and the optically thin transition of $\rm H ^{13}CO^+$ (solid blue) are also shown in the figure. The blue component of the $\rm HCO^+$ line shows a clear presence of broadened wing likely to be due to outflow. To avoid contamination from this outflow component, we restrict the velocity range between $\rm 16.1-21.0~km~s^{-1} $ while fitting the model. 
The model derived parameters are listed in {\tab}\ref{two_layer}. The model fitted values are fairly consistent (slightly smaller) with our previous estimates. 

\subsubsection{Outflow feature} \label{outflow}

%%%%%%%%%%%%%%%%%%%%%%%%%%%% moment %%%%%%%%%%%%%%%%%%%%%%%%%%
\begin{figure*}
\centering
%\vspace*{0.9cm}
%\hspace*{0.22cm}
\includegraphics[scale=0.20]{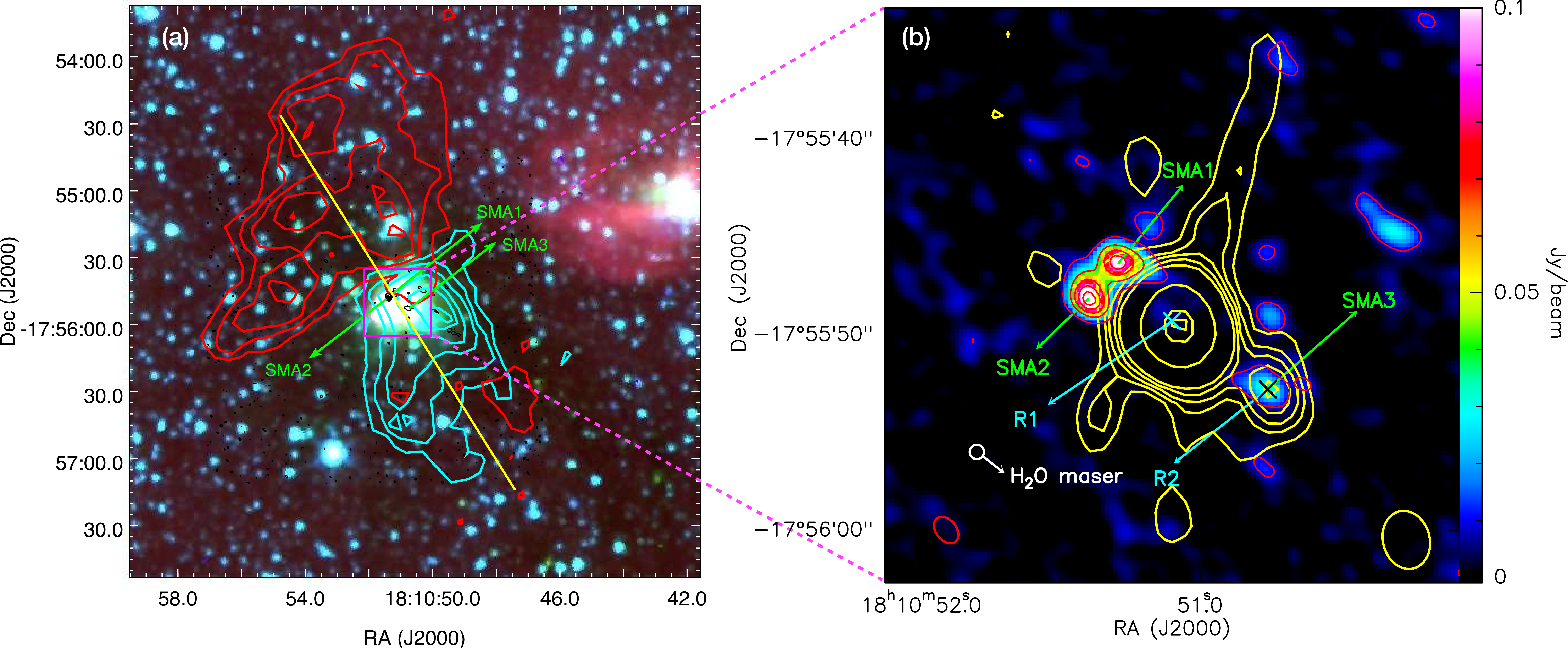} 
\caption{(a) The {\it Spitzer} IRAC colour composite image of {\g12}, overlaid with the SMA 1.1~mm  emission contours in black, with the contour levels same as in {\fig}\ref{FIR}(k). The $\rm ^{12}CO~(3-2) $ emission integrated from the peak of the blueshifted profile to the blue wings ($\rm 9.3-15.8~km~s^{-1} $) is represented using blue contours and from the peak of the red profile to the red wings ($\rm 20.8-27.3~km~s^{-1} $) is represented using red contours. The contours start from the $5~\sigma$ level for both the red and blue lobes and increases in steps of $3~\sigma$ and $4~\sigma$, respectively ($\rm \sigma=2.7~K~km~s^{-1}$ for red lobe and $\rm \sigma=2.3~K~km~s^{-1}$ for blue lobe ). The yellow line defines the cut along which the position-velocity (PV) diagram is made. The cut is selected in such a way that it passes through the red and blue lobes and also through the extended green emission. The red and blue lobes of the molecular outflow lie along a similar axis as the ionized jet. (b) The colour scale represents the 1.1~mm continuum emission from SMA observed towards {\g12} with the contour levels same as in {\fig}\ref{Dust_SED}(k). Radio emission at 1390~MHz is represented by yellow contours with the contour levels same as that in {\fig}\ref{radio}(a). The restoring beams of the 1390~MHz map and 1.1~mm map are indicated at the bottom- right and left of the image, respectively. The `x's indicate the positions of R1 and R2. The white circle marks the position of the $\rm H_2O$ maser in the vicinity of {\g12}.}
\label{outflow_moment}
\end{figure*}
%%%%%%%%%%%%%%%%%%%%%%%%%%%%%5%%%%%%%%%%%%%
%%%%%%%%%%%%%%%%%%%%%%%%%%%%pv %%%%%%%%%%%%%%%%%%%%%%%%%%
\begin{figure}
\centering
%\vspace*{0.9cm}
%\hspace*{-1.2cm}
\includegraphics[scale=0.21]{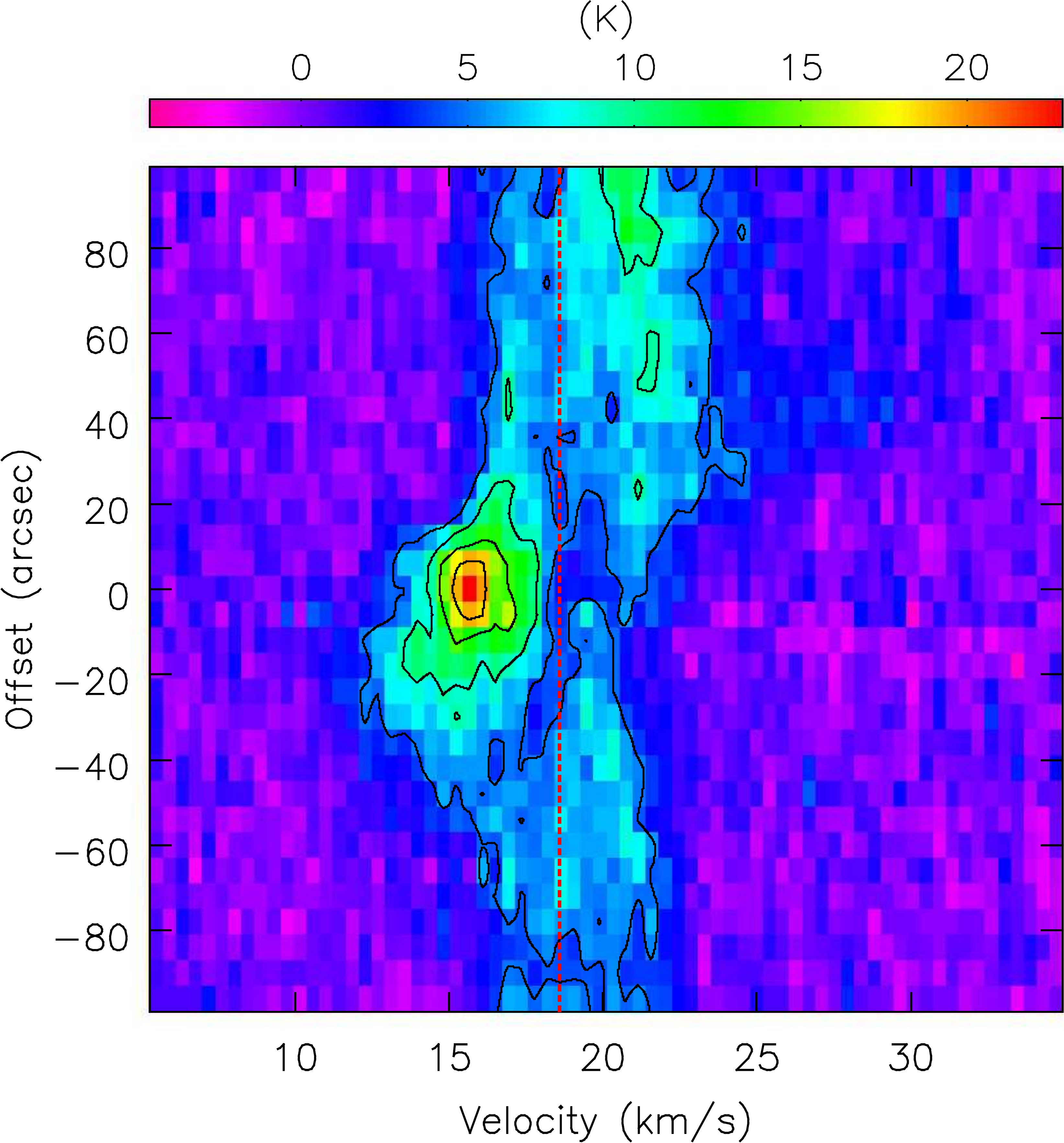} 
\caption{ The PV diagram of the $\rm ^{12}CO~(3-2) $ transition along the cut shown in yellow in {\fig}\ref{outflow_moment}(a) at a position angle of 32$^\circ$. The contour levels are 4, 9, 14 and 18 times $\sigma$ ($\sigma \sim 1.0~\rm K$). The zero offset in the PV diagram corresponds to the position of the central coordinate of {\g12} ($\rm \alpha_{J2000}= 18^{h}10^{m}51.1^s, \delta_{J2000} = -17\degree 55\arcmin 50\arcsec$). The LSR velocity, $\rm 18.3~km~s^{-1} $, is represented by the dashed red line. }
\label{pv}
\end{figure}
%%%%%%%%%%%%%%%%%%%%%%%%%%%%%5%%%%%%%%%%%%%
%%%%%%%%%%%%%%%%%%%%%%%%%%%% channel %%%%%%%%%%%%%%%%%%%%%%%%%%
\begin{figure}
\centering 
\includegraphics[scale=0.45]{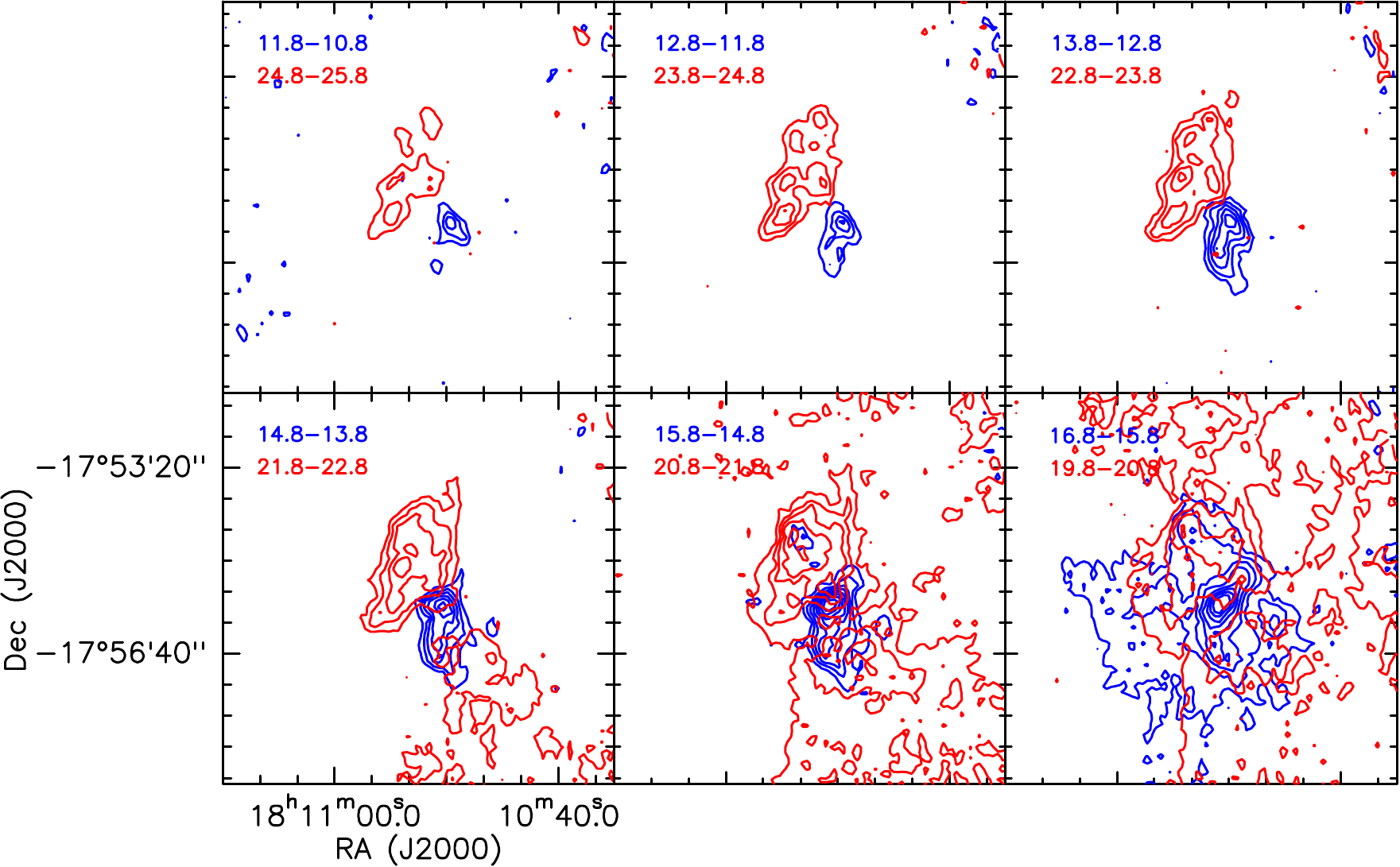} 
\caption{Channel maps of $\rm ^{12}CO~(3-2)$ line associated with molecular cloud harbouring {\g12}. Each box contains a pair of maps corresponding to the red- and blueshifted emission at the same offset from the LSR velocity. The channel widths are indicated at the top left of each map. The red contours correspond to the red wing and the blue contours correspond to the blue wing. The contours start from the 3$\sigma$ level of each map and increases in steps of 3$\sigma$.}
\label{channel_fwhm}
\end{figure}
%%%%%%%%%%%%%%%%%%%%%%%%%%%%%%%%%%%%%%%%%%%%%%%%%%%%5

Massive molecular outflows are ubiquitous in star forming regions \citep{2002A&A...383..892B} and often co-exist with ionized jets \citep[e.g.][]{{1996ASPC...93....3A},{2016MNRAS.460.1039P}}. The jets are believed to entrain the gas and dust from the ambient molecular cloud, thus driving molecular outflows.  
According to several studies, broad wings of the optically thick lines like $\rm HCO^+$ are well accepted signatures of outflow activity \citep[e.g.][]{{2007ApJ...663.1092K},{2010A&A...520A..49S},{2013ApJ...771...24S}}.
As mentioned in the previous section, broadening of the blue wing of the infall tracer line of $\rm HCO^+$ is seen in {\g12}. Given the association with an EGO and the alignment with a large scale CO outflow features, the origin of the broad blue wing can be attributed to be due to the outflow. Alternate scenarios like unresolved velocity gradients \citep{2014A&A...565A.101T} or gravo-turbulent fragmentation \citep{2005ApJ...620..786K} have been invoked for broadened wings but are less likely to be the case here. 
In this section, we focus on the   
rotational transition lines of CO that are well known tracers of molecular outflow, and investigate the outflow kinematics of the molecular cloud associated with {\g12} using the archival data of the isotopologues of $\rm CO~(3-2)$ transition from JCMT and $\rm ^{13}CO~(1-0)$ observation from TRAO. 

\par The red and blueshifted velocity profiles of the CO transitions shown in {\fig}\ref{CO_plot}(a) can be attributed to emission arising from distinct components of the CO gas that are moving in opposite directions away from the central core. We note that the peaks have different shifts with respect to the LSR velocity with the $\rm ^{12}CO~(3-2)$ line showing the maximum shift and  $\rm C^{18}O~(3-2)$ line has the minimum shift.
The peaks of the red component of $\rm ^{12}CO~(3-2)$, $\rm ^{13}CO~(3-2)$ and $\rm C^{18}O~(3-2)$ transitions are shifted by 2.5, 1.6 and $\rm 1.0~km~s^{-1}$ from the LSR velocity. For the blue component the shifts are 2.5, 1.7 and $\rm 0.9~km~s^{-1}$, respectively. $\rm ^{12}CO$ molecule, having the lowest critical density among the three, effectively traces the outer envelope of the molecular cloud, hence showing the maximum shift and the $\rm C^{18}O$ molecule, the densest among the three species is a tracer of the dense core of the molecular cloud and thus shows the minimum shift. 

\par In order to map the outflow in the vicinity of {\g12}, we construct the zeroth moment map of the two components using the task, {\tt IMMOMENTS} in CASA. The zeroth moment map is the integrated intensity map that gives the intensity distribution of a molecular species within the specified velocity range. The $\rm ^{12}CO~(3-2)$ emission is integrated from the peak of the blueshifted profile to the blue wings that corresponds to the lower velocity channels ranging from $\rm 9.3-15.8~km~s^{-1} $ for the blue component and from the peak of the redshifted profile to the red wing that corresponds to the higher velocity channels ranging from $\rm 20.8-25.3~km~s^{-1} $ for the red component.
The contours are shown overlaid on the {\it Spitzer} IRAC colour composite image in 
{\fig}\ref{outflow_moment}(a). The figure reveals the presence of two distinct, spatially separated red and blue lobes. High-velocity gas is also seen towards the tail of the blue component. The location of the 1.1~mm dense cores, SMA1, SMA2 and SMA3 are also marked in this figure. The central part covering the brightest portion of the IRAC emission (location of the EGO), is shown in {\fig}\ref{outflow_moment}(b) with the spatial distribution of the ionized gas overlaid on the 1.1~mm dust emission. To corroborate with the zeroth moment map showing the outflow lobes, in {\fig}\ref{pv}, we show the position-velocity (PV) diagram constructed along the outflow direction (position angle of $\rm \sim 32^\circ$; east of north) highlighted in {\fig}\ref{outflow_moment}(a). The direction along which the PV diagram is made is chosen such that both the red and blue lobes are sampled and it also covers the region of extended 4.5~{\um} emission of {\g12}.  The zero offset in the PV diagram corresponds to the position of the central coordinate of the EGO, {\g12} ($\rm \alpha_{J2000}= 18^{h}10^{m}51.1^s, \delta_{J2000} = -17\degree 55\arcmin 50\arcsec$). As expected, the PV diagram also clearly reveals distinct red and blue components of the $\rm ^{12}CO~(3-2)$ emission from the LSR velocity of the cloud represented by a red dashed line. Towards the lower region of the PV diagram we can trace a weaker redshifted $\rm ^{12}CO~(3-2)$ component consistent with the high-velocity tail seen in the zeroth moment map.  

\par To further probe the velocity structure of the cloud associated with {\g12}, we generate channel maps of the $\rm ^{12}CO~(3-2) $ emission following the method outlined in \citet{2014A&A...565A..34S}. 
To define suitable velocity ranges and identify the blue and redshifted outflow emission, we set the inner limits of the velocity at $\sim V_{\rm LSR} \pm {\rm FWHM}/2$, where FWHM is the $\rm H^{13}CO^+$  linewidth. Taking the offset to be  $\rm \pm 1.5~km~s^{-1}$ from the LSR velocity, the inner velocity limit for the red and the blueshifted lobes are estimated to be $\rm 19.8~km~s^{-1}$ and  $\rm 16.8~km~s^{-1}$, respectively. The channel maps constructed are shown in {\fig}\ref{channel_fwhm}. Each grid displays a pair of maps with a velocity width of $\rm 1~km~s^{-1}$. 
Prominent outflow features begin to appear at velocities $\rm \sim 13.8$ and $\rm 21.8~km~s^{-1}$ from the blue and red components, respectively. Beyond these velocities there is no contribution from the central core. Closer to the LSR velocity, the emission is rather complex making it difficult for the outflow features to be discernible. This is understandable since near the LSR velocity, the emission from the outflow components is likely to be contaminated by the infall motion and contribution from the diffuse gas. The channel maps are comparable to those presented by \citet{2009ApJ...696...66Q} for the region G240.31+0.07 with similar complex velocity structure near the LSR. The channel maps also show the presence of a redshifted component between velocities  $\rm 20.8-22.8~km~s^{-1}$ overlaping with the blue lobe towards the south-west. As will be discussed later in Section \ref{hub_filament}, such a velocity distribution can be indicative of accretion through filaments.  

\par Morphologically, the spatially separated red and blue lobes associated with {\g12} resembles a wide-angle bipolar outflow seen in the star forming region G240.31+0.07 studied by \citet{2009ApJ...696...66Q}. These authors suggest the wide-angle bipolar outflow as the ambient gas being swept up by an underlying wide-angle wind and is driven by one of the three mm peaks located close to the geometric centre of the bipolar outflow. 
Only a handful of studies have found the presence of wide-angle bipolar molecular outflows associated with high mass star formation \citep[e.g.][]{{1998ApJ...507..861S},{2009ApJ...696...66Q}}. 
\citet{1998ApJ...507..861S} investigate the likely driving source of the poorly collimated molecular outflow associated with G192.16. Coexistence of wide-angle CO outflow with shock-excited {\h2} emission, prompted them to conclude that the G192.16 outflow is powered by the combination of a disk-wind and a jet. 
Given the likely association of {\g12} with an ionized jet supported by the presence of shock-excited NIR {\h2} lines, we propose a similar picture of coexistence of disk-wind and a jet, where the wide-angle bipolar CO outflow is likely to be driven by the underlying wide-angle wind. 

\par Of crucial importance is the identification of the driving source for this outflow. \citet{2008A&A...488..579M} elucidates about the star forming region IRAS~18151-1208, where two detected bipolar outflows are shown to be powered by two mm sources, MM1 and MM2. They have also detected a third mm core, MM3 that does not show any outflow activity. \citet{2002A&A...383..892B}, in their statistical study of massive molecular outflows, state that a large fraction of their sample show bipolar outflow and these are seen to be associated with massive mm sources and in most cases, are centred on the mm peaks. 
As seen in {\fig}\ref{outflow_moment}, two mm cores (SMA1 and SMA2) are located towards the centroid of the bipolar outflow associated with {\g12}. These are shown to be potential high-mass star forming cores,  Further, the absence of radio emission and IR sources imply a very early evolutionary phase. We further investigate whether SMA1 and/or SMA2 are the powering sources of the CO bipolar outflow in {\g12}.   

 \par Following \citet{2009ApJ...696...66Q}, the dynamical timescale of the outflow seen associated with {\g12} is computed using the expression, $T_{dyn} = L_{flow}/v_{max} $, where $ L_{flow} \sim 1.2~\rm pc$ is the half length of the end-to-end extension of the flow and $v_{max} \sim 9.0~\rm km~s^{-1}$ is the maximum flow velocity from the LSR velocity of the cloud. This yields a dynamical timescale of the outflow that is $\rm 1.3 \times 10^5$~yr. Comparing with the results obtained by \citet{2009ApJ...696...66Q}, \citet{1998ApJ...507..861S} and \citet{2003ApJ...584..882S}, our estimated value is in agreement with massive molecular outflow from an UC {\hii} region. This result supports our unfolding picture of radio component R1, where coexistence of an UC {\hii} region and an ionized thermal jet is seen and the large-scale CO outflow can also be attributed to the UC {\hii} region. However, it should also be kept in mind that unlike \citet{2009ApJ...696...66Q}, there is no SMA core coinciding with R1. Moreover, their results are based on interferometric observations, whereas, the JCMT CO outflow data used here are from single-dish measurements. In these single-dish observations, the inner outflow jets are mostly unresolved and the measurements are less sensitive to high-velocity outflow emission resulting in an overestimation of the dynamical ages. Hence, one cannot rule out the possibility of SMA1 and/or SMA2 being the outflow driving sources similar to the case of AFGL 5142 \citep{2016ApJ...824...31L}. If indeed the binary cores, SMA1 and SMA2 are the outflow driving cores, then the nature of the radio emission in R1 (and R2) can be thought of as ionized jet emission driven by the mm cores which also drives the large scale CO outflows detected. Possibility of multiple outflows from the mm cores and the UC {\hii} region also exists \citep{{2002A&A...387..931B},{2003A&A...408..601B}}. This advocates for high-resolution observations for a better understanding.

\subsubsection{Hub-filament system} \label{hub_filament}
%%%%%%%%%%%%%%%%%%%%%%%%%%%%hub-filament%%%%%%%%%%%%%%%%%%%%%%%%%%
\begin{figure}
\centering 
\includegraphics[scale=0.35]{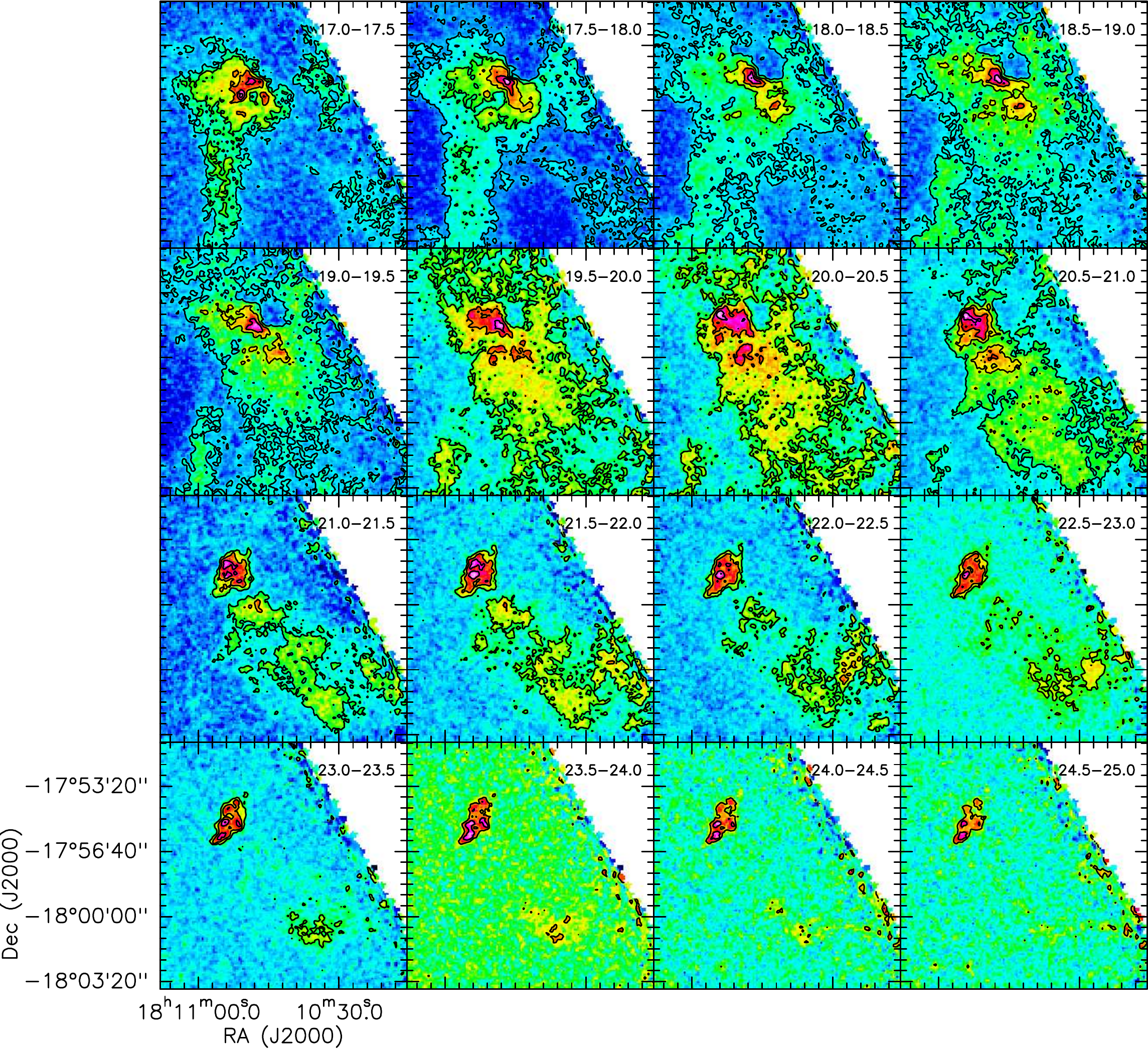} 
\caption{Channel maps of $\rm ^{12}CO$ emission is shown here with each channel having a velocity width of $\rm 0.5~km~s^{-1} $. For each map, the black contours represent the $\rm ^{12}CO$ emission starting from the $3\sigma$ level and increasing with a step of $3\sigma$.}
\label{filament}
\end{figure}
%%%%%%%%%%%%%%%%%%%%%%%%%%%%%%%%%%%%%%%%%%%%%%%%%%%%%%%%
%%%%%%%%%%%%%%%%%%%%%%%%%%%%hub-filament%%%%%%%%%%%%%%%%%%%%%%%%%%
\begin{figure}
\centering 
\includegraphics[scale=0.35]{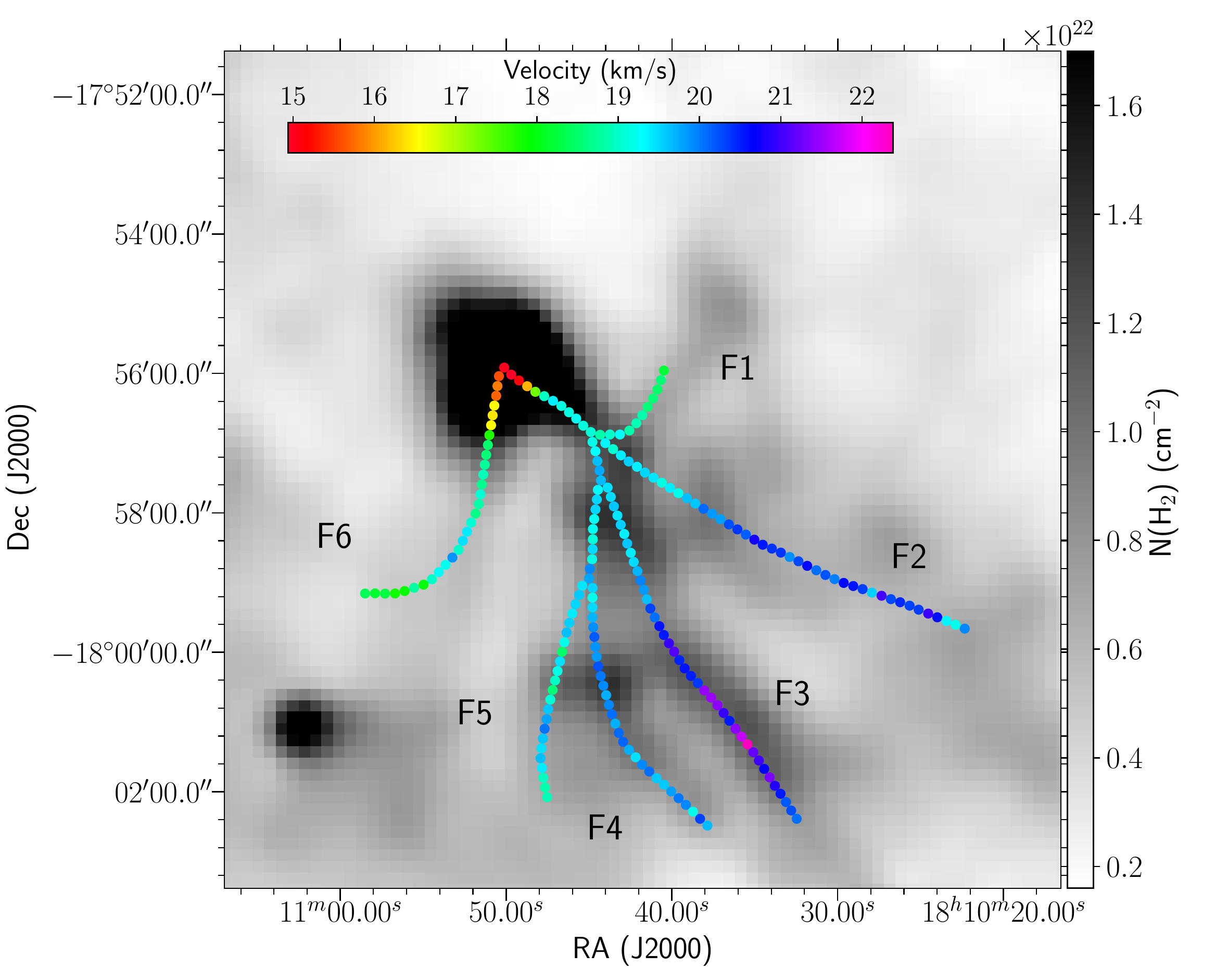} 
\caption{The velocity peaks of $\rm ^{12}CO~(3-2) $ extracted along the filaments is overlaid on the column density map of the region associated with {\g12} shown in grey scale. The positions of all the filaments are also labelled. }
\label{velocity_filament}
\end{figure}
%%%%%%%%%%%%%%%%%%%%%%%%%%%%%%%%%%%%%%%%%%%%%%%%%%%%%%%%

From the above discussion we hypothesize a picture of global collapse of the molecular cloud harbouring the EGO, {\g12}. Interestingly, {\fig}\ref{FIR}(d) and (h), unfold the presence of large scale filamentary structures along the south-west direction of {\g12}, all merging at the location of the clump, C1 enveloping {\g12}. As mentioned earlier, the concurrence of a collapsing cloud with converging filaments suggest a hub-filament system. In literature, hub-filament systems are common in sites of high-mass star formation \citep[e.g.][]{{2013A&A...555A.112P},{2016ApJ...824...31L},{2018ApJ...852...12Y}}. In such systems, converging flows are detected where matter funnels in through the filaments into the hub, where accretion is most pronounced. Morphologically, the molecular cloud system associated with {\g12} resembles the hub-filament system associated with the star forming region, G22 \citep{2018ApJ...852...12Y} and the IRDC, SDC335 ($\rm SDC335.579-0.292$) \citep{2013A&A...555A.112P}.
\par To delve deeper into this picture, we investigate the velocity structure of the filaments. To proceed, we construct the channel maps of the $\rm ^{12}CO$ emission which are illustrated in {\fig}\ref{filament}. The velocity ranges are selected by examining the JCMT $\rm ^{12}CO$ data cube and choosing the range where the $\rm ^{12}CO$ emission is detected. The velocity width of each channel is chosen to be $\rm 0.5~km~s^{-1}$ similar to that used by \citet{2019A&A...621A.130L} to investigate the $\rm C^{18}O~(2-1)$ emission associated with IRDC, G351.776-0.527.
The spatial coincidence of the $\rm ^{12}CO$ emission with the filaments is remarkable. The gas associated with the filaments is consistently redshifted with respect to the LSR velocity of the clump, C1. The velocity of the cloud along the filaments peak in the velocity range of $\rm \sim18-23~km~s^{-1}$. The variation in velocity suggests bulk gas motion along the filaments. 
From the channel maps we can see that the velocity of the molecular gas along the filament decreases as it approaches the central core, with the maximum velocity at the south-west end of the filament. It has to be noted here that the $\rm ^{12} CO~(3-2)$ transition is also a tracer of molecular outflow and as shown in Section \ref{outflow}, {\g12} is also an outflow source. Hence, the decrease in velocity near the clump, C1 may be attributed to the interaction with the molecular outflow. 
\par To further elaborate on the velocity structure, we extract the spectra of $\rm ^{12}CO~(3-2)$ along the filaments with a step size of half the angular resolution ($ \sim 15''/2$) of the JCMT-HARP observation. The peak velocities estimated by fitting 1D Gaussian profiles to the spectra are shown in {\fig}\ref{velocity_filament} as colour-coded circles overlaid on the {\h2} column density map. Along the filaments, F1 and F6 it can be seen that the velocity is within the range $ \rm 17-19~km~s^{-1} $, closer to the LSR velocity, and increasing towards the clump, C1. However, along the filaments F2, F3, and F4 the velocity is on the higher side of the LSR velocity, ranging from $ \rm 19-21~km~s^{-1} $ and decreasing towards the central clump, C1. But in the case of filament, F5, we do not clearly notice any velocity gradient, and the values are seen to vary in the range $ \rm 19-20~km~s^{-1} $. Similar velocity gradients along filaments are detected in star forming regions such as SDC335 \citep{2013A&A...555A.112P}, G22 \citep{2018ApJ...852...12Y} and AFGL 5142 \citep{2016ApJ...824...31L}. Following these authors, we also attribute the velocity gradients to be due to gas inflow through filaments. 
A number of other mechanisms have also been proposed, which include filamentary collapse, filament collision, rotation, expansion and wind-acceleration to explain the observed velocity distributions \citep{2014A&A...561A..83P}. 
If the velocity distribution were to be explained by expansion scenario, we should have observed red-skewed velocity profiles of optically thick lines ($\rm HCO^+$, HCN and HNC) towards the clump, C1 \citep{2018ApJS..234...28L}. On the contrary, we observe blue-skewed velocity profiles, hence rendering the expansion picture to be unlikely. Further, we do not observe any cloud collision signature of enhanced line-widths at the junction \citep{2018ApJ...852...12Y}. The molecular outflow likely to be driven by the thermal jet in the clump, C1, cannot possibly explain the red-skewed velocity along the filaments, $ \rm F2-F5 $, since these fall along the blue lobe of the detected outflow. Also, in general, outflows are located between filaments \citep{2016ApJ...824...31L}. Hence, on comparison with earlier studies and the lack of evidence to prove otherwise, we infer that the observed velocity gradient is a result of gas inflow along the filaments, although there could be contribution from the outflowing gas. Nonetheless, $\rm ^{12}CO~(3-2)$ being an optically thick molecular line transition, cannot effectively probe the velocity distribution within the filaments. Thus, high resolution observations of optically thin lines would give a better picture. 
As discussed in Section \ref{clump_text}, 11 eleven clumps, $ \rm C2-C12$, lie along the identified filaments. It is also seen that along the filaments the dust temperature is lower than C1. The mass estimates show that these are less massive than the central core (clump, C1). \citet{2018ApJ...852...12Y}, in the study of the hub-filament system associated with G22, finds two clumps more massive than the other clumps along the filament. These clumps,that dominate the emission at wavelengths longer than 24~{\um} are the most active star forming regions in G22. This concurs well with {\g12}, where the most massive clump, C1, is an active star forming clump.  

\section{SUMMARY} \label{summary}

We carried out a comprehensive multiwavelength study towards the EGO, {\g12} and its associated environment. Our main results are summarized as follows

\begin{enumerate}

\item The radio continuum emission mapped at 1390~MHz reveals a linear structure extended in the north-east and south-west direction with the presence of two compact radio components, R1 and R2 that are unresolved at 610~MHz. The peak emission at 610~MHz is coincident with the component R1.

\item We explore different scenarios to explain the nature of the ionized emission. Under the UC {\hii} framework, assuming the emission at 1390~MHz to be optically thin, the observed Lyman continuum flux translates to an ionizing source of spectral type of $\rm B1-B0.5$. An alternative picture of ionized thermal jet is examined, given various observed characteristics including the spectral index values of $0.3-0.9$ in the region. We are prompted to consider the co-existence of the UC {\hii} region with an ionized jet being powered by the same YSO.  IRAS~18079-1756 (2MASS J18105109-1755496), a deeply embedded Class II YSO is likely to be the driving source.

\item Presence of shock-excited {\h2} and [FeII] line emission is confirmed from NIR narrow-band imaging and spectroscopy in concurrence with the jets/outflows picture.

\item A massive central clump, C1 is identified from the 870~{\um} map which envelopes the detected radio and the enhanced and extended 4.5~{\um} emission. The clump has a mass 1375~M{\sun} and total luminosity $\rm 2.8 \times 10^4~L_\odot$. Two-component modeling shows the presence of an inner warm component
surrounded by an extended outer, cold envelope traced mostly by the FIR wavelengths

\item Seven molecular species from the MALT90 survey are detected towards the EGO, {\g12}. The optically thick lines, $\rm HCO^+$ and HCN show signatures of protostellar infall. From the blue-skewed profile of the $\rm HCO^+$ line, infall velocity and mass infall rate are estimated to be $\rm 1.8~km~s^{-1}$ and $\rm 9.9 \times 10^{-3}~M_\odot yr^{-1}$. 

\item From the line observations of the $J=3-2$ transition of the molecular species $\rm ^{12}CO$, $\rm ^{13}CO$ and $\rm C^{18}O$, we detect the presence of a wide-angle bipolar outflow. From the dynamical age, $\rm 1.3 \times 10^5~yr$, of the bipolar outflow it seems likely that the UC {\hii} drives the same though the possibility of the SMA cores (SMA1 and SMA2) being the powering source(s) cannot be ruled out. 

\item Signature of a hub-filament system is seen in the 8.0~{\um} and FIR images and is supported by the constructed column density and dust temperature maps. A detailed study of the gas kinematics agrees with bulk motion in the filaments and suggest a likely picture of gas inflow along the filaments to C1.   

\item A conjectured hypothesis of the EGO, {\g12}, satisfying the multiwavelength observations, could be an active star forming complex where very early evolutionary cores (SMA1 and SMA2) are seen. Apart from this, an accreting (likely through filaments) MYSO in an initial phase of an UC {\hii} region and driving a large-scale molecular outflow entrained by a likely ionized thermal jet is detected. 

\end{enumerate}

\section*{ACKNOWLEDGEMENTS}
We thank the referee for critically going through the manuscript and giving valuable suggestions. We thank the staff of the GMRT, who made these observations possible. GMRT is run by the National Centre for Radio Astrophysics of the Tata Institute of Fundamental Research. We also thank the staff of UKIRT for their assistance in the observations. UKIRT is owned by the University of Hawaii (UH) and operated by the UH Institute for Astronomy. When some of the data reported here were acquired, UKIRT was supported by NASA and operated under an agreement among the University of Hawaii, the University of Arizona, and Lockheed Martin Advanced Technology Center; operations were enabled through the cooperation of the East Asian Observatory. This research made use of NASA/IPAC Infrared Science Archive, which is operated by the Jet Propulsion Laboratory, Caltech under contract with NASA. This publication also made use of data products from {\it Hersche}l (ESA space observatory) and the Millimetre Astronomy Legacy Team 90 GHz (MALT90) survey. We also made use of the ATLASGAL data products. The ATLASGAL project is a collaboration between the Max-Planck-Gesellschaft, the European Southern Observatory (ESO) and the Universidad de Chile.

\bibliography{reference}

\end{document}